\documentclass[aps,twocolumn,nofootinbib]{revtex4}

\usepackage{graphicx}
\usepackage{dcolumn}
\usepackage{bm}

\usepackage{hyperref}
\hypersetup{
    colorlinks=true,
    linkcolor=blue,
    citecolor=blue,
    urlcolor=blue
}

\usepackage{amssymb,amsfonts}
\usepackage{epsfig}
\usepackage{epstopdf}
\usepackage[all]{xy}
\usepackage{amsthm}
\usepackage{dcolumn}
\usepackage{hyperref}
\usepackage{url}
\usepackage{dsfont}
\usepackage{slashed}
\usepackage{mathrsfs}

\usepackage{bbm}
\usepackage{tikz}
\usepackage{physics}
\usepackage{enumitem}
\usepackage{soul}
\usepackage{float}
\usepackage{graphicx}
\usepackage{booktabs}

\newcommand{\be}{\begin{equation}}
\newcommand{\ee}{\end{equation}}
\newcommand{\bea}{\begin{eqnarray}}

\newcommand{\eea}{\end{eqnarray}}

\def\inbar{\,\vrule height1.5ex width.4pt depth0pt}
\def\IR{\relax{\rm I\kern-.18em R}}
\def\IC{\relax\hbox{$\inbar\kern-.3em{\rm C}$}}

\begin{document}

\title{de Sitter Scalar Discrete Series: Gupta-Bleuler Structure and Holography}

\author{Maryam Bajalan$^{1}$\footnote{maryam.bajalan@math.bas.bg}}

\author{Jean-Pierre Gazeau$^{2,3}$\footnote{gazeau@apc.in2p3.fr, j.gazeau@uwb.edu.pl}}

\author{Hamed Pejhan$^{1}$\footnote{pejhan@math.bas.bg}}

\affiliation{$^1$Institute of Mathematics and Informatics, Bulgarian Academy of Sciences, Acad. G. Bonchev Str. Bl. 8, 1113, Sofia, Bulgaria\\
$^2$Universit\'e Paris Cit\'e, CNRS, Astroparticule et Cosmologie, F-75013 Paris, France\\$^3$Faculty of Mathematics, University of Bia\l ystok, 15-245 Bia\l ystok, Poland}

\date{\today}

\begin{abstract}
    {We show that scalar discrete-series unitary irreducible representations (UIRs) $\Pi_{p,0}$ ($p=1,2,\cdots$) of the de Sitter (dS) group $\mathrm{SO}_0(1,4)$ admit a dS-covariant Krein realization on the dS hyperboloid, endowed with a dS-invariant non-degenerate Klein-Gordon (KG) sesquilinear form, in which the group action is indecomposable and organizes naturally into a Gupta-Bleuler triplet. The positive- and negative-norm sectors are already present in the underlying Krein space, whereas a null sector emerges only at an intermediate stage, where the induced KG form becomes degenerate and its radical leads canonically to the physical quotient carrying the UIR $\Pi_{p,0}$. We further show that suitable limits of the bulk theory at the ``future'' and ``past'' conformal boundaries ${\mathcal{I}}^\pm$ give rise to dS-invariant boundary realizations endowed with induced kernel inner products. While the bulk negative-norm sector admits no independent boundary counterpart, the boundary realization retains the physical and gauge structures inherited from the bulk. The resulting boundary module nevertheless remains indecomposable, with its physical quotient carrying the discrete-series representation $\Pi_{p,0}$. The antipodal symmetry provides a natural relation between the realizations on ${\mathcal{I}}^+$ and ${\mathcal{I}}^-$, ensuring the consistency of the boundary construction and its geometric interpretation. At the heart of the analysis lies a Fourier-type bulk-boundary transform that provides a dS-covariant identification of the bulk and boundary physical sectors, establishing a one-to-one intertwining correspondence between the bulk and boundary realizations of $\Pi_{p,0}$ while preserving reflection positivity.}
\end{abstract}

\maketitle


\setcounter{equation}{0} 
\section{Introduction}

Elementary systems in dS spacetime should be defined not by analogy with flat Minkowski spacetime, but intrinsically through the UIRs of the dS relativity group $\mathrm{SO}_0(1,4)$. This is Wigner's principle in dS form \cite{Wigner1939, Wigner1949, Gazeau2022, GazPejTod2026}. Throughout this work, we adopt the global representation-theoretic framework of Refs. \cite{Gazeau2022, GazPejTod2026}, in which elementary systems in dS and anti-dS spacetimes are canonically constructed from the UIRs of their respective relativity groups. Here, we focus on a particularly subtle scalar family within the dS discrete series UIRs.

The representation theory of $\mathrm{SO}_0(1,4)$ is known \cite{Gazeau2022, GazPejTod2026, Thomas, Newton, Takahashi, Dixmier, Martin1974, Ottoson, Schwarz, Letsios-2025, Lipsman-1974} to organize the dS UIRs into three main families: the principal, complementary, and discrete series, the latter of which constitutes the square-integrable sector. In addition to these genuine series representations, one encounters certain limiting representations that are traditionally grouped with the discrete series and are often referred to as scalar members of the discrete family. These representations occupy a distinguished borderline position in the classification; they lie at the edge of the square-integrable sector and therefore do not share all the structural features of the genuine discrete series (see, for instance, Ref. \cite{Lipsman-1974}).

In the notation adopted here, these scalar representations are denoted by $\Pi_{p,0}$, with $p=1,2,\cdots$. We show that they admit a dS-covariant Krein realization on the dS hyperboloid, carried by a dS-invariant Krein space endowed with a non-degenerate Hermitian KG sesquilinear form of indefinite signature. The dS action on this space is indecomposable and gives rise naturally to a Gupta-Bleuler triplet, namely a hierarchy of invariant subspaces rather than a dS-covariant decomposition into Hilbert and anti-Hilbert sectors. Within this hierarchy, a null sector emerges at an intermediate stage when the restriction of the KG sesquilinear form becomes degenerate; its radical is canonically identified with the gauge sector, and the corresponding quotient yields the physical realization space carrying the representation $\Pi_{p,0}$. In this way, the null sector is not introduced as an external gauge artifact, but arises intrinsically from the restricted Krein structure and constitutes an essential ingredient in the dS-covariant construction of the physical quotient representation.

We further show that the scalar discrete-series family considered here also admits a natural holographic realization. The idea of relating a quantum theory in dS spacetime to a Euclidean theory on its conformal boundary was originally motivated by the dS/CFT proposal \cite{Strominger2001, Witten2001}. A central difficulty in this setting is that the standard Euclidean boundary construction does not, in general, admit a reflection-positive inner product \cite{Strominger2001, Bros2002}. Since elementary systems in dS spacetime are described, in Wigner's sense, by UIRs of the dS group, this issue is fundamentally representation-theoretic as well as quantum-field-theoretic; a satisfactory dS holographic framework should relate bulk and boundary realizations of the relevant representations in a manner compatible with their invariant inner products. This viewpoint has motivated recent developments emphasizing the group-theoretical structure of the dS group and its role in the problem of reflection positivity \cite{GazeauOlmoPejhan2024, 0001, 0002, 0003}. A concrete realization of this idea was achieved in Ref. \cite{GazeauOlmoPejhan2024}, where a bulk-boundary correspondence was established for scalar principal-series representations.

The present work extends this framework to the scalar discrete-series family $\Pi_{p,0}$, $p=1,2,\cdots$, whose bulk realization exhibits a structure fundamentally different from that of the principal series. We show that suitable limits of the bulk modes at the future and past conformal boundaries ${\mathcal{I}}^{\pm}$ give rise to boundary realizations endowed with induced kernel inner products. While the bulk negative-norm sector admits no independent boundary counterpart, the boundary realization retains the physical and gauge structures inherited from the bulk and remains indecomposable. Its physical quotient furnishes a unitary irreducible realization of $\Pi_{p,0}$, thereby providing a boundary realization of the same representation in a substantially simpler setting. Furthermore, antipodal symmetry establishes a natural relation between the realizations on ${\mathcal{I}}^{+}$ and ${\mathcal{I}}^{-}$, showing that they furnish equivalent boundary realizations of the same representation. At the heart of this construction lies a Fourier-type bulk-boundary transform that provides a dS-covariant identification of the bulk and boundary physical sectors, establishing a one-to-one intertwining correspondence between the bulk and boundary realizations of $\Pi_{p,0}$ while preserving reflection positivity. The resulting framework, therefore, realizes the scalar discrete series simultaneously in the bulk and at conformal infinity, with the two physical realizations explicitly linked by the holographic transform.

The remainder of the paper is organized as follows. Section \ref{Sec. Machinery} reviews the necessary dS geometry, the relevant group-theoretical framework, and the representation-theoretic preliminaries underlying the analysis. Section \ref{Sec. Modes} derives the scalar discrete-series wave equation from the quadratic dS Casimir operator, constructs its mode solutions, and studies the associated KG sesquilinear structure. It then develops the corresponding dS-covariant Krein realization, exhibits the resulting indecomposable representation of the dS group, and shows how its invariant subspace structure gives rise naturally to a Gupta-Bleuler triplet whose physical quotient carries the UIR $\Pi_{p,0}$. Section \ref{Sec. Boundary} constructs the induced boundary realizations on the conformal boundaries ${\mathcal{I}}^{\pm}$, establishes their dS-invariant kernel inner products and indecomposable structures, and analyzes the antipodal correspondence between the future and past boundaries. Section \ref{Sec. Fourier} establishes a Fourier-type bulk-boundary transform and the resulting dS-covariant holographic correspondence between the physical sectors realizing the UIR $\Pi_{p,0}$ in the bulk and on the boundary. Finally, Section \ref{Sec. conclusion} summarizes the main results and discusses their implications. Technical derivations and supplementary calculations are collected in five appendices.

The main conventions of our notation are:
\begin{enumerate}
    \item{We set $\mathbb{N}_0:=\{0,1,2,\cdots\}$ and $\mathbb{N}:=\{1,2,\cdots\}$.}
    
    \item{We reserve the indices $a,b,c,d$ to run over $0,\cdots,4$, the indices $\mu,\nu,\lambda$ to run over $0,\cdots,3$, and the indices $i,j$ to run over $1,\cdots,4$.}
    
    \item{If $\{\mathfrak a\}$ is a family of complex functions, we write $\mathcal{A}:= \operatorname{span}_{\mathbb C} \{\mathfrak a\}$, and, when convenient, denote this complex linear span by $\{\mathfrak a\}_{\mathrm{span}}$. Whenever the resulting space is endowed with a Hilbert, anti-Hilbert, or Krein structure, its completion with respect to the corresponding induced norm is understood.}

    \item{We define $\overline{{\mathcal{A}}\,} :=\{\overline{\mathfrak{a}}\}_{\mathrm{span}}$, where the overline denotes point-wise complex conjugation.}
    
    \item{Various sesquilinear forms are employed throughout this manuscript. We denote them generically by $\left\langle\cdot,\cdot\right\rangle_{\mathfrak{z}}$, where the subscript $\mathfrak{z}$ specifies the particular sesquilinear form under consideration.}

    \item{Let ${\mathcal{A}}$ and $\mathcal{B}$ be subspaces of a vector space equipped with a sesquilinear form $\left\langle\cdot,\cdot\right\rangle_{\mathfrak{z}}$. We write ${\mathcal{A}}\perp_{\mathfrak{z}} {\mathcal{B}}$ if $\left\langle\mathfrak{a},\mathfrak{b}\right\rangle_{\mathfrak{z}}=0$, for all $\mathfrak{a}\in{\mathcal{A}}$ and $\mathfrak{b}\in{\mathcal{B}}$; otherwise, we write ${\mathcal{A}}\not\perp_{\mathfrak{z}}{\mathcal{B}}$.}

    \item{The angular quantum number $L$ is understood to range over all admissible values $L\in\mathbb N_0$ unless stated otherwise. Whenever $L$ is restricted to a subset of its admissible range, we indicate the restriction directly on the $L$-label. For instance,
    \begin{align*}
        \phi^{(1)}_{L^{}_{\geqslant p}lm}(\rho,u) := N^{(1)}_{L^{}_{\geqslant p}}\, \lambda^{(1)}_{L^{}_{\geqslant p}}(\rho)\, Y^{}_{L^{}_{\geqslant p}lm}(u)
    \end{align*}
    denotes the family of modes $\phi^{(1)}_{Llm}$ with $L\geqslant p$, for a given $p=1,2,\cdots$. Occasionally, for emphasis, we explicitly indicate the unrestricted range by writing, for example, $\phi^{(1)}_{L^{}_{\geqslant 0}lm}$, even though the condition $L\geqslant 0$ imposes no restriction beyond the admissible domain of $L$.}
\end{enumerate}

\section{dS Geometry, Representations, and Casimir Structure}\label{Sec. Machinery}

\subsection{Ambient Realization of dS UIRs}

The four-dimensional dS manifold may be realized as the one-sheeted hyperboloid
\begin{align}
    \mathrm{dS} = \left\{ x = (x^a) \in {\mathbb{R}}^{1,4} \;;\; (x)^2 := x\cdot x = \eta_{ab}x^a x^b = -R^2 \right\} \,,
\end{align}
embedded in the five-dimensional Minkowski spacetime $\mathbb{R}^{1,4}$ endowed with metric $\eta_{ab}=\mathrm{diag}(1,-1,-1,-1,-1)$. Here,  $a,b=0,1,2,3,4$, and the positive constant $R$ represents the dS radius. Topologically, dS is diffeomorphic to $\mathbb{R}\times\mathbb{S}^3$, with the factor $\mathbb{R}$ corresponding to a timelike coordinate.

The connected dS relativity group is $\mathrm{SO}_0(1,4)$, or equivalently its universal covering group $\mathrm{Sp}(2,2)$ \cite{Gazeau2022, GazPejTod2026}. The associated Lie algebra is generated by the ten Killing vector fields
\begin{align}
    K_{ab} = x_a\partial_b - x_b\partial_a \,,
\end{align}
which are tangent to the hyperboloid and therefore act intrinsically on dS spacetime.

At the representation-theoretic level, a given dS UIR may admit several equivalent realizations. Besides the spacetime realization on suitable spaces of (spinor-)tensor fields over the dS hyperboloid, it may also be realized on homogeneous spaces or on coadjoint orbits of the dS group. The latter perspective is particularly relevant to the boundary constructions developed in subsequent sections. To keep the present discussion as direct as possible, however, we restrict ourselves here to the spacetime realization. A detailed treatment of the homogeneous-space and orbital realizations can be found in Ref. \cite{Gazeau2022}.

Specifically, in the spacetime realization considered here, and restricting attention to integer-spin representations, the dS UIRs are realized on suitable spaces of rank-$r$ tensor fields $\phi^{(r)}_{a_1\cdots a_r}$ on the dS hyperboloid. These fields are assumed to satisfy the following conditions:
\begin{enumerate}
    \item{\emph{Symmetry:} 
    \begin{align}
        \phi^{(r)}_{a_1\cdots a_m a_n\cdots a_r} = \phi^{(r)}_{a_1\cdots a_n a_m\cdots a_r} \,.
    \end{align}
    We restrict attention to totally symmetric tensor fields, which furnish the conventional ambient-space realization of bosonic dS representations.}

    \item{\emph{Homogeneity:}
    \begin{align} \label{Homogeneity}
        x\cdot\partial\,\phi^{(r)} = \varrho\,\phi^{(r)} \,.
    \end{align}
    Homogeneity constrains the radial dependence of the ambient representative. Indeed, an intrinsic tensor field on the dS hyperboloid is determined only by the values of its ambient representative on the constraint surface $(x)^2=-R^2$. Consequently, if $\phi^{(r)}(x)$ is an ambient representative, then so is 
    \begin{align}
        \phi^{(r)}(x) + (x^2+R^2)\, \chi^{(r)}(x) \,,
    \end{align}
    for any sufficiently regular ambient tensor field $\chi^{(r)}$, since both fields coincide on the hyperboloid. The ambient extension is therefore highly non-unique. Imposing the homogeneity condition \eqref{Homogeneity} restricts this freedom by selecting ambient representatives of fixed degree $\varrho$, thereby providing a distinguished homogeneous realization of the intrinsic dS tensor field.}

    \item{\emph{Transversality:}
    \begin{align}
        x^{a_m}\phi^{(r)}_{a_1\cdots a_m\cdots a_r}=0 \,,
    \end{align}
    for all $m=1,2,\cdots,r$. This condition removes components normal to the dS hyperboloid and ensures that the ambient tensor field defines an intrinsic tensor field on dS spacetime.}

    \item{\emph{Regularity:} The tensor fields $\phi^{(r)}_{a_1\cdots a_r}$ are assumed to be sufficiently regular for the differential operators realizing the dS Lie algebra, together with the associated Casimir operators, to act on a common invariant domain.}

    \item{\emph{Invariant sesquilinear form:} The carrier space is endowed with an invariant inner product, or more generally with an invariant sesquilinear form $\left\langle\cdot , \cdot\right\rangle$. This invariance expresses dS covariance at the level of the representation space.}

    \item{\emph{Non-degeneracy:} The invariant sesquilinear form is assumed to be non-degenerate, meaning that the map $\phi \;\longmapsto\; \left\langle \phi,\cdot\right\rangle$ is injective. Equivalently, if $\left\langle \phi,\psi\right\rangle=0$ for all $\psi$, then $\phi=0$. Thus, the form separates vectors and faithfully encodes the structure of the underlying carrier space. If the form were degenerate, non-zero vectors in its radical would be orthogonal to every vector and hence undetectable by the form; the natural space associated with the form would then be the corresponding quotient by the radical. As will be seen later, such degenerate invariant forms arise naturally on certain invariant subspaces and play a central role in the indecomposable representation-theoretic structures developed in this work.}
\end{enumerate}

In this realization, the Killing vector fields $K_{ab}$ are represented by the essentially self-adjoint generators
\begin{align}
    K^{}_{ab} \quad\longmapsto\quad L^{(r)}_{ab} = M^{}_{ab} + S^{(r)}_{ab} \,,
\end{align}
where the orbital part is
\begin{align}
    M^{}_{ab} := -\mathrm{i}\left(x^{}_a\partial^{}_b - x^{}_b\partial^{}_a\right) \,,
\end{align}
and the spin part acts on the tensor indices as
\begin{align}
    S^{(r)}_{ab}\phi^{(r)}_{a_1\cdots a_r}
    := -\mathrm{i}\sum_{m=1}^r \Big( \eta_{a a_m}\phi^{(r)}_{a_1\cdots(a_m\mapsto b)\cdots a_r} - (a\rightleftharpoons b) \Big) \,.
\end{align}

In the scalar case ($r=0$), the total generators coincide with the orbital generators, namely, $L^{(r=0)}_{ab}=M^{}_{ab}$. Moreover, the homogeneity condition introduced above is preserved by the dS action. Indeed, since $[M_{ab} \,,\, x\cdot\partial]=0$, and since the spin part $S^{(r)}_{ab}$ acts only on tensor indices, one has $[L^{(r)}_{ab} \,,\, x\cdot\partial]=0$. Consequently, if $x\cdot\partial\,\phi^{(r)}=\varrho\,\phi^{(r)}$, then
\begin{align}
    x\cdot\partial\left(L^{(r)}_{ab}\phi^{(r)}\right) = \varrho\,L^{(r)}_{ab}\phi^{(r)} \,,
\end{align}
so the dS generators preserve the space of homogeneous tensor fields of degree $\varrho$.

The operators $L^{(r)}_{ab}$ satisfy the dS Lie-algebra relations
\begin{align}\label{adjoint commutator}
    \left[L^{(r)}_{ab},L^{(r)}_{cd}\right]
    = \mathrm{i}\left(
    \eta_{ac}L^{(r)}_{bd}
    +\eta_{bd}L^{(r)}_{ac}
    -\eta_{ad}L^{(r)}_{bc}
    -\eta_{bc}L^{(r)}_{ad}
    \right)\,.
\end{align}
Moreover, these generators satisfy 
\begin{align}\label{12}
    \left[L^{(r)}_{ab},(x)^2\right]=0\,,
\end{align}
so that they preserve the defining constraint $(x)^2=-R^2$ of the dS hyperboloid. Consequently, they induce well-defined differential operators on the hyperboloid itself.

Within this framework, the enveloping algebra of the dS generators admits two independent Casimir operators, given by
\begin{align} \label{Casimir 2}
    Q^{(1)}_r &= -\frac{1}{2}L^{(r)}_{ab}L^{(r)ab} \quad \text{(quadratic)} \,, \\[3pt]
    Q^{(2)}_r &= -W^{(r)}_a W^{(r)a} \quad \text{(quartic)} \,,
\end{align}
where $W^{(r)}_a=-\frac{1}{8}{\cal E}_{abcde} L^{(r)bc}L^{(r)de}$ defines the dS analogue of the Pauli-Lubanski vector, with ${\cal E}_{abcde}$ the totally antisymmetric Levi-Civita symbol in five dimensions. Owing to the fact that the generators $L^{(r)}_{ab}$ act tangentially to the dS hyperboloid (see Eq. \eqref{12}), these Casimir operators are themselves intrinsically defined on dS spacetime and satisfy
\begin{align}
    \left[Q^{(1,2)}_r,(x)^2\right] = 0 \,.
\end{align}
Note that $Q^{(1,2)}_r$ denotes either of the dS Casimir operators $Q^{(1)}_r$ or $Q^{(2)}_r$. Furthermore, as central elements of the algebra, they commute with all generators,
\begin{align}
    \left[Q^{(1,2)}_r,L^{(r)}_{ab}\right] = 0 \,,
\end{align}
for all $a,b=0,1,\cdots,4$.

On an irreducible dS representation, Schur's lemma implies that the Casimir operators act as scalar multiples of the identity. Specifically, for every tensor field $\phi^{(r)}_{a_1\cdots a_r}$ belonging to the carrier space of a given dS UIR, one has
\begin{align}\label{wa.eq}
    Q^{(1,2)}_r \; \phi^{(r)}_{a_1\cdots a_r} = \left\langle Q^{(1,2)}_r\right\rangle \; \phi^{(r)}_{a_1\cdots a_r} \,,
\end{align}
where $\left\langle Q^{(1,2)}_r\right\rangle$ denotes the corresponding Casimir eigenvalue. Following Dixmier's parametrization \cite{Dixmier}, these eigenvalues are labeled by a pair $(p,q)$, with $p\in\mathbb N/2$ and $q\in\mathbb C$, and explicitly read as
\begin{align}
    \label{Casimir rank 2}
    \left\langle Q^{(1)}_r\right\rangle &= -p(p+1)-(q+1)(q-2) \,, \\[3pt]
    \label{Casimir rank 4} 
    \left\langle Q^{(2)}_r\right\rangle &= -p(p+1)q(q-1) \,.
\end{align}

Accordingly, the Casimir eigenvalue equations \eqref{wa.eq} determine the representation labels. More precisely, the admissible values of the pair $(p,q)$ classify the UIRs of the dS group. These representations fall into the following three families \cite{Gazeau2022, GazPejTod2026, Thomas, Newton, Takahashi, Dixmier, Martin1974, Ottoson, Schwarz, Letsios-2025}:
\begin{enumerate}
    \item{\emph{Principal series}: The principal series splits into two cases:
    \begin{enumerate}
        \item{$p=s=0,1,2,\cdots$ and $q=\frac{1}{2}\pm \mathrm{i}\nu$, with $\nu\in{\mathbb{R}}$;}
        
        \item{$p=s=\frac{1}{2},\frac{3}{2},\cdots$ and $q=\frac{1}{2}\pm \mathrm{i}\nu$, with $\nu\in{\mathbb{R}}\setminus\{0\}$.}
    \end{enumerate}
    Here, $p=s$ carries the spin interpretation, and the representations labeled by $\nu$ and $-\nu$ are equivalent \cite{Gazeau2022}.}

    \item{\emph{Complementary series}: The complementary series likewise splits into two cases:
    \begin{enumerate}
        \item{$p=s=0$ and $q=\frac{1}{2}\pm\nu$, with $\nu\in{\mathbb{R}}$ and $0<|\nu|<\frac{3}{2}$;}
        
        \item{$p=s=1,2,\cdots$ and $q=\frac{1}{2}\pm\nu$, with $\nu\in{\mathbb{R}}$ and $0<|\nu|<\frac{1}{2}$.}
    \end{enumerate}
    Again, $p=s$ has the meaning of spin, and the two signs of $\nu$ label equivalent representations \cite{Gazeau2022}.}

    \item{\emph{Discrete series}: The discrete series consists of two distinct families:
    \begin{enumerate}
        \item{$p=1,2,\cdots$ and $q=0$;}
        
        \item{$p=\frac{1}{2},1,\frac{3}{2},\cdots$ and $q=p,p-1,\cdots,1$ or $\frac{1}{2}$.}
    \end{enumerate}
    In the case $p=s=q$, the label has the usual spin/helicity interpretation \cite{GazPejTod2026}. Throughout this work, we employ Dixmier's notation $\Pi_{p,q}$ for the discrete-series representations \cite{Dixmier}. In particular, special attention will be devoted to the scalar representations $\Pi_{p,0}$, $p=1,2,\cdots$, traditionally grouped with the discrete series, which form the main subject of the present work.}
\end{enumerate}

\paragraph{Massiveness and masslessness in dS representation theory.} In the flat, or null-curvature, limit, the principal-series UIRs of the dS group contract to the massive UIRs of the Poincar\'{e} group, thereby reproducing the entire massive sector of the Wigner classification. Consequently, the principal series is interpreted as describing massive elementary systems in dS spacetime \cite{Gazeau2022, Mickelsson, Garidi}. There is, however, no dS UIR whose contraction gives the massless infinite-spin representations of the Poincar\'{e} group. By contrast, the scalar complementary-series representation $(p=s=0 \,,\, q=1)$, together with the lower-end higher-spin ($s>0$) discrete-series representations $(p=s \,,\, q=s)$, forms a distinguished class of dS UIRs admitting a unique extension to UIRs of the conformal group $\mathrm{SO}_0(2,4)$ \cite{GazPejTod2026, Barut}. This extension coincides with the conformal extension of the usual massless Poincar\'{e} representations \cite{Todorov, Mack}. For this reason, these dS representations are called massless. The remaining dS UIRs either have no meaningful Poincar\'{e} contraction limit or contract to representations without a standard physical interpretation.

\paragraph{Beyond the flat-spacetime correspondence.} It should be emphasized that the preceding statement does not imply that dS UIRs without a direct Minkowskian counterpart are physically irrelevant. On the contrary, every dS UIR admits a consistent study in its own right, both at the mathematical level of group representation theory and at the physical level of covariant field quantization. This observation applies in particular to the scalar family $\Pi_{p,0}$, $p=1,2,\cdots$, which is the focus of the present work. As will be shown below, the entire scalar discrete-series family $\Pi_{p,0}$ exhibits a rich indecomposable representation-theoretic structure of Gupta-Bleuler type. For instance, a particularly distinguished member of this family is $\Pi_{1,0}$, which plays a fundamental structural role in dS-covariant quantum gravity; the graviton field is constructed from a spin-two projector acting on a scalar structure function carrying precisely this representation \cite{Pejhan2019}. Remarkably, the indecomposable structure underlying this scalar sector is subsequently reflected in the corresponding spin-two theory, where it gives rise to one of the two intertwined Gupta-Bleuler hierarchies of the dS graviton, the second being associated with the ordinary spin-two gauge invariance; for a relevant discussion, see Ref. \cite{Gibbons}. Consequently, the scalar discrete-series family is not merely a representation-theoretic curiosity lacking a flat-space counterpart, but constitutes an essential component of the representation-theoretic framework underlying gauge fields in dS spacetime.

\paragraph{From Casimir operators to wave equations.} Within the spacetime realization described above, a useful structural observation is that, for a given dS UIR, the corresponding representation space, equipped with an invariant non-degenerate inner product or, more generally, an invariant non-degenerate sesquilinear form, contains a dense invariant subspace spanned by sufficiently regular tensor fields $\phi^{(r)}_{a_1\cdots a_r}$ satisfying the Casimir eigenvalue equations \eqref{wa.eq}, with eigenvalues determined by the representation labels \cite{Gazeau2022}. From the viewpoint of quantum field theory, the quartic Casimir operator $Q^{(2)}_r$ typically gives rise to a higher-order differential equation whose solution space is not minimal for the characterization of the representation. Consequently, the quadratic Casimir equation, being of lower differential order while still encoding the representation labels, assumes the primary role and is naturally interpreted as the field, or wave, equation in the present group-theoretical framework,
\begin{align}\label{fieldEq}
    \left(Q^{(1)}_r + p(p+1) + (q+1)(q-2)\right) \phi^{(r)}_{a_1\cdots a_r}(x) = 0 \,.
\end{align}
The admissible values of the pair $(p,q)$ are precisely those characterizing the principal, complementary, and discrete series of dS UIRs.

\paragraph{Notation.} From this point onward, the quadratic Casimir operator $Q_r^{(1)}$ will simply be denoted by $Q^{}_r$; in particular, $Q_0^{(1)}$ will be written as $Q^{}_0$.

\subsection{Ambient-Space Realization of the Scalar Discrete Series}\label{Sec. AmbientCoordinate}

We now turn to the scalar discrete-series sector that forms the focus of the present work. A dS scalar field will be denoted by $\phi(x):=\phi^{(r=0)}(x)$ and, for simplicity, assumed to be homogeneous of degree zero in the ambient space, namely $x\cdot\partial\, \phi(x) = 0$. The corresponding representation space is equipped with the Klein-Gordon (KG) sesquilinear form
\begin{align} \label{eq:KG_inner_product}
    \left\langle \phi^{}_1(x) \,,\, \phi^{}_2(x) \right\rangle_{\vphantom{}_{\text{KG}}} = \mathrm{i}\int_{\Sigma} \overline{\phi^{}_1(x)} \stackrel{\leftrightarrow}{\partial}_{s} \phi^{}_2(x)\, \mathrm{d}\sigma^s \,,
\end{align}
where $\Sigma$ is a Cauchy surface, $\mathrm{d}\sigma^s$ is the corresponding directed surface element, and the bidirectional derivative $\stackrel{\leftrightarrow}{\partial}_{s}$ is defined by
\begin{align}\label{miriam2}
    f\stackrel{\leftrightarrow}{\partial}_{s}g := f(\partial_s g) - g(\partial_s f) \,.
\end{align} 
This KG sesquilinear form \eqref{eq:KG_inner_product} is invariant under the action of the dS group \cite{Birrell-1982, Higuchi-1987}. Moreover, it obeys the conjugate symmetry
\begin{align}\label{miriamCNK}
    \left\langle \phi_1(x) \;,\; \phi_2(x) \right\rangle_{\vphantom{}_{\text{KG}}} &= \overline{\left\langle \phi_2(x) \;,\;\phi_1(x) \right\rangle}_{\vphantom{}_{\text{KG}}} \nonumber\\[3pt]
    &= -\left\langle \overline{\phi_2(x)}\,,\,\overline{\phi_1(x)}\right\rangle_{\vphantom{}_{\text{KG}}} \,.
\end{align}

For scalar fields, the quartic Casimir operator vanishes identically, whereas the quadratic Casimir operator $Q_0$ is given by \cite{Gazeau2022}
\begin{align}\label{Q_0}
    Q_0 = -\frac{1}{2}M^{}_{ab}M^{ab}
    = -R^2 ({\partial}^{\texttt{T}})^2 \,,
\end{align}
where ${\partial}^{\texttt{T}}_a := \theta_{ab}\partial^b = \partial_a + R^{-2}x_a\,x\cdot\partial$ is the transverse derivative, and $\theta_{ab} := \eta_{ab} + R^{-2} x_a x_b$ is the transverse projector. By construction, $x^a\theta_{ab}=0=x^b\theta_{ab}$ and therefore $x\cdot{\partial}^{\texttt{T}}=0$. The transverse derivative also satisfies ${\partial}^{\texttt{T}}_a x^{}_b = \theta^{}_{ab}$ and ${\partial}^{\texttt{T}}_a \,(x)^2 = 0$. The latter identity shows that ${\partial}^{\texttt{T}}_a$ commutes with the ambient constraint $(x)^2=-R^2$, and is therefore intrinsically defined on the dS hyperboloid.

Finally, specializing the scalar Casimir (wave) equation to the discrete-series representations $\Pi_{p,0}$, $p=1,2,\cdots$, which constitute the focus of the present work, one obtains
\begin{align}\label{wave.eq.111}
    \left( Q_0 + (p-1)(p+2) \right) \phi(x) = 0 \,.
\end{align}

\subsection{From Ambient to Intrinsic Formulation}\label{Sec. Link}

In this subsection, we establish the relation between the ambient-space description and the local (intrinsic) coordinate formulation, and then specialize the resulting expressions to the scalar case of interest.

The intrinsic field $\phi^{(r)}_{\mu_1\cdots\mu_r}(X)$ is locally represented by the ambient field $\phi^{(r)}_{a_1\cdots a_r}(x)$ according to
\begin{eqnarray}\label{tt}
    \phi^{(r)}_{\mu_1\cdots\,\mu_r}(X) = x^{a_1}_{\,\,,\,\mu_1}\cdots\, x^{a_r}_{\,\,,\,\mu_r} \,\phi^{(r)}_{a_1\cdots\, a_r}\big(x(X)\big) \,,
\end{eqnarray}
where $x^{a_i}_{\,\,,\,\mu_i}:=\partial x^{a_i}/\partial X^{\mu_i}$, while $X^\mu$ with $\mu=0,1,2,3$ denote local coordinates on the dS hyperboloid. The intrinsic metric $g_{\mu\nu}$ is obtained by restricting the ambient metric of $\mathbb{R}^{1,4}$ to dS,
\begin{align}
    \mathrm{d}s^2 = \eta_{ab}\,\mathrm{d}x^a \mathrm{d}x^b\Big|_{(x)^2=-R^2} = g_{\mu\nu}\,\mathrm{d}X^\mu \mathrm{d}X^\nu \,.
\end{align}
In view of Eq. \eqref{tt}, the transverse projector $\theta_{ab}$ is the unique symmetric transverse ambient tensor associated with the intrinsic dS metric, namely
\begin{align}
    g_{\mu\nu}
    =
    x^a_{\,\,,\,\mu} x^b_{\,\,,\,\nu} \;\theta_{ab} \,.
\end{align}
Within this framework, intrinsic covariant derivatives are related to ambient tangential derivatives by
\begin{align}
    \nabla_\mu \nabla_\nu \cdots\, \nabla_\rho \,\phi^{(r)}_{\lambda_1\cdots\, \lambda_r} =&\; x^a_{\,\,,\,\mu} x^b_{\,\,,\,\nu} \cdots\, x^c_{\,\,,\,\rho} \; x^{d_1}_{\,\,,\,\lambda_1} \cdots\, x^{d_r}_{\,\,,\,\lambda_r} \nonumber\\[3pt] 
    &\times \big(\mathfrak{T} {\partial}^{\texttt{T}}_a\big) \big(\mathfrak{T}{\partial}^{\texttt{T}}_b\big) \cdots\, \big(\mathfrak{T}{\partial}^{\texttt{T}}_c\big) \phi^{(r)}_{d_1\cdots\, d_r} \,,
\end{align}
where the operator $\mathfrak{T}$, defined by
\begin{eqnarray}
    (\mathfrak{T} \phi)^{(r)}_{a_1 \;\cdots\; a_r}(x) \equiv \left( \prod_{i=1}^r \theta^{b_i}_{a_i} \right) \phi^{(r)}_{b_1 \;\cdots\; b_r}(x) \,,
\end{eqnarray}
guarantees the transversality in each tensorial index.

For the scalar fields considered in this work, the d'Alembertian becomes
\begin{align}
    \square_{\text{dS}} \phi &= g^{\mu\nu}\nabla_\mu\nabla_\nu\phi \nonumber\\[3pt]
    &= g^{\mu\nu} x^a_{\,\,,\,\mu} x^b_{\,\,,\,\nu} \Big( {\partial}^{\texttt{T}}_a{\partial}^{\texttt{T}}_b - R^{-2} x_b{\partial}^{\texttt{T}}_a \Big)\phi \nonumber\\[3pt]
    &= \theta^{ab} \Big( {\partial}^{\texttt{T}}_a{\partial}^{\texttt{T}}_b - R^{-2} x_b{\partial}^{\texttt{T}}_a \Big)\phi = ({\partial}^{\texttt{T}})^2\phi \,.
\end{align}
Together with Eq. \eqref{Q_0}, this gives $Q_0=-R^2 \square_{\text{dS}}$. Consequently, the discrete-series scalar wave equation \eqref{wave.eq.111} reduces to
\begin{align}\label{wave.eq.1}
    \left(R^2 \square_{\text{dS}} - (p-1)(p+2)\right)\phi(X)=0 \,,
\end{align}
where $p=1,2,\cdots$. 

For each fixed value of $p=1,2,\cdots$, Eq. \eqref{wave.eq.1} provides the intrinsic dS realization of the ambient-space wave equation \eqref{wave.eq.111}. The ambient Casimir eigenvalue condition is seen to be equivalent to a KG-type field equation on the dS hyperboloid. From the representation-theoretic perspective, however, the discrete-series UIR $\Pi_{p,0}$ is not associated with an individual solution of Eq. \eqref{wave.eq.1}, but rather with a suitable dS-invariant space of sufficiently regular solutions endowed with a non-degenerate KG sesquilinear form. This solution space furnishes the spacetime realization of $\Pi_{p,0}$ that will serve as the starting point for the constructions developed in the following sections. Although both the solutions and the corresponding representation space depend on the parameter $p$, this dependence will not be displayed explicitly in the notation for $\phi$, to avoid unnecessary clutter.

\section{Scalar Discrete-Series Modes and Indecomposable Krein-Gupta-Bleuler Realization on the dS Hyperboloid} \label{Sec. Modes}

\subsection{Global Conformal Coordinates, Mode Decomposition, and KG Structure} \label{Sec. ConformalCoor}

The construction of the representation space is intrinsically coordinate-independent. For an explicit realization, however, we use global conformal coordinates and write the ambient-space embedding as
\begin{align}\label{confo-coor}
    X: = X(\rho,u) = \left(R\tan \rho,\, R(\sec\rho) \,u\right) \,,
\end{align}
where $\rho\in\left(-\frac{\pi}{2},\frac{\pi}{2}\right)$ and $u=(u^1,u^2,u^3,u^4)\in{\mathbb{S}}^3\subset\mathbb R^4$ is parametrized by $(\alpha,\theta,\psi)\in[0,\pi]\times[0,\pi]\times[0,2\pi)$ as
\begin{align} \label{kha48}
    \begin{cases}
        u_{}^1 = \sin \alpha\, \sin \theta\, \cos \psi\,,\\
        u_{}^2 = \sin \alpha\, \sin \theta\, \sin \psi\,,\\
        u_{}^3 = \sin \alpha\, \cos \theta\,,\\
        u_{}^4 = \cos \alpha \,.
    \end{cases}
\end{align}
The standard $\mathrm{O}(4)$-invariant measure on $\mathbb{S}_{}^3$ is 
\begin{align}\label{measure}
    \mathrm{d}u = \sin_{}^2\alpha\,\sin\theta\, \mathrm{d}\alpha\, \mathrm{d}\theta\, \mathrm{d}\psi \,.
\end{align}

With the dS hyperboloid embedded in $\mathbb{R}^{1,4}$ through $X(\rho,u)$, the KG form on the Cauchy surface $\rho=0$ is
\begin{align} \label{KG.def}
    \left\langle \phi_1^{}(\rho,u) \,,\, \phi_2^{}(\rho,u) \right\rangle_{\vphantom{}_{\text{KG}}} = \mathrm{i} R^2 \int_{\rho=0} \overline{\phi_1^{}(\rho,u)} \,\stackrel{\leftrightarrow}{\partial}_{\rho}^{} \phi_2^{}(\rho,u)\, \mathrm{d}u \,.
\end{align}

In these coordinates, the wave equation \eqref{wave.eq.1} becomes
\begin{align} \label{wave.eq.cc}
    \left[\cos_{}^{4}\rho \, \frac{\partial}{\partial\rho} \left(\cos_{}^{-2}\rho \, \frac{\partial}{\partial\rho}\right) - \cos_{}^{2}\rho \, \Delta_{3}^{} \right.& \nonumber \\ 
    \left. - (p-1)(p+2)\vphantom{\cos_{}^{4}\rho \, \frac{\partial}{\partial\rho}\left(\cos_{}^{-2}\rho \, \frac{\partial}{\partial\rho}\right) -  \cos_{}^{2}\rho \, \Delta_{3}^{}}\right]&\phi(\rho,u) =0\,,
\end{align}
where $\Delta_3^{}$ denotes the Laplace-Beltrami operator on $\mathbb{S}_{}^3$,
\begin{align} \label{LB-S3}
    \Delta_3 =&\, \frac{\partial^2}{\partial\alpha^2} +2\cot\alpha\,\frac{\partial}{\partial\alpha}\nonumber\\
    &+ \frac{1}{\sin^2\alpha}\left(\frac{\partial^2}{\partial\theta^2}+\cot\theta\,\frac{\partial}{\partial\theta}+\frac{1}{\sin^2\theta}\frac{\partial^2}{\partial\psi^2}\right)\,.  
\end{align}

For a given $p=1,2,\cdots$, we solve Eq. \eqref{wave.eq.cc} by means of separation of variables, decomposing the dependence on the global conformal time $\rho$ and the angular coordinates $u\in{\mathbb{S}}^3$. Accordingly, we consider solutions of the form
\begin{align}\label{psi.sep}
    \phi(\rho,u) = N\, \lambda(\rho) \,Y_{Llm}(u) \,,
\end{align}
where $N$ is a real normalization constant to be determined from the KG sesquilinear form \eqref{KG.def}, $\lambda(\rho)$ is the conformal-time mode function, and $Y_{Llm}(u)$ are the orthonormal hyperspherical harmonics on ${\mathbb{S}}^3$. The latter constitute a complete orthonormal basis of $L^2_{\mathbb{C}}({\mathbb{S}}^3)$ and are eigenfunctions of the Laplace-Beltrami operator $\Delta_3$ \cite{Avery},
\begin{align}\label{S3-eigenvalue}
    \Delta_3 Y_{Llm}(u) = -L(L+2)\,Y_{Llm}(u) \,,
\end{align}
where $L,l\in\mathbb N_0$ and $m\in\mathbb Z$ obey $0\leqslant l\leqslant L$, $|m|\leqslant l$. Further details on the hyperspherical harmonics on ${\mathbb{S}}^3$, including their definition and relevant properties, are collected in Appendix \ref{A-Hyp-Har}.

\paragraph{Relation to Higuchi's global dS modes.} Scalar mode solutions on global dS spacetime obtained by analytic continuation from scalar spherical harmonics were previously constructed by Higuchi \cite{Higuchi-1987}. The present construction proceeds directly from the dS wave equation and retains both linearly independent radial solutions. The resulting solution space provides the starting point for the subsequent analysis of the KG structure and the associated indecomposable Krein-Gupta-Bleuler realization.

Substituting the ansatz \eqref{psi.sep} into the field equation \eqref{wave.eq.cc} and making use of the eigenvalue relation \eqref{S3-eigenvalue}, we obtain the ordinary differential equation
\begin{align} \label{eqrad}
    &\left[\cos_{}^4 \rho \,\frac{\mathrm{d}}{\mathrm{d}\rho} \left( \cos_{}^{-2} \rho \,\frac{\mathrm{d}}{\mathrm{d}\rho} \right) \right. + L(L+2) \cos_{}^2 \rho \nonumber \\[3pt]
    &\hspace{2.5cm} \left. \vphantom{\frac{\mathrm{d}}{\mathrm{d}\rho}} - (p-1)(p+2) \right] \lambda(\rho) = 0 \,, 
\end{align}
which governs the conformal-time dependence of the mode functions. As derived in Appendices \ref{details-on-se.3} and \ref{App. Linear Independence of the Two Solutions}, Eq. \eqref{eqrad} admits two linearly independent solutions, expressed in terms of hypergeometric functions understood via distributional analytic continuation:
\begin{align}
    \lambda_L^{(1)}(\rho) :=&\, 2^{1-p}\,e_{}^{-\mathrm{i}(L-p+1)\rho}\,(\cos\rho)_{}^{1-p}\nonumber\\[3pt]
    &\times\, {}_2F_1^{}\left(-p,\,L+1-p;\,L+2;\,-e_{}^{-2\mathrm{i}\rho}\right) , \label{chi_p-1}\\[6pt]
    \lambda^{(2)}_L(\rho) :=&\, 2^{p+2}\,e^{-\mathrm{i}(L+p+2)\rho}\,(\cos\rho)^{p+2}\nonumber\\[3pt]
    &\times\, {}_2F_1^{}\left(p+1,\,L+p+2;\,2p+2;\, 2e^{-\mathrm{i}\rho}\cos\rho \right) . \label{chi_p-2}
\end{align}
The latter solution is valid in the region where $\cos\rho <\frac{1}{2}$. It follows that the wave equation \eqref{wave.eq.cc} admits the corresponding family of linearly independent analytic mode solutions
\begin{align} \label{ModeNaive}
    \phi^{(1)}_{Llm}(\rho,u) &:= N_L^{(1)}\, \lambda_L^{(1)}(\rho)\,Y_{Llm}^{}(u)\,, \nonumber\\[3pt]
    \phi^{(2)}_{Llm}(\rho,u) &:= N_L^{(2)}\, \lambda_L^{(2)}(\rho)\,Y_{Llm}^{}(u) \,,
\end{align}
where $N_L^{(1)}$ and $N_L^{(2)}$ are real normalization constants, whose values are determined by imposing normalization with respect to the KG sesquilinear form \eqref{KG.def}.

\paragraph{Implicit dependence on the parameter $p$.} It should be understood that the mode solutions $\phi^{(1)}_{Llm}(\rho,u)$ and $\phi^{(2)}_{Llm}(\rho,u)$ depend on the discrete-series parameter $p$. However, to avoid overburdening the notation, this dependence is left implicit throughout and is only indicated explicitly when required for clarity.

\paragraph{Complex-conjugate mode solutions.} Since Eq. \eqref{wave.eq.cc} has real-valued coefficients, the complex conjugates $\overline{\phi^{(1)}_{Llm}}$ and $\overline{\phi^{(2)}_{Llm}}$ are also solutions. Furthermore, because $\left\{ \phi^{(1)}_{Llm} \,,\, \phi^{(2)}_{Llm} \right\}$ forms a fundamental set of solutions, the complex-conjugate modes must admit expansions in this basis. More precisely (see Appendix \ref{complex-con-sol}),
\begin{align}
    \overline{\phi^{(1)}_{Llm}} &= 
    \begin{cases}
        A \,\phi^{(1)}_{Ll,-m} + B \,\phi^{(2)}_{Ll,-m}\,, \quad &L \geqslant p \,,\\[7pt]
        A\, \phi^{(1)}_{Ll,-m}\,, \quad &L < p \,,
    \end{cases} \label{complex(1)}\\[4pt]
    \overline{\phi^{(2)}_{Llm}} &= - A \,\phi^{(2)}_{Ll,-m}\,, \label{complex(2)}
\end{align}
where
\begin{align}
    A &= (-1)^{L+p+1+m} \,, \\[3pt]
    B &= {(-1)^{L+1+m}} \,\frac{\Gamma(p+1) \, \Gamma(L+2)}{\Gamma(L-p+1)\, \Gamma(2p+2)} \,.
\end{align}

\paragraph{Notation for restricted mode families.} Whenever the angular quantum number $L$ is restricted to a subset of its admissible values, we indicate this restriction directly on the $L$-label. For example,
\begin{align}
    \phi^{(1)}_{L^{}_{\geqslant p}lm}(\rho,u) := N^{(1)}_{L^{}_{\geqslant p}}\, \lambda^{(1)}_{L^{}_{\geqslant p}}(\rho)\, Y^{}_{L^{}_{\geqslant p}lm}(u)
\end{align}
denotes the family of modes $\phi^{(1)}_{Llm}$ with $L\geqslant p$. Conversely, in the absence of an explicit restriction, as in $\phi^{(1)}_{Llm}$, the quantum number $L$ is understood to range over all admissible values $L\in\mathbb N_0$. Occasionally, for emphasis, we explicitly indicate the unrestricted range by writing, for example, $\phi^{(1)}_{L^{}_{\geqslant 0}lm}$, even though the condition $L\geqslant 0$ imposes no restriction beyond the admissible domain of $L$.

\paragraph{Alternative complete basis of solutions.} By virtue of the first relation in Eq. \eqref{complex(1)}, the modes $\phi^{(2)}_{L_{\geqslant p}lm}$ may be replaced by $\overline{\phi^{(1)}_{L^{}_{\geqslant p}lm}}$ without changing the solution space. Consequently, the family
\begin{align}\label{50}
    \left\{ \phi^{(1)}_{L^{}_{\geqslant p}lm} \,,\, \overline{\phi^{(1)}_{L^{}_{\geqslant p}lm}} \,,\, \phi^{(1)}_{L^{}_{< p}lm} \,,\, \phi^{(2)}_{L^{}_{< p}lm} \right\}
\end{align}
provides an alternative complete basis of the solution space generated by
\begin{align}
    \left\{ \phi^{(1)}_{L^{}_{\geqslant 0}lm} \,,\, \phi^{(2)}_{L^{}_{\geqslant 0}lm} \right\} \,.
\end{align}
This equivalent basis will prove useful in the constructions that follow.

\begin{figure*}[!t]
    \centering
    \includegraphics[width=0.66\linewidth]{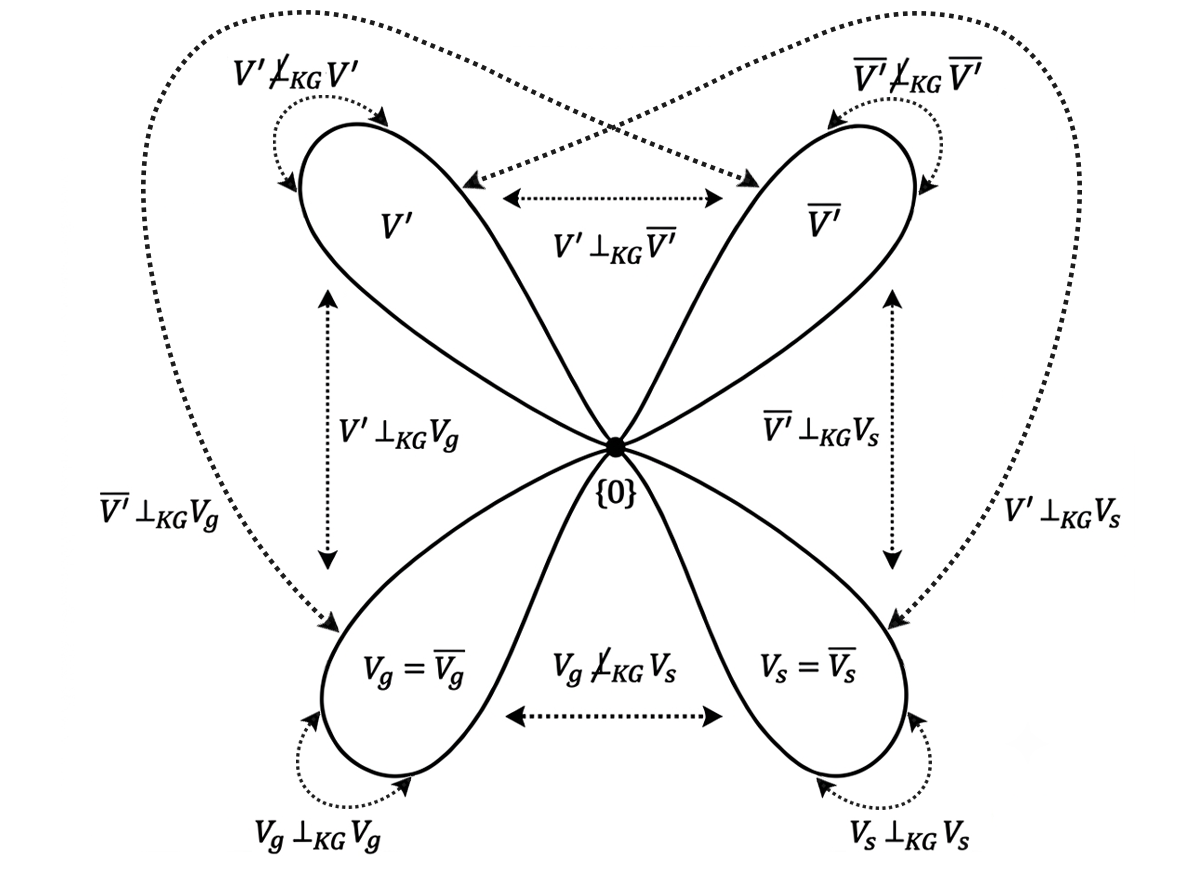}
    \caption{Schematic illustration of the KG orthogonality structure of the aforementioned subspaces, as induced by the orthogonality relations of their corresponding mode bases.}
    \label{fig}
\end{figure*}

Having obtained the complete set of mode solutions, we now identify, for a fixed $p=1,2,\cdots$, those families of modes that span non-degenerate subspaces with respect to the KG sesquilinear form. Such subspaces provide natural candidates for the realization of the corresponding representation space. To this end, we evaluate the KG products among the fundamental mode solutions \eqref{50}. A direct calculation yields (see Appendix \ref{App. KGNorm})
\begin{align}
    \left\langle \phi^{(1)}_{L^{}_{\geqslant p}lm} \;,\; \phi^{(1)}_{L^\prime_{\geqslant p}l^\prime_{}m^\prime_{}}\right\rangle_{\vphantom{}_{\text{KG}}} &= \delta_{LL_{}^\prime} \,\delta_{ll_{}^\prime} \,\delta_{mm_{}^\prime} \,, \label{kha30}\\[3pt]
    \left\langle \overline{\phi^{(1)}_{L^{}_{\geqslant p}lm}} \;,\; \overline{\phi^{(1)}_{L^\prime_{\geqslant p}l^\prime_{}m^\prime_{}}}\right\rangle_{\vphantom{}_{\text{KG}}} &= - \delta_{LL_{}^\prime} \,\delta_{ll_{}^\prime} \,\delta_{mm_{}^\prime} \,, \label{kha30'}\\[3pt]
    \left\langle \phi^{(1)}_{L^{}_{\geqslant p}lm} \;,\; \overline{\phi^{(1)}_{L^\prime_{\geqslant p}l^\prime_{}m^\prime_{}}}\right\rangle_{\vphantom{}_{\text{KG}}} &= 0 \,, \label{miriamDBD}\\[3pt]
    \left\langle \phi^{(1)}_{L_{<p}^{}lm}\;,\;  \phi^{(2)}_{L_{<p}^\prime l_{}^\prime m_{}^\prime} \right\rangle_{\vphantom{}_{\text{KG}}} &= \delta_{LL_{}^\prime} \,\delta_{ll_{}^\prime} \,\delta_{mm_{}^\prime} \,, \label{kha27} \\[3pt]
    \left\langle \phi^{(2)}_{L_{<p}^{}lm}\;,\;  \phi^{(1)}_{L_{<p}^\prime l_{}^\prime m_{}^\prime} \right\rangle_{\vphantom{}_{\text{KG}}} &= \delta_{LL_{}^\prime} \,\delta_{ll_{}^\prime} \,\delta_{mm_{}^\prime} \,, \label{kha27'} \\[3pt]
    \left\langle \phi^{(1)}_{L_{<p}^{}lm}\;,\;  \phi^{(1)}_{L_{<p}^\prime l_{}^\prime m_{}^\prime} \right\rangle_{\vphantom{}_{\text{KG}}} &= 0 \,, \label{thelastone} \\[3pt]
    \left\langle \phi^{(2)}_{L_{<p}^{}lm}\;,\;  \phi^{(2)}_{L_{<p}^\prime l_{}^\prime m_{}^\prime} \right\rangle_{\vphantom{}_{\text{KG}}} &= 0 \,, \label{thelastone'}
\end{align}
with the normalization constants chosen as
\begin{align}
    N_{L^{}_{\geqslant p}}^{(1)} := &\frac{1}{2\sqrt{2}\, R}\frac{\sqrt{\Gamma(L-p+1) \,\Gamma(L+p+2)}}{\Gamma(L+2)}\,, \label{normalization constants1} \\[3pt]
    N_{L_{<p}^{}}^{(1)} :=& \frac{1}{\sqrt{2}\, R} \,, \label{normalization constants2} \\[3pt]
    N_{L_{<p}^{}}^{(2)} :=& \frac{(-1)^{p+1}}{4\sqrt{2}\, R}\, \frac{\Gamma(p+1) \,\Gamma(L+p+2)}{\Gamma(2p+2)\, \Gamma(L+2)} \,. \label{normalization constants3}
\end{align}
The KG products not explicitly displayed in Eqs. \eqref{kha30}-\eqref{thelastone'} either vanish identically because the corresponding modes belong to disjoint admissible ranges of the quantum number $L$, namely $L\geqslant p$ and $L<p$ (see Appendix \ref{App. KGNorm}), or are determined by the above relations together with the conjugate-symmetry property \eqref{miriamCNK} and the conjugation identities \eqref{complex(1)} and \eqref{complex(2)}. The resulting KG structure is illustrated schematically in Fig. \ref{fig}.

\paragraph{Alternative derivation.} It is worth noting that the above relations agree with those obtained previously in Ref. \cite{Gazeau2010}. The derivation presented there, however, follows a different route, relying on the expansion of the dS plane waves associated with the scalar discrete series in terms of hypergeometric functions. By contrast, the present approach derives these relations directly from the explicit mode solutions of the field equation, thereby providing an independent and complementary derivation.

We organize the corresponding fundamental mode families into ``true'', ``gauge'', and ``supplementary'' sectors by defining
\begin{align}
    \phi^{}_{L^{}_{\geqslant p}lm} := \phi^{(1)}_{L^{}_{\geqslant p}lm} \,, & \quad\mbox{and}\quad \overline{\phi^{}_{L^{}_{\geqslant p}lm}} := \overline{\phi^{(1)}_{L^{}_{\geqslant p}lm}} \,, \label{true>=p} \\[3pt]
    \phi^{(g)}_{L_{<p}^{}lm} := \phi^{(1)}_{L_{<p}^{}lm} \,, \label{gauge} & \\[3pt]
    \phi^{(s)}_{L_{<p}^{}lm} := \phi^{(2)}_{L_{<p}^{}lm}\,. & \label{supp} 
\end{align}
The terminology ``true'', ``gauge'', and ``supplementary'' anticipates the role played by these sectors in the Gupta-Bleuler structure associated with the representation $\Pi_{p,0}$, as will become clear below.

Among these mode solutions, the modes $\phi^{}_{L^{}_{\geqslant p}lm}$ and their complex conjugates $\overline{\phi^{}_{L^{}_{\geqslant p}lm}}$ span two distinct non-degenerate subspaces with respect to the KG form. We shall refer to these modes collectively as the ``true modes''. The corresponding subspaces are denoted by
\begin{align}
    V^\prime &:= \left\{\phi^{}_{L^{}_{\geqslant p}lm}\right\}_{\mathrm{span}}\,, \label{V'}\\[3pt] \overline{V^\prime} &:= \left\{\overline{\phi^{}_{L^{}_{\geqslant p}lm}}\right\}_{\mathrm{span}}\,. \label{V'bar}
\end{align}
Equations \eqref{kha30} and \eqref{kha30'} show that the KG form on $V^\prime$ is positive-definite, while on $\overline{V^\prime}$ it is negative-definite. Hence,
\begin{align}\label{tru-con}
    V^\prime \cap\overline{V^\prime}=\{0\}\,.
\end{align}
As will become apparent below, these non-degenerate subspaces play a distinguished role in the representation-theoretic structure of the theory.

Neither the family of ``gauge modes'' nor the family of ``supplementary modes'' spans a non-degenerate subspace individually; see Eqs. \eqref{thelastone} and \eqref{thelastone'}. However, a suitable linear combination of the modes $\phi^{(g)}_{L_{<p}^{}lm}$ and $\phi^{(s)}_{L_{<p}^{}lm}$ gives rise to a family of true modes for $L<p$, spanning a non-degenerate subspace with respect to the KG form. More precisely, we define
\begin{align}
    \phi^{}_{L^{}_{<p}lm} := \phi^{(g)}_{L^{}_{<p}lm} + \frac{1}{2} \phi^{(s)}_{L^{}_{<p}lm}\,.\label{eq:up_un_modes1}
\end{align}
By Eqs. \eqref{complex(1)} and \eqref{complex(2)}, its complex conjugate is given by
\begin{align}\label{dada'}
    \overline{\phi^{}_{L^{}_{<p}lm}} &= \overline{\phi^{(g)}_{L^{}_{<p}lm}} + \frac{1}{2} \;\overline{\phi^{(s)}_{L^{}_{<p}lm}} \nonumber\\[3pt]
    &= (-1)^{L+p+1+m}\left(\phi^{(g)}_{L^{}_{<p},l,-m} -\frac{1}{2}\; \phi^{(s)}_{L^{}_{<p},l,-m}\right)\,. 
\end{align}
Equations \eqref{kha27} and \eqref{kha27'} then immediately imply that the KG inner products among these modes are
\begin{align}
    \left\langle \phi^{}_{L_{<p}^{}lm} \;,\; \phi^{}_{L_{<p}^\prime l_{}^\prime m_{}^\prime} \right\rangle_{\vphantom{}_{\text{KG}}} &= \delta_{LL_{}^\prime} \,\delta_{ll_{}^\prime} \,\delta_{mm_{}^\prime} \,,\\[3pt]
    \left\langle\, \overline{\phi^{}_{L_{<p}^{}lm}} \;,\; \overline{\phi^{}_{L_{<p}^\prime l_{}^\prime m_{}^\prime}} \,\right\rangle_{\vphantom{}_{\text{KG}}} &= -\delta_{LL_{}^\prime} \,\delta_{ll_{}^\prime} \,\delta_{mm_{}^\prime} \,,\\[3pt]
    \left\langle\, \phi^{}_{L_{<p}^{}lm} \;,\; \overline{\phi^{}_{L_{<p}^\prime l_{}^\prime m_{}^\prime}} \,\right\rangle_{\vphantom{}_{\text{KG}}} &= 0 \,.
\end{align}

Hence, the modes $\phi_{L^{}_{<p}lm}$, together with their complex conjugates $\overline{\phi_{L^{}_{<p}lm}}$, form a complete family of true modes for $L<p$. Their positive- and negative-norm sectors are distinct and orthogonal:
\begin{align}
    \left\{\phi_{L^{}_{<p}lm}\right\}_{\mathrm{span}} \;&\cap\; \left\{\overline{\phi_{L^{}_{<p}lm}}\right\}_{\mathrm{span}} = \{0\} \,, \label{direcsum2} \\[3pt]
    \left\{\phi_{L^{}_{<p}lm}\right\}_{\mathrm{span}} \;&\!\!\!\perp_{\vphantom{}_{\text{KG}}} \;\left\{\overline{\phi_{L^{}_{<p}lm}}\right\}_{\mathrm{span}}\,. \label{direcsum1}
\end{align}

According to the KG orthogonality relations among the subspaces introduced above, summarized schematically in Fig. \ref{fig}, the total non-degenerate space of mode solutions then takes the form
\begin{align}\label{eq:Vtotal_Krein}
    V_{\text{tot}} = \underbrace{\left\{ \phi^{}_{L^{}_{\geqslant 0}lm} \right\}_{\mathrm{span}}}_{=:\; H_+} 
    \oplus\;\underbrace{\left\{ \overline{\phi^{}_{L^{}_{\geqslant 0}lm}} \right\}_{\mathrm{span}}}_{=:\; H_-}\,,
\end{align}
where
\begin{align} 
    \left\{ \phi^{}_{L^{}_{\geqslant 0}lm} \right\}_{\mathrm{span}} &=\, \underbrace{\left\{ \phi_{L^{}_{\geqslant p}lm} \right\}_{\mathrm{span}}}_{=:V^\prime} \oplus\; \left\{\phi^{}_{L^{}_{<p}lm}\right\}_{\mathrm{span}} \,, \label{Hilbert}\\[3pt]
    \left\{ \overline{\phi^{}_{L^{}_{\geqslant 0}lm}} \right\}_{\mathrm{span}} &=\, \underbrace{\left\{ \overline{\phi_{L^{}_{\geqslant p}lm}} \right\}_{\mathrm{span}}}_{=:\overline{V^\prime}} \oplus\; \left\{\overline{\phi^{}_{L^{}_{<p}lm}}\right\}_{\mathrm{span}} \,. \label{Anti-Hilbert}
\end{align}
The space $V_{\text{tot}} = H_{+} \oplus H_{-}$ is naturally endowed with the structure of a Krein space; the KG form is positive-definite on $H_{+}$ and negative-definite on $H_{-}$, so that $H_{+}$ and $H_{-}$ constitute the Hilbert and anti-Hilbert sectors of the Krein space $V_{\text{tot}}$, respectively.

Henceforth, for a fixed $p=1,2,\cdots$, our objective is to identify the smallest subspace of $V_{\text{tot}}$ that is invariant under the dS action and carries the representation $\Pi_{p,0}$. Among the non-degenerate subspaces appearing in \eqref{eq:Vtotal_Krein}, the space $V^\prime$ is the natural starting point. It is positive-definite with respect to the KG form, in accordance with the usual physical interpretation of this structure. Moreover, unlike the positive-definite subspace $\left\{\phi^{}_{L^{}_{<p}lm}\right\}_{\mathrm{span}}$, which is finite-dimensional and therefore cannot support the infinite-dimensional representation $\Pi_{p,0}$, the space $V^\prime$ is infinite-dimensional and thus provides a natural candidate from which to construct a realization of $\Pi_{p,0}$.

Before proceeding further, and for later reference, let us explicitly define the corresponding degenerate subspaces spanned by the gauge and supplementary modes
\begin{align}
    V_g &:= \left\{\phi^{(g)}_{L^{}_{<p}lm}\right\}_{\mathrm{span}}\,, \label{Vg}\\[3pt] 
    V_s &:= \left\{\phi^{(s)}_{L^{}_{<p}lm}\right\}_{\mathrm{span}}\,. \label{Vs}
\end{align}
Note that
\begin{align}\label{null-over}
    V_g\cap V_s &= \{0\} \,, \nonumber\\[3pt]
    V^\prime\cap V_g &= \{0\} = \overline{V^\prime}\cap V_g\,, \nonumber\\[3pt]
    V^\prime\cap V_s &= \{0\} = \overline{V^\prime}\cap V_s\,.
\end{align}
Moreover, according to Eqs. \eqref{complex(1)} and \eqref{complex(2)}, both $V_g$ and $V_s$ are invariant under point-wise complex conjugation. Consequently,
\begin{align}\label{barVgs=Vgs}
    \overline{V_g} = V_g \,, \quad \overline{V_s} = V_s \,.  
\end{align}

\begin{table*}[t]
    \centering
    \setlength{\tabcolsep}{12pt}
    \begin{tabular}{ccc}
    \hline\hline
    \text{Space} & \text{dS invariance} & \text{KG form} \\[3pt]
    \hline
    $V_g$                            & $\checkmark$ & $0$          \\[3pt]
    $V_s$                            & $\times$     & $0$          \\[3pt]
    $V^\prime$                       & $\times$     & $+$          \\[3pt]
    $\overline{V^\prime}$            & $\times$     & $-$          \\[3pt]
    $V = V^\prime\oplus V_g$             & $\checkmark$ & $\geqslant0$ \\[3pt]
    $H_+$                            & $\times$     & $+$          \\[3pt]
    $H_-$                            & $\times$     & $-$          \\[3pt]
    $V_{\mathrm{tot}} = H_+\oplus H_-$ & $\checkmark$ & $\pm$         \\[3pt]
    \hline
    \end{tabular}
    \caption{dS invariance and definiteness properties of the KG form on the relevant bulk spaces.}
\end{table*}

\subsection{Representation Space(s) and Krein-Gupta-Bleuler Structure} \label{Sec. RepSpaces}

\subsubsection{Successive dS-Invariant and Non-Degenerate Extensions}\label{Sec. dSAction}

In this subsection, for a fixed $p=1,2,\cdots$, we examine the total non-degenerate space of analytic mode solutions $V_{\text{tot}}$ \eqref{eq:Vtotal_Krein} and seek, within it, the minimal dS-invariant structure associated with the infinite-dimensional UIR $\Pi_{p,0}$, starting from the positive-definite sector selected by the KG form. This construction will ultimately lead to the physical carrier space of the theory.

The first natural candidate is the space $V^\prime$ defined in \eqref{V'}. However, $V^\prime$ does not by itself furnish a representation space, since it is not preserved by the action of the dS group. Indeed, under the dS action, one finds (see Appendix \ref{App. Action})
\begin{align}\label{mix.action}
    V^\prime \;\;\xrightarrow[]{\text{ dS action }}\;\; V^\prime \;\oplus\; V_g \,.
\end{align}
Thus, $V^\prime$ is not dS-invariant. 

The minimal enlargement of $V^\prime$ compatible with the dS action is obtained by adjoining the degenerate gauge subspace $V_g$. This yields the space 
\begin{align}
    V := V^\prime \oplus V_g \,.
\end{align}
As it turns out, $V$ is dS-invariant
\begin{align}
    V\;\; \xrightarrow[]{\text{ dS action }}\;\;V\,,
\end{align}
a consequence of the fact that the gauge subspace is itself stable under the dS action (see Appendix \ref{App. Action})
\begin{align}\label{gau-ds}
    V_g\;\; \xrightarrow[]{\text{ dS action }}\;\;V_g \,.
\end{align}

However, the presence of the gauge subspace $V_g$ renders $V$ degenerate (see Fig. \ref{fig}). Indeed, every vector in $V_g$ is KG orthogonal to all vectors in $V$, so that $V_g$ coincides with the radical of $V$. To obtain a non-degenerate representation space, it is therefore necessary to enlarge $V$ once more.

As discussed in the preceding subsection (see Eqs. \eqref{Hilbert} and \eqref{eq:up_un_modes1}), the minimal non-degenerate extension of $V$ is obtained by adjoining to the gauge modes $\phi^{(g)}_{L^{}_{<p}lm}$ their corresponding paired modes $\phi^{(s)}_{L^{}_{<p}lm}$,
\begin{align}
    V_g \;\;\xrightarrow[]{\text{ non-degenerate completion }}\;\; \left\{\phi^{}_{L^{}_{<p}lm}\right\}_{\mathrm{span}} \,.
\end{align}
Accordingly, we get
\begin{align} \label{87}
    &\left\{ \phi^{}_{L^{}_{\geqslant 0}lm} \right\}_{\mathrm{span}} = \nonumber\\[3pt]
    &\quad\quad V^\prime \oplus \left\{\phi^{}_{L^{}_{<p}lm} := \phi^{(g)}_{L^{}_{<p}lm} + \frac{1}{2} \phi^{(s)}_{L^{}_{<p}lm}\right\}_{\mathrm{span}} \,,
\end{align}
which, by construction, is positive-definite and non-degenerate with respect to the KG form.

However, although the inclusion of the supplementary modes $\phi^{(s)}_{L^{}_{<p}lm}$ removes the degeneracy of $V$, the resulting KG positive-definite subspace \eqref{87} is no longer invariant under the dS action. Indeed, one finds (see Appendix \ref{App. Action})
\begin{align}\label{supp.ds}
    \left\{ \phi^{(s)}_{L^{}_{<p}lm} \right\}_{\mathrm{span}} \;\; & \xrightarrow[]{\text{ dS action }}\;\;\left\{\phi^{(s)}_{L^{}_{<p}lm}\right\}_{\mathrm{span}} \; \oplus \; V^\prime \; \oplus \; \overline{V^\prime}\,.
\end{align}
Consequently,
\begin{align}
   \left\{\phi^{}_{L^{}_{\geqslant 0}lm}\right\}_{\mathrm{span}} \;\; \xrightarrow[]{\text{ dS action }} \;\; \left\{\phi^{}_{L^{}_{\geqslant 0}lm}\right\}_{\mathrm{span}} \; \oplus \; \overline{V^\prime}\,.
\end{align}
Therefore, the space $\left\{\phi^{}_{L^{}_{\geqslant 0}lm}\right\}_{\mathrm{span}}$ fails to be invariant under the dS action. Indeed, the dS generators necessarily mix this space with the negative-definite sector $\overline{V^\prime}$, thereby precluding the existence of a dS-invariant positive-definite realization.

Therefore, the next minimal candidate is the non-degenerate, albeit no longer KG positive-definite, subspace $\left\{\phi^{}_{L^{}_{\geqslant 0}lm}\right\}_{\mathrm{span}} \; \oplus \; \overline{V^\prime}$. Yet, this space is not invariant under the dS action (see Appendix \ref{App. Action}),
\begin{align}
    \left\{\phi^{}_{L^{}_{\geqslant 0}lm}\right\}_{\mathrm{span}} \; \oplus \; \overline{V^\prime} & \nonumber\\[3pt]
    \xrightarrow[]{\text{ dS action }}\;\; & \left\{\phi^{}_{L^{}_{\geqslant 0}lm}\right\}_{\mathrm{span}} \; \oplus \; \overline{V^\prime} \oplus \overline{V_g} \,.
\end{align}
Notably, no further sectors are generated from the enlarged space,
\begin{align}\label{miriamBP}
    \left\{\phi^{}_{L^{}_{\geqslant 0}lm}\right\}_{\mathrm{span}} \; \oplus \; \overline{V^\prime} \oplus \overline{V_g} & \nonumber\\[3pt]
    \xrightarrow[]{\text{ dS action }}\;\; & \left\{\phi^{}_{L^{}_{\geqslant 0}lm}\right\}_{\mathrm{span}} \; \oplus \; \overline{V^\prime} \oplus \overline{V_g} \,.
\end{align}
Hence, $\left\{\phi^{}_{L_{\geqslant 0}lm}\right\}_{\mathrm{span}} \oplus \overline{V^\prime} \oplus \overline{V_g}$ is the minimal dS-invariant enlargement of $\left\{\phi^{}_{L_{\geqslant 0}lm}\right\}_{\mathrm{span}} \oplus \overline{V^\prime}$.

Nevertheless, the dS-invariant subspace \eqref{miriamBP} remains degenerate due to the presence of the null sector $\overline{V_g}$ (see Fig. \ref{fig}). As in the preceding discussion, removing this degeneracy necessitates a further enlargement of the space. The unique non-degenerate extension is obtained by adjoining to each gauge mode $\overline{\phi^{(g)}_{L^{}_{<p}lm}}$ its corresponding paired mode $\overline{\phi^{(s)}_{L^{}_{<p}lm}}$. Equivalently, one completes the family of modes with $L<p$ by including the corresponding complex-conjugate true modes:
\begin{align}
    \overline{V_g}
    \;\;\xrightarrow[]{\text{ non-degenerate completion }}\;\; \left\{\overline{\phi^{}_{L^{}_{<p}lm}}\right\}_{\mathrm{span}} \,.
\end{align}

Consequently, the minimal non-degenerate dS-invariant subspace is the total space of mode solutions itself,
\begin{align}
    V_{\text{tot}} &= \left\{\phi^{}_{L_{\geqslant 0}lm}\right\}_{\mathrm{span}} \oplus \overline{V^\prime} \oplus \left\{\overline{\phi^{}_{L^{}_{<p}lm}}\right\}_{\mathrm{span}} \nonumber\\[4pt]
    &\hspace{2cm}\xrightarrow[]{\text{ dS action }}\;\; V_{\text{tot}} \,.
\end{align}

\subsubsection{Krein-Gupta-Bleuler Structure and Physical Carrier Space}

\begin{figure*}[!t]
    \centering
    \includegraphics[width=0.55\linewidth]{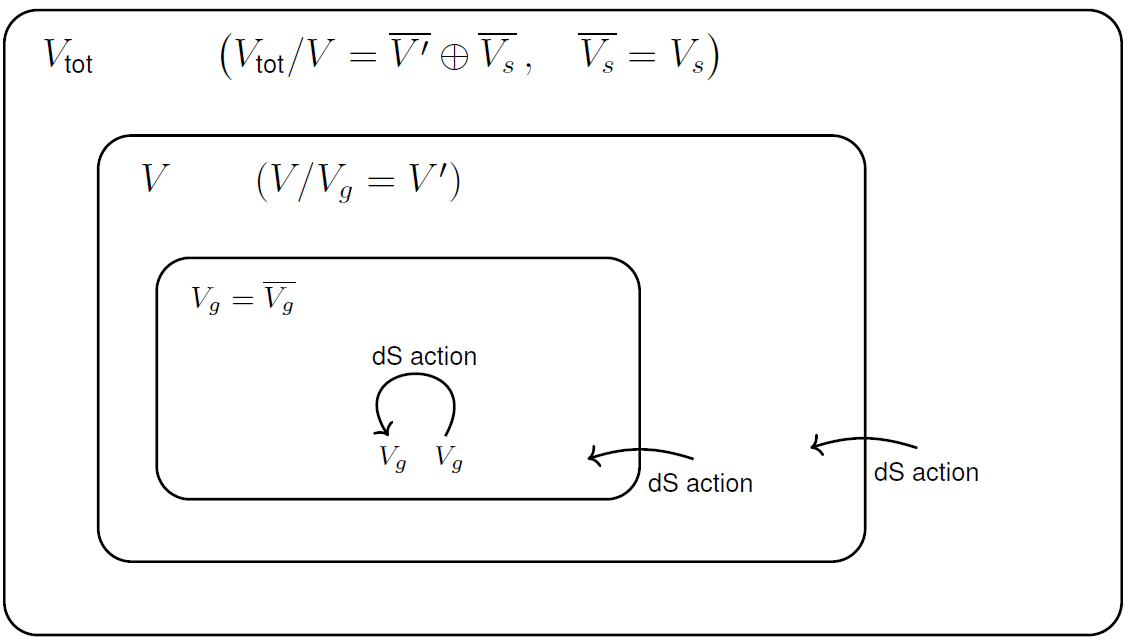}
    \caption{Euler-type schematic of the nested Krein-Gupta-Bleuler structure. The invariant chain is $V_g\subset V\subset V_{\text{tot}}$. The physical quotient is $V/V_g \simeq V^\prime$, while the remaining quotient is $V_{\text{tot}}/V \simeq \overline{V^\prime}\oplus V_s$. The identities $\overline{V_g}=V_g$ and $\overline{V_s}=V_s$ express the conjugation stability of the gauge and supplementary sectors. Arrows indicate leakage under the dS action.} 
    \label{fig:Gupta-structure}
\end{figure*}

In the preceding analysis, we showed that, starting from the KG positive-definite sector $V^\prime$, the combined requirements of dS invariance and non-degeneracy lead, through a sequence of canonical enlargements, to the total dS-invariant Krein space $V_{\mathrm{tot}}$. The resulting hierarchy of representation spaces is summarized schematically in Fig. \ref{fig:Gupta-structure}. We now show that it admits a natural interpretation in terms of a Gupta-Bleuler construction. Indeed, the invariant chain
\begin{align}\label{K-G-B}
    V_g \subset V \subset V_{\text{tot}}
\end{align}
naturally realizes a Gupta-Bleuler triplet.

The subspace $V_g$ is a dS-invariant null sector and coincides with the radical of the KG form restricted to $V$. The intermediate space $V$ contains both physical and gauge states, but remains degenerate owing to the presence of $V_g$. The physical carrier space is therefore obtained by passing to the quotient
\begin{align}
    V_{\mathrm{phys}} := V^\prime \simeq V/V_g \, .
\end{align}
The resulting space is an infinite-dimensional KG positive-definite Hilbert space. Under the above identification, the true modes $\phi^{}_{L^{}_{\geqslant p}lm}$ provide a complete set of representatives of the physical states, and their span carries the UIR $\Pi_{p,0}$; see Section \ref{Sec. IndeRepSpac}.

The ambient space of the Gupta-Bleuler construction is provided by the non-degenerate dS-invariant Krein space $V_{\text{tot}}$. In this realization, the quotient
\begin{align}
    V_{\text{tot}}/V \simeq \overline{V^\prime} \oplus V_s
\end{align}
contains the negative-definite sector together with the supplementary modes required for the non-degenerate completion of the gauge sector.

Furthermore, the decomposition displayed in Fig. \ref{fig:Gupta-structure} is compatible with complex conjugation. In particular, $\overline{V_g}=V_g$ and $\overline{V_s}=V_s$ (see Eq. \eqref{barVgs=Vgs}), while the positive- and negative-definite sectors are interchanged under complex conjugation, $V^\prime \longleftrightarrow \overline{V^\prime}$ (see Eqs. \eqref{kha30} and \eqref{kha30'}).

A noteworthy feature of the present construction is that the gauge sector does not arise as a null component of the underlying Krein space from the outset. Rather, the positive- and negative-norm sectors are already present in the non-degenerate dS-invariant Krein realization $V_{\mathrm{tot}}$, whereas the gauge sector emerges only upon passing to the intermediate invariant subspace $V$, where the restriction of the KG form becomes degenerate and admits $V_g$ as its radical. In this sense, the Gupta-Bleuler triplet is not imposed a priori; rather, it emerges intrinsically from the structure of the dS-covariant Krein realization itself.

\subsubsection{Indecomposable Representation Structure}\label{Sec. IndeRepSpac}

As shown in Section \ref{Sec. dSAction}, for each fixed $p=1,2,\cdots$, the dS action links the physical, gauge, supplementary, and negative-definite sectors through the leakage mechanisms displayed in Fig. \ref{fig:Gupta-structure}, thereby endowing $V_{\text{tot}}$ with the structure of an indecomposable dS module. At the representation-theoretic level, this structure can be summarized schematically as
\begin{align}
    \underbrace{\Pi_{p,0}}_{V_{\text{tot}}/V}
    \;\;\xrightarrow[]{\text{ dS action }}
    \underbrace{\Pi_{p,0}}_{V_{\mathrm{phys}}\,\simeq\, V/V_g}
    \xrightarrow[]{\text{ dS action }}\;\;
    \underbrace{\pi_{p,0}}_{V_g}\,.
\end{align}
The quotient spaces $V_{\mathrm{tot}}/V$ and $V/V_g$ are both infinite-dimensional and furnish realizations of the scalar discrete-series UIR $\Pi_{p,0}$, endowed respectively with negative- and positive-definite KG structures. By contrast, the dS-invariant gauge sector $V_g$ is finite-dimensional. A direct counting of the gauge modes gives
\begin{align}\label{dim}
    \dim(V_g) = \frac{1}{6}\,p(p+1)(2p+1)\,.
\end{align}
Thus, $V_g$ carries the corresponding finite-dimensional non-unitary irreducible representation $\pi_{p,0}$ of the dS group. In the notation of Ref. \cite{Gazeau2022}, this representation is labelled by
\begin{align}\label{IR}
    (n_1 \,,\, n_2) = (0 \,,\, p-1)\,.
\end{align}
The finite-dimensional non-unitary irreducible representation $\pi_{p,0}$ is Weyl equivalent to the discrete-series UIR $\Pi_{p,0}$. More precisely, for each $p=1,2,\cdots$, both representations are characterized by the same eigenvalue of the quadratic Casimir operator and hence belong to the same Weyl equivalence class \cite{Gazeau2022}.

The appearance of $\Pi_{p,0}$ and $\pi_{p,0}$ in the above diagram should not be interpreted as a decomposition of $V_{\mathrm{tot}}$ into independent dS subrepresentations. Indeed, for each $p=1,2,\cdots$, the physical, gauge, supplementary, and negative-definite sectors are linked by the dS action through the leakage mechanisms displayed in Fig. \ref{fig:Gupta-structure}. Although the gauge sector $V_g$ is itself dS-invariant and carries the finite-dimensional non-unitary irreducible representation $\pi_{p,0}$, the scalar discrete-series UIR $\Pi_{p,0}$ does not arise as an invariant subspace of $V_{\mathrm{tot}}$. Rather, it is realized only through the quotient structures
\begin{align}
    V/V_g \,, \quad\mbox{and}\quad V_{\mathrm{tot}}/V \,.
\end{align}
As a matter of fact, $V_{\mathrm{tot}}$ is an indecomposable dS-covariant Krein-Gupta-Bleuler module rather than a direct sum of irreducible dS representations.

\paragraph{Consistency check.} As a concluding remark of the present discussion and as a consistency check, we note that, in the special case $p=1$, the constructions developed in Sections \ref{Sec. ConformalCoor} and \ref{Sec. RepSpaces} reduce precisely to those previously established in Ref. \cite{Gazeau2000}.

\paragraph{Relation to Higuchi's construction.} An important part of the representation-theoretic structure underlying the present construction was already identified by Higuchi \cite{Higuchi-1987}. Working with global modes obtained by analytic continuation from scalar spherical harmonics, Higuchi established the dS invariance of the KG form, identified the positive- and zero-norm modes, and showed that the zero-norm sector—corresponding to the gauge sector in the present formulation—is invariant under the dS action, with the corresponding quotient carrying a unitary representation. The present construction incorporates this quotient structure into a larger dS-invariant indecomposable framework obtained from both linearly independent radial solutions of the wave equation. Within this framework, the gauge and supplementary sectors are identified explicitly, and the physical quotient is embedded into the non-degenerate Krein space $V_{\mathrm{tot}}$, which carries the resulting indecomposable dS representation, through the Gupta-Bleuler triplet \eqref{K-G-B}.

\subsection{Antipodal Symmetry of the Mode Solutions}

The antipodal map on $\mathrm{dS}$ is defined in the ambient space $\mathbb{R}^{1,4}$ coordinates by $x\longmapsto -x$. In global conformal coordinates $X(\rho,u) = (R\tan\rho,\, R(\sec\rho)\,u)$ \eqref{confo-coor}, this map acts as $(\rho,u)\longmapsto(-\rho,-u)$ which follows immediately from
\begin{align}
    X(\rho,u) \;\longmapsto\; -X(\rho,u) = X(-\rho,-u) \,.
\end{align}

By construction, the radial solutions \eqref{chi_p-1} and \eqref{chi_p-2} satisfy 
\begin{align}\label{radial-anti}
    \lambda_L^{(1)}(-\rho) = \overline{\lambda_L^{(1)}(\rho)}\,,\quad \lambda_L^{(2)}(-\rho) = \overline{\lambda_L^{(2)}(\rho)}\,. 
\end{align}
Furthermore, under the antipodal map $u\longmapsto -u$, we have (see Appendix \ref{A-Hyp-Har}), 
\begin{align}\label{antipodal-Yllm}
    Y_{Llm}(-u) = (-1)^L\, Y_{Llm}(u) \,.
\end{align}

Combining Eqs. \eqref{radial-anti} and \eqref{antipodal-Yllm} with the conjugation formula for $Y_{Llm}$ (see Appendix \eqref{A-Hyp-Har}), the antipodal transformation of the true modes explicitly reads
\begin{align}\label{antipodal-all-true}
    &\phi^{}_{L^{}_{\geqslant 0}lm}(\rho,u) \nonumber\\
    &\;\xrightarrow{\text{ antipodal }}\; \phi^{}_{L^{}_{\geqslant 0}lm}(-\rho,-u) = (-1)^{L+m}\,\overline{\phi^{}_{L^{}_{\geqslant 0}l,-m}(\rho,u)} \,.
\end{align}
The same computation applies to the gauge and supplementary modes. Upon using Eqs. \eqref{complex(1)} and \eqref{complex(2)} to express the complex-conjugate modes $\overline{\phi^{(g)}_{L^{}_{<p},l,-m}}$ and $\overline{\phi^{(s)}_{L^{}_{<p},l,-m}}$ in terms of the original mode basis, one finds
\begin{align}
    \phi^{(g)}_{L^{}_{<p}lm}(-\rho,-u) &= (-1)^{p+1}\,\phi^{(g)}_{L^{}_{<p}lm}(\rho,u) \,, \label{antipodal-gauge}\\[3pt]
    \phi^{(s)}_{L^{}_{<p}lm}(-\rho,-u) &= (-1)^{p}\,\phi^{(s)}_{L^{}_{<p}lm}(\rho,u) \,. \label{antipodal-supp}
\end{align}

The role of the antipodal relations derived above becomes apparent in the next section, where they provide the bridge between the boundary realizations at ${\mathcal{I}}^+$ and ${\mathcal{I}}^-$ and clarify the corresponding boundary representation structure.

\paragraph{Possible physical interpretation.} A noteworthy consequence of the antipodal structure is that, at the level of the physical sector for a given $p=1,2,\cdots$, the antipodal map \eqref{antipodal-all-true} exchanges the positive-definite realization $V^\prime = \left\{ \phi^{}_{L^{}_{\geqslant p}lm} \right\}_{\mathrm{span}}$ with its complex-conjugate counterpart $\overline{V^\prime} = \left\{ \overline{\phi^{}_{L^{}_{\geqslant p}lm}} \right\}_{\mathrm{span}}$. Although both spaces furnish equivalent realizations of the same discrete-series representation $\Pi_{p,0}$, this exchange suggests that antipodal symmetry may encode information beyond its purely representation-theoretic role. In particular, it bears a suggestive resemblance to the observer-dependent particle-antiparticle correspondence proposed recently for quantum fields on dS spacetime \cite{GazeauEPL}, according to which antipodally related regions, endowed with opposite time orientations, give rise to complementary particle and antiparticle descriptions of the same underlying system. The present work does not address this question at the level of quantum field theory. Nevertheless, the emergence of the exchange $V^\prime \longleftrightarrow \overline{V^\prime}$ as an intrinsic feature of the discrete-series realization suggests that the interplay between antipodal symmetry, time orientation, and particle interpretation may possess a deeper representation-theoretic origin. Clarifying this connection and its possible implications for quantum field theory in dS spacetime constitutes an interesting direction for future investigation.

\section{Boundary Realization}\label{Sec. Boundary}

In this section, we construct the boundary realization of the bulk Krein-Gupta-Bleuler representation space by analyzing its asymptotic behavior at the future and past conformal boundaries, ${\mathcal{I}}^+$ and ${\mathcal{I}}^-$. This is achieved by taking the boundary limits $\rho\longrightarrow\pm\frac{\pi}{2}$ of the bulk modes and determining the induced representation structure on the conformal boundaries.

The antipodal relations established in Eq. \eqref{antipodal-all-true} imply that the boundary realizations on ${\mathcal{I}}^+$ and ${\mathcal{I}}^-$ are not independent; the boundary data on ${\mathcal{I}}^-$ is completely determined by that on ${\mathcal{I}}^+$ through complex conjugation together with the transformation $m\longmapsto -m$. Consequently, it is sufficient to construct the boundary realization on ${\mathcal{I}}^+$. Nevertheless, for consistency, we shall also examine the corresponding realization on ${\mathcal{I}}^-$ and verify explicitly that the two are related by the antipodal symmetry.

\subsection{The Boundary Function Spaces}

In the global conformal coordinates $X=X(\rho,u)$ \eqref{confo-coor}, the dS hyperboloid is parametrized by $(\rho,u)\in\left(-\frac{\pi}{2},\frac{\pi}{2}\right)\times{\mathbb{S}}^3$. The future and past conformal boundaries are reached in the limits $\rho\longrightarrow \pm\frac{\pi}{2}$, respectively. Both boundaries are naturally identified with copies of ${\mathbb{S}}^3$, the identification being provided by the angular coordinate $u$ \eqref{kha48}.

More precisely, the maps
\begin{align}\label{eq:I+-diffeo}
    {\mathcal{I}}^\pm &\;\;\longrightarrow\;\; {\mathbb{S}}^3 \,, \nonumber\\[3pt]
    \lim_{\rho\longrightarrow\pm\frac{\pi}{2}}X(\rho,u) &\;\;\longmapsto\;\; u \,,
\end{align}
are diffeomorphisms. Under these identifications, the natural measure on ${\mathcal{I}}^\pm$, induced by the conformal compactification, coincides with the standard $\mathrm O(4)$-invariant measure ${\mathrm{d}}u$ on ${\mathbb{S}}^3$ introduced in Eq. \eqref{measure}. Consequently, functions on ${{\mathcal{I}}}^{\pm}$ may be identified canonically with functions on ${\mathbb{S}}^3$, and integrals over ${{\mathcal{I}}}^{\pm}$ reduce to integrals over ${\mathbb{S}}^3$.

We therefore equip the boundary function space with the standard $L^2$ inner product
\begin{align}\label{eq:L2-inner}
    \left\langle f_1\,,\, f_2\right\rangle^{}_{L^2} := \int_{\mathbb{S}^3} \overline{f_1(u)}\,f_2(u)\,\mathrm{d}u\,.
\end{align}
The corresponding Hilbert space is $L^2_{\mathbb C}({\mathbb{S}}^3)$, while the boundary modes constructed below belong to its dense subspace $C^\infty({\mathbb{S}}^3)$.

The hyperspherical harmonics ${Y_{Llm}}(u)$ form a complete orthogonal basis of $L^2_{\mathbb C}({\mathbb{S}}^3)$ with respect to the $L^2$ inner product \eqref{eq:L2-inner}. Consequently, every function $f\in L^2_{\mathbb C}({\mathbb{S}}^3)$ admits the expansion
\begin{align}
    f = \sum_{L\geqslant 0} \sum_{l,m} \hat{f}_{Llm}\,Y_{Llm}
\end{align}
where $\hat{f}_{Llm} = \left\langle Y_{Llm}\,,\,f\right\rangle^{}_{L^2}$.

For each $p=1,2,\cdots$, the space $L^2_{\mathbb C}({\mathbb{S}}^3)$ admits the orthogonal decomposition
\begin{align}\label{eq:L2-decomp2}
    L^2_{\mathbb{C}}(\mathbb{S}^3) = H_{p-1}\oplus V_{p-1}\,,
\end{align}
where 
\begin{align}
    V_{p-1} &:= \left\{Y_{L^{}_{<p}lm}\right\}_{\mathrm{span}}\,, \label{eq:Vp-1-def}\\[3pt]
     H_{p-1}  &:= \left\{f\in L^2_{\mathbb{C}}(\mathbb{S}^3)\;;\;\left\langle Y_{L^{}_{<p}lm}\,,\,f\right\rangle^{}_{L^2} = 0\right\} \,. \label{eq:Hp-1-def}
\end{align}
As shown by Takahashi \cite{Takahashi}, the space $V_{p-1}$ is finite-dimensional, with dimension given by Eq. \eqref{dim}, and carries the irreducible representation $\pi_{p,0}$ of $\mathrm{SO}_0(1,4)$, which is Weyl equivalent to the discrete-series UIR $\Pi_{p,0}$; see Eq. \eqref{IR} and the related discussion. The orthogonal complement $H_{p-1}$ is infinite-dimensional, and can be identified, as a vector space, with the quotient $L^2_{\mathbb C}({\mathbb{S}}^3)/V_{p-1}$.

To endow this quotient realization with its natural $\mathrm{SO}_0(1,4)$-invariant Hilbert structure, we must go beyond the $L^2$ inner product. Indeed, the inner product \eqref{eq:L2-inner} is invariant under the natural action of the compact subgroup $\mathrm O(4)$ on ${\mathbb{S}}^3$. However, it is not the invariant Hilbert structure associated with the full $\mathrm{SO}_0(1,4)$ realization of the discrete-series UIR $\Pi_{p,0}$ \cite{Takahashi}. To describe this realization, we introduce the $\mathrm{SO}_0(1,4)$-invariant sesquilinear form \cite{Takahashi}
\begin{align}\label{eq:Hp-1-inner}
    \left\langle f_1\,,\, f_2 \right\rangle_{p-1} &= \frac{(-1)^{p+1}}{4\pi^2(2p-1)!} \iint_{\mathbb{S}^3\times\mathbb{S}^3} \overline{f_1(u_1)}\,f_2(u_2) \nonumber\\[3pt]
    &\quad\times|u_1-u_2|^{2(p-1)}\log|u_1-u_2|^{-2}\,\mathrm{d}u_1\,\mathrm{d}u_2\,.
\end{align}
The subspace $V_{p-1}$ lies in the radical of this form. The form descends to a positive-definite $\mathrm{SO}_0(1,4)$-invariant inner product on the quotient $C^\infty({\mathbb{S}}^3)/V_{p-1}$, thereby yielding the Hilbert-space realization of $\Pi_{p,0}$. Via the decomposition \eqref{eq:L2-decomp2}, this quotient is naturally identified with $H_{p-1}$.

An orthonormal basis of $H_{p-1}$ with respect to the induced inner product \eqref{eq:Hp-1-inner} is given by
\begin{align}\label{eq:5.78}
    \psi^{}_{L^{}_{\geqslant p}lm}(u) := \sqrt{\frac{\Gamma(L-p+1)}{\Gamma(L+p+2)}}\; Y_{L^{}_{\geqslant p}lm}(u) \,,
\end{align}
which satisfies  
\begin{align}
    \left\langle \psi^{}_{L^{}_{\geqslant p}lm} \;,\; \psi^{}_{L^{\prime}_{\geqslant p}l^\prime m^\prime} \right\rangle_{p-1} &= \delta_{LL^\prime} \,\delta_{ll^\prime} \,\delta_{mm^\prime} \label{orthonormality} \,.
\end{align}
For convenience, we also introduce
\begin{align} \label{gaugeBoundaryModes}
    \psi^{}_{L^{}_{<p}lm}(u) := Y_{L_{<p}lm}(u) \,,
\end{align}
which belong to the radical of the sesquilinear form
\eqref{eq:Hp-1-inner}. Consequently,
\begin{align}\label{orthonormality'}
    \left\langle \psi^{}_{L^{}_{\geqslant p}lm} \;,\; \psi^{}_{L^{\prime}_{< p}l^\prime m^\prime} \right\rangle_{p-1} = 0 \,.
\end{align}

\paragraph{Absence of an independent complex-conjugate sector.} Notably, unlike the bulk KG structure, the complex-conjugate modes $\overline{\psi^{}_{L^{}_{\geqslant p}lm}}$ possess the same norm as their unconjugated counterparts in $(H_{p-1},\langle\cdot,\cdot\rangle_{p-1})$:
\begin{align}\label{eq:bdry-norm}
    \left\langle\overline{\psi^{}_{L^{}_{\geqslant p}lm}} \;,\;  \overline{\psi_{L^\prime_{\geqslant p} l^\prime m^\prime}}\right\rangle_{p-1} &= \delta_{LL^\prime} \,\delta_{ll^\prime} \,\delta_{mm^\prime}\,.
\end{align}
The two families are not mutually orthogonal. Rather, they satisfy
\begin{align}\label{eq:bdry-cross}
    \left\langle\overline{\psi^{}_{L^{}_{\geqslant p}lm}} \;,\; \psi_{L^\prime_{\geqslant p} l^\prime m^\prime}\right\rangle_{p-1} &= (-1)^m \,\delta_{LL^\prime} \,\delta_{ll^\prime} \,\delta_{m^\prime,-m}\,.
\end{align}
Note that the above relations follow readily from Eq. \eqref{cop}. Consequently, complex conjugation does not generate an independent orthogonal sector in the boundary realization. This is further reflected in the fact that both $H_{p-1}$ and the finite-dimensional space $V_{p-1}$ are invariant under point-wise complex conjugation (see, again, Eq. \eqref{cop}). More precisely,
\begin{align}
    \overline{H_{p-1}} = H_{p-1}\,, \quad \overline{V_{p-1}} = V_{p-1} \,.
\end{align}

\subsection{Boundary Limits of the Bulk Modes}

For a given $p=1,2,\cdots$, using the asymptotic formulas collected in Appendix \ref{App. Linear Independence of the Two Solutions}, the true modes $\phi^{}_{L^{}_{\geqslant p}lm}$ \eqref{true>=p} and the gauge modes $\phi^{(g)}_{L^{}_{<p}lm}$ \eqref{gauge}, both built from $\lambda_L^{(1)}$, behave as $(\cos\rho)^{1-p}$ as $\rho\longrightarrow\pm\frac{\pi}{2}$ and are therefore singular, whereas the supplementary modes $\phi^{(s)}_{L^{}_{<p}lm}$ \eqref{supp}, built from $\lambda_L^{(2)}$, behave as $(\cos\rho)^{p+2}$ and therefore tend to zero. To extract finite boundary representatives from the singular mode families, we introduce the scale and phase factors
\begin{align}
    \mathfrak{S}^{}_s &:= R\, \frac{2^{(p+{1}/{2})}\, \Gamma(p+1)}{\Gamma(2p+1)}\; (\cos\rho)^{p-1}\,, \\[3pt]
    \mathfrak{S}^{\pm}_p &:= e^{\pm\mathrm{i}(L-p+1)\frac{\pi}{2}} \,,
\end{align}
and define the boundary representatives through the corresponding rescaled limits
\begin{align}\label{eq:L-def}
    \!\mathcal{L}_\pm \,\colon\;\, V_{\text{tot}} &\;\longrightarrow\; L^2_{\mathbb{C}}(\mathbb{S}^3)\,, \nonumber\\[3pt]
    \phi(\rho,u) &\;\longmapsto\; \mathcal{L}_\pm\left(\phi(\rho,u)\right) := \lim_{\rho\longrightarrow\pm\frac{\pi}{2}} \left(\mathfrak{S}^{}_s\,\mathfrak{S}^\pm_p\, \phi(\rho,u)\right) .
\end{align}

In particular, combining the results of Appendix \ref{App. Linear Independence of the Two Solutions} with Eqs. \eqref{normalization constants1}-\eqref{normalization constants3} and \eqref{cop}, one obtains
\begin{align}
    \mathcal{L}_\pm\left(\phi^{}_{L^{}_{\geqslant p}lm}(\rho,u)\right) =&\, \lim_{\rho\longrightarrow \pm\frac{\pi}{2}}\left(\mathfrak{S}^{}_s\,\mathfrak{S}^\pm_p\, \phi^{}_{L^{}_{\geqslant p}lm}(\rho,u)\right)\nonumber\\[3pt]
    =&\, \psi^{}_{L^{}_{\geqslant p}lm}(u) \;\in H_{p-1}\,, \label{eq:L-complete-final}
\end{align}
\begin{align}
    \mathcal{L}_\pm\left(\phi^{}_{L^{}_{< p}lm}(\rho,u)\right) =&\, \lim_{\rho\longrightarrow \pm\frac{\pi}{2}} \left(\mathfrak{S}^{}_s\,\mathfrak{S}^\pm_p\, \phi^{}_{L^{}_{<p}lm}(\rho,u)\right)\nonumber\\[3pt]
    =&\, \lim_{\rho\longrightarrow \pm\frac{\pi}{2}} \left(\mathfrak{S}^{}_s\,\mathfrak{S}^\pm_p\, \phi^{(g)}_{L^{}_{<p}lm}(\rho,u)\right)\nonumber\\[3pt]
    =&\, \mathcal{N}_{L_{< p}} \,\psi^{}_{L^{}_{<p}lm}(u) \;\in V_{p-1}\,, \label{bdry-true2}
\end{align}
\begin{align}
    \!\!\!\mathcal{L}_\pm \left(\overline{\phi^{}_{L^{}_{\geqslant p}lm}(\rho,u)}\right) =&\, \lim_{\rho\longrightarrow \pm\frac{\pi}{2}}\left(\mathfrak{S}^{}_s\,\mathfrak{S}^\pm_p\, \overline{\phi^{}_{L^{}_{\geqslant p}lm}(\rho,u)}\right)\nonumber\\[3pt]
    =&\, (-1)^{L-p+1+m}\, \psi^{}_{L^{}_{\geqslant p},l,-m}(u) \nonumber\\[3pt]
    &\hspace{1.5cm}\in H_{p-1}\,, \label{eq:L-complete-final'}
\end{align}
\begin{align}
    \!\!\mathcal{L}_\pm\left(\overline{\phi^{}_{L^{}_{< p}lm}(\rho,u)}\right) \!=& \lim_{\rho\longrightarrow \pm\frac{\pi}{2}} \left(\mathfrak{S}^{}_s\,\mathfrak{S}^\pm_p\, \overline{\phi^{}_{L^{}_{<p}lm}(\rho,u)}\right)\nonumber\\[3pt]
    =& \lim_{\rho\longrightarrow \pm\frac{\pi}{2}} \left(\mathfrak{S}^{}_s\,\mathfrak{S}^\pm_p\, \overline{\phi^{(g)}_{L^{}_{<p}lm}(\rho,u)}\right)\nonumber\\[3pt]
    =&\, (-1)^{L-p+1+m}\, \mathcal{N}_{L_{< p}} \,\psi^{}_{L^{}_{<p},l,-m}(u) \nonumber\\[3pt]
    &\hspace{1.5cm} \in V_{p-1}\,, \label{bdry-true2'}
\end{align}
where $\mathcal{N}_{L^{}_{< p}} := {\Gamma(L+2)}/{\Gamma(L+p+2)}$. Note that, although the phase factors are physically immaterial, we retained them explicitly throughout the above derivation of the boundary limits in order to make the limiting procedure transparent and to enable the reader to follow the correspondence between the bulk and boundary realizations.

\paragraph{Absence of boundary representatives for the supplementary modes.} It follows directly from the relations established in Appendix \ref{App. Linear Independence of the Two Solutions} that
\begin{align}
    \mathcal{L}_\pm\left(\phi^{(s)}_{L^{}_{<p}lm}(\rho,u)\right) =\! \lim_{\rho \longrightarrow \pm\frac{\pi}{2}} \left(\mathfrak{S}^{}_s\,\mathfrak{S}^\pm_p \,\phi^{(s)}_{L^{}_{<p}lm}(\rho,u)\right) = 0 \,.
\end{align}
Thus, the supplementary modes admit no non-trivial boundary representatives at ${\mathcal{I}}^\pm$ and therefore play no further role in the boundary construction. 

\paragraph{Absence of an independent boundary negative-definite sector.} The boundary limits obtained above make manifest a feature already encoded in the Hilbert-space structure of $H_{p-1}$. Indeed, for a given $p=1,2,\cdots$, Eqs. \eqref{eq:L-complete-final}-\eqref{bdry-true2'} show that the images of the positive- and negative-definite bulk sectors under ${\mathcal{L}}_\pm$ coincide, the corresponding boundary representatives differing only by the replacement $m\longmapsto -m$ (up to a phase). Thus, the negative-definite bulk sector acquires no independent boundary counterpart, and the resulting boundary realization contains no separate negative-definite sector.

\paragraph{Indecomposable structure of the boundary realization.} For a given $p=1,2,\cdots$, using the material presented in {Appendix \ref{App. Action} and consistently with the discussion below Eqs. \eqref{eq:Vp-1-def} and \eqref{eq:Hp-1-def}, the action of the dS group on \begin{align}
    L^2_{\mathbb C}({\mathbb{S}}^3) = H_{p-1}\oplus V_{p-1}
\end{align}
satisfies
\begin{align}
    H_{p-1} \;\; &\xrightarrow[]{\text{ dS action }} \;\; H_{p-1}\oplus V_{p-1} \,, \label{actionHp-1} \\[3pt]
    V_{p-1} \;\; &\xrightarrow[]{\text{ dS action }} \;\; V_{p-1} \,.\label{actionVp-1}
\end{align}
Thus, $V_{p-1}$ is an invariant subspace, whereas $H_{p-1}$ is not. Consequently, the dS action on $L^2_{\mathbb C}({\mathbb{S}}^3)$ is indecomposable. This boundary indecomposable structure should be compared with its bulk counterpart. In the bulk, the dS action organizes the representation into three indecomposable constituents associated with the gauge, physical, and negative-definite sectors (see Fig. \ref{fig:Gupta-structure}). By contrast, the boundary construction retains only two independent constituents. Indeed, as established above, the negative-definite bulk sector admits no independent boundary representative, its image under the boundary maps ${\mathcal{L}}_\pm$ coinciding with that of the positive-definite sector. Consequently, the boundary realization preserves the gauge and physical structures inherited from the bulk, while the Krein doubling present in the bulk realization disappears. The resulting boundary representation therefore remains non-semisimple and indecomposable, but with a simpler two-step structure in which the invariant subspace $V_{p-1}$ plays the role of the gauge sector, while $H_{p-1}$ provides its non-invariant complement carrying the physical degrees of freedom.
  
\paragraph{Antipodal compatibility of the boundary realization.} As a non-trivial consistency check, Eqs. \eqref{eq:L-complete-final}-\eqref{bdry-true2'} are fully compatible with the bulk antipodal relation \eqref{antipodal-all-true}. As illustrated in Fig. \ref{Fig. antipodal}, the antipodal map in the bulk sends the mode $\phi^{}_{L_{\geqslant 0}lm}(\rho,u)$ to its antipodal partner $\phi^{}_{L_{\geqslant 0}lm}(-\rho,-u) = (-1)^{L+m}\, \overline{\phi^{}_{L_{\geqslant 0}l,-m}(\rho,u)}$. The boundary maps, either ${\mathcal L}_+$ or ${\mathcal L}_-$, then associate these antipodally related bulk modes with boundary representatives living on opposite conformal boundaries. More precisely, following ${\mathcal L}_+$, the mode $\phi^{}_{L_{\geqslant 0}lm}(\rho,u)$ is represented on ${{\mathcal{I}}}^+$, while the boundary representative of its antipodal partner is realized on ${{\mathcal{I}}}^-$. Following ${\mathcal L}_-$, the same correspondence is obtained with the roles of ${{\mathcal{I}}}^+$ and ${{\mathcal{I}}}^-$ interchanged. Independently, the horizontal arrows on the upper and lower rows implement the antipodal involution $u\longmapsto -u$ on each boundary sphere ${{\mathbb{S}}}^3$; $\psi^{}_{L^{}_{\geqslant 0} lm} (-u) = (-1)^L \,\psi^{}_{L^{}_{\geqslant 0} lm} (u)$ (see Eq. \eqref{antipodal-Yllm1}). Figure \ref{Fig. antipodal}, therefore, exhibits two mutually compatible antipodal structures: the bulk antipodal map relates pairs of modes whose boundary representatives lie on opposite conformal boundaries, while the boundary antipodal map relates antipodal representatives within each individual boundary realization. Consequently, the realizations on ${{\mathcal{I}}}^+$ and ${{\mathcal{I}}}^-$ are not independent. Rather, they constitute mirror realizations of the same discrete-series representation, each mode on one boundary being paired with a unique antipodal counterpart on the other. Thus, the future and past boundary realizations carry equivalent information, differing only through the antipodal organization inherited from the bulk theory.

\begin{figure}[h]
    \begin{center}
    \begin{tikzpicture}[scale=0.97,
        arr/.style ={latex-latex, thick, line width=1pt},
        sarr/.style={-latex, thick, line width=1pt},
        lbl/.style ={font=\scriptsize, fill=white, inner sep=2pt},]
 
    \node (A) at (0,  6)
        {$\mathcal{N}^{}_{L^{}_{\geqslant 0}}
        \psi^{}_{L^{}_{\geqslant 0}lm}(u)$};
    \node (B) at (6,  6)
        {$\mathcal{N}^{}_{L^{}_{\geqslant 0}}
        \psi^{}_{L^{}_{\geqslant 0}lm}(u)$};
    \node (C) at (0,  3)
        {$\phi^{}_{L^{}_{\geqslant 0}lm}(\rho,u)$};
    \node (D) at (6,  3)
        {$\overline{\phi^{}_{L^{}_{\geqslant 0}l,-m}(\rho,u)}$};
    \node (E) at (0,  0)
        {$\mathcal{N}^{}_{L^{}_{\geqslant 0}}
        \psi^{}_{L^{}_{\geqslant 0}lm}(u)$};
    \node (F) at (6,  0)
        {$\mathcal{N}^{}_{L^{}_{\geqslant 0}}
        \psi^{}_{L^{}_{\geqslant 0}lm}(u)$};
 
    \draw[arr] (A) -- (B)
        node[lbl, midway, below] {boundary ${\mathcal{I}}^+$}
        node[lbl, midway, above]{$u \longleftrightarrow-u$};
    \draw[arr] (C) -- (D)
        node[lbl, midway, below] {dS bulk}
        node[lbl, midway, above]{$(\rho,u)\longleftrightarrow(-\rho,-u)$};
    \draw[arr] (E) -- (F)
        node[lbl, midway, below] {boundary ${\mathcal{I}}^-$}
        node[lbl, midway, above]{$u\longleftrightarrow -u$};
 
    \draw[sarr] (C) -- (A)
        node[lbl, midway, left]{$\mathcal{L}_+$};
    \draw[sarr] (C) -- (E)
        node[lbl, midway, left]{$\mathcal{L}_-$};
    \draw[sarr] (D) -- (B)
        node[lbl, midway, right]{$\mathcal{L}_-$};
    \draw[sarr] (D) -- (F)
        node[lbl, midway, right]{$\mathcal{L}_+$};
    \end{tikzpicture}
    \end{center}
    \caption{Interplay between the bulk antipodal map and the boundary realizations on ${\mathcal{I}}^{\pm}$, \textbf{modulo phase factors}. Here, $\mathcal{N}^{}_{L^{}_{\geqslant p}}=1$, while $\mathcal{N}^{}_{L^{}_{<p}}=\Gamma(L+2)/\Gamma(L+p+2)$.}
    \label{Fig. antipodal}
\end{figure}
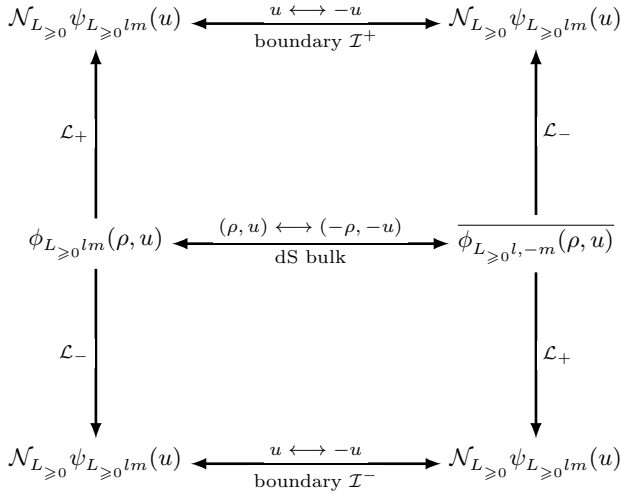

\section{Fourier-Type Bulk-Boundary Intertwiner and Unitary Equivalence}\label{Sec. Fourier}

We now turn to the bulk-boundary correspondence associated with the scalar discrete-series representation $\Pi_{p,0}$, for fixed $p=1,2,\cdots$. As established in the preceding sections, the bulk and boundary realizations exhibit markedly different indecomposable structures. The bulk theory is carried by the indecomposable Krein-Gupta-Bleuler module $V_{\mathrm{tot}}$, whose physical content $\Pi_{p,0}$ is encoded in the quotient space $V/V_g \simeq V^\prime$ (see Fig. \ref{fig:Gupta-structure}). By contrast, the boundary realization possesses a simpler two-step indecomposable structure $H_{p-1}\oplus V_{p-1}$, where $H_{p-1}$ furnishes the natural carrier space of the discrete-series representation $\Pi_{p,0}$. Unlike its bulk counterpart, the boundary construction contains neither an independent negative-definite sector nor non-trivial supplementary modes; its only non-physical component is the finite-dimensional invariant subspace $V_{p-1}$.

Consequently, the correspondence developed below is not defined between the full bulk and boundary modules, but between the respective physical sectors supporting the representation $\Pi_{p,0}$. We shall show that it is implemented by a Fourier-type transform $\mathcal{F}$ whose restriction to these sectors defines an inner-product-preserving bijection $V^\prime \,\longleftrightarrow\, H_{p-1}$, while intertwining the corresponding dS group actions. In this sense, $\mathcal{F}$ furnishes a precise holographic map relating the bulk and boundary realizations of the discrete-series representation $\Pi_{p,0}$.

We recall that a bulk dS point $X=X(\rho,u)$ is parametrized by the global conformal coordinates $(\rho,u)$ introduced in Eq. \eqref{confo-coor}. Let $v_\infty\in{{\mathbb{S}}}^3$ denote an independent boundary variable. Since the antipodal map relates the future and past conformal boundaries (see Fig. \ref{Fig. antipodal}), ${{\mathbb{S}}}^3$ may be regarded as a common realization of either boundary, and no distinction will be made between them at this stage. For each $(X,v_\infty)\in \mathrm{dS}\times{{\mathbb{S}}}^3$ and a fixed $p=1,2,\cdots$, we introduce the bulk-boundary intertwining kernel
\begin{align}\label{eq:K-ds-corrected}
    K=K(X,v_{\infty}) := \sum_{L=p}^{\infty}\sum_{l,m} \phi^{}_{L^{}_{\geqslant p}lm}(\rho,u)\; \overline{\psi^{}_{L^{}_{\geqslant p}lm} (v_{\infty})} \,.
\end{align}

Consider the map
\begin{align}\label{eq:F-def-corrected}
    \mathcal{F} \;\colon\;\; V_{\text{tot}} &\;\longrightarrow\; H_{p-1} \oplus V_{p-1} \,,\nonumber\\[3pt]
    \phi_{Llm}^{}\left(X(\rho,u)\right) &\;\longmapsto\; \left\langle \, K(X, v_{\infty})\,,\, \phi_{Llm}^{}\left(X(\rho,u)\right) \,\right\rangle_{\vphantom{}_{KG}}\,.
\end{align}
For a given $p=1,2,\cdots$, using Eqs. \eqref{kha30}-\eqref{thelastone'}, one finds that its action on the basis vectors is
\begin{align}\label{F-action}
    V^\prime \;\ni\; \phi^{}_{L^{}_{\geqslant p}lm}(\rho,u) \;\;\longmapsto&\;\; {\mathcal{F}} \left(\phi^{}_{L^{}_{\geqslant p}lm}(\rho,u)\right) \nonumber\\[3pt]
    &\quad= \psi^{}_{L^{}_{\geqslant p}lm}(v_\infty) \;\in\; H_{p-1}\,, \\[3pt]
    \phi^{}_{L^{}_{<p}lm}(\rho,u) \;\;\longmapsto&\;\; {\mathcal{F}} \left(\phi^{}_{L^{}_{<p}lm}(\rho,u) \right) = 0 \,, \\[3pt]
    \overline{\phi^{}_{L^{}_{\geqslant p}lm}(\rho,u)} \;\;\longmapsto&\;\;{\mathcal{F}} \left(\overline{\phi^{}_{L^{}_{\geqslant p}lm}(\rho,u)} \right) = 0 \,, \\[3pt]
    \overline{\phi^{}_{L^{}_{<p}lm}(\rho,u)} \;\;\longmapsto&\;\;{\mathcal{F}} \left(\overline{\phi^{}_{L^{}_{<p}lm}(\rho,u)} \right) = 0\,.
\end{align}
Hence, $\mathcal F$ is a well-defined linear map, although it is not injective.

Conversely, consider the map
\begin{align}\label{eq:F-inv-L2}
    \mathcal{G}\colon H_{p-1}\oplus V_{p-1} & \;\longrightarrow\;V_{\text{tot}}\,,\nonumber\\[3pt]
    \psi_{Llm}^{}(v_\infty) & \;\longmapsto\; \left\langle \overline{K(X,v_\infty)},\,\psi_{Llm}^{}(v_\infty)\right\rangle_{p-1} \,.
\end{align}
For a given $p=1,2,\cdots$, using the orthonormality relations \eqref{orthonormality} and \eqref{orthonormality'}, its action on the basis vectors is
\begin{align}\label{inversaction}
    H_{p-1} \;\ni\; \psi^{}_{L^{}_{\geqslant p}lm}(v_\infty) \;\;\longmapsto&\;\; {\mathcal{G}} \left(\psi^{}_{L^{}_{\geqslant p}lm}(v_\infty)\right) \nonumber\\[3pt]
    &\quad= \phi^{}_{L^{}_{\geqslant p}lm}(\rho,u) \;\in\; V^\prime\,, \\[3pt] 
    V_{p-1} \;\ni\; \psi^{}_{L^{}_{< p}lm}(v_\infty) \;\;\longmapsto&\;\; {\mathcal{G}} \left(\psi^{}_{L^{}_{<p}lm}(v_\infty)\right) = 0 \,.
\end{align}
Thus, $\mathcal G$ is likewise well defined, but not injective.

The maps $\mathcal{F}$ and $\mathcal{G}$ are not mutually inverse on their full domains. However, Eqs. \eqref{F-action} and \eqref{inversaction} imply that
\begin{align}
    \mathcal{G}\circ\mathcal{F} \Big|_{V^\prime} = \mathrm{id}\,, \quad\mbox{and}\quad \mathcal{F}\circ\mathcal{G} \Big|_{H_{p-1}} = \mathrm{id} \,.
\end{align}
This shows that, upon restriction to the physical sector $V^\prime$, the map $\mathcal F$ becomes a bijection onto $H_{p-1}$, with inverse given by $\mathcal G$. Hence,
\begin{align}
    \mathcal{F} &\;\colon\;\; \;\;V^\prime \quad\longrightarrow\; H_{p-1} \,, \nonumber\\[3pt]
    \mathcal{F}^{-1} :=\mathcal{G} &\;\colon\;\; H_{p-1} \;\longrightarrow\; V^\prime 
\end{align}
define a bijection between the two carrier spaces $V^\prime$ and $H_{p-1}$.

Furthermore, using Eqs. \eqref{kha30} and \eqref{orthonormality}, one finds
\begin{align}
    &\left\langle \mathcal{F}\left(\phi^{}_{L^{}_{\geqslant p}lm}\right) \,,\, \mathcal{F}\left(\phi_{L^\prime_{\geqslant p} l^\prime m^\prime}\right)\right\rangle_{p-1}
    \nonumber\\[3pt]
    &\hspace{3cm}= \left\langle\phi^{}_{L^{}_{\geqslant p}lm} \,,\, \phi_{L^\prime_{\geqslant p} l^\prime m^\prime}\right\rangle_{\vphantom{}_{KG}}\,.
\end{align}
Since the modes $\phi^{}_{L^{}_{\geqslant p}lm}$ form an orthonormal basis of $V^\prime$, the above identity extends by linearity and continuity to all vectors in $V^\prime$. Consequently, $\mathcal{F}$ preserves the inner product and defines a bijection
\begin{align}\label{Fourier}
    \mathcal{F} \;\colon\;\; \left(V^\prime \;,\;\left\langle\cdot\,,\,\cdot\right\rangle_{\vphantom{}_{KG}} \right) \;\;\xrightarrow{\;\simeq\;}\;\; \left(H_{p-1} \;,\;\left\langle\cdot \,,\,\cdot\right\rangle_{p-1} \right)\,.
\end{align}

For a given $p=1,2,\cdots$, the space $V^\prime$ carries the bulk physical realization of the discrete-series representation $\Pi_{p,0}$, endowed with the KG inner product, while $H_{p-1}$ carries the corresponding boundary realization on $\mathbb{S}^{3}$, equipped with the invariant inner product $\langle\cdot,\cdot\rangle_{p-1}$. The Fourier-type transform $\mathcal{F}$ establishes a bijective correspondence between these two spaces and preserves their respective inner products. Moreover, as shown in Appendix \ref{App. Action}, $\mathcal{F}$ intertwines the induced actions of the dS algebra on the corresponding physical quotient spaces. Consequently, $\mathcal{F}$ is simultaneously an isometric isomorphism and an intertwining operator, thereby establishing the equivalence of the bulk and boundary realizations of the scalar discrete-series representation $\Pi_{p,0}$. In this sense, $\mathcal{F}$ provides the dS-covariant bulk-boundary correspondence associated with $\Pi_{p,0}$.

\section{Conclusion and Outlook}\label{Sec. conclusion}

In this paper, we developed a unified framework for the scalar discrete-series representations $\Pi_{p,0}$, $p=1,2,\cdots$, of the dS group $\mathrm{SO}_0(1,4)$, encompassing both their bulk realization in dS spacetime and their boundary realization at conformal infinity. Within this framework, we established the indecomposable dS-invariant Krein-Gupta-Bleuler triplet structure of the bulk realization together with the corresponding indecomposable dS-invariant boundary douplet structure at conformal infinity, identified the antipodal correspondence between the future and past boundary realizations, and derived a natural holographic correspondence between the physical sectors of the bulk and boundary realizations. The main results are summarized below.

\subsection{Conclusion}

\paragraph{Bulk realization of the scalar discrete series.} For each $p=1,2,\cdots$, we constructed an explicit spacetime realization of the scalar discrete-series representation $\Pi_{p,0}$ by solving its quadratic Casimir eigenvalue equation in global conformal coordinates. Interpreting this equation as a dS-covariant wave equation, we obtained a complete mode decomposition and identified the associated true, gauge, and supplementary sectors. While the true modes and their complex conjugates span non-degenerate invariant subspaces carrying equivalent realizations of $\Pi_{p,0}$, the gauge and supplementary sectors emerge as indispensable constituents of the full indecomposable dS-covariant realization.

Specifically, we showed that the complete mode space carries a natural dS-invariant Krein structure on which the dS group acts indecomposably, with the true, gauge, and supplementary sectors forming a Gupta-Bleuler triplet. The invariant KG sesquilinear form is degenerate on a distinguished invariant subspace whose radical coincides with the gauge sector, while the corresponding quotient canonically realizes the physical Hilbert space carrying the UIR $\Pi_{p,0}$.

\paragraph{Antipodal structure.} The antipodal symmetry of dS spacetime plays a fundamental role in the present framework. We showed that it acts consistently on all sectors of the bulk realization and is fully compatible with its Krein-Gupta-Bleuler structure. Beyond being a symmetry of the bulk theory, the antipodal map provides the geometric link between the future and past conformal boundaries of dS spacetime. It thereby induces a natural correspondence between the boundary realizations at ${\mathcal I}^+$ and ${\mathcal I}^-$, showing that they furnish equivalent realizations of the same scalar discrete-series representation.

\paragraph{Boundary realization at conformal infinity.} We showed that suitable limits of the bulk modes at the conformal boundaries ${\mathcal I}^{\pm}$ give rise to dS-covariant boundary realizations of the scalar discrete series. Although the boundary theory inherits the physical and gauge sectors of the bulk construction, its indecomposable structure is considerably simpler. In particular, the bulk supplementary modes admit no non-trivial boundary representatives, while the KG negative-definite sector has no independent counterpart at conformal infinity.

As a result, the boundary realization reduces to an indecomposable two-step structure consisting solely of gauge and physical sectors. The corresponding physical quotient continues to carry the UIR $\Pi_{p,0}$, thereby realizing the same physical representation as in the bulk. These results show that the full Krein-Gupta-Bleuler triplet is an intrinsically bulk feature required for a fully dS-covariant spacetime realization, whereas its boundary counterpart retains only the essential structure needed to encode the physical representation.

\paragraph{Bulk-boundary correspondence.} For each $p=1,2,\cdots$, we constructed an explicit bulk-boundary correspondence between the physical sectors realizing the scalar discrete-series representation $\Pi_{p,0}$. The correspondence is implemented by a Fourier-type transform that is simultaneously an isometric isomorphism and an intertwining operator for the dS action, thereby preserving both the invariant inner products and the unitary representation. Consequently, it provides an explicit dS-covariant identification of the bulk and boundary physical realizations, allowing the physical representation encoded in the bulk Krein-Gupta-Bleuler triplet to be reconstructed entirely from its simpler boundary counterpart. Together with the antipodal correspondence between ${\mathcal I}^{+}$ and ${\mathcal I}^{-}$, this yields a complete representation-theoretic holographic framework for the scalar discrete series.

\subsection{Outlook}

\paragraph{Cohomological interpretation of the Gupta-Bleuler structure.} The Krein-Gupta-Bleuler structures established in the present work naturally admit a cohomological interpretation in terms of extensions of representations of the dS group. In this framework, the invariant subspaces, the physical quotient space, and the indecomposable representation structure identified here can be described in terms of the cohomology of group representations, following the general Gupta-Bleuler formalism developed by Araki \cite{Araki-1985-97}; see also \cite{Pierotti-1990-31, Simon-1980}. Such a formulation is expected to provide an intrinsic classification of the non-trivial extension classes underlying the scalar discrete series and to further clarify their interplay with the bulk and boundary realizations developed in the present work. A detailed cohomological treatment, together with its extension to higher-spin discrete-series representations, will be presented in a forthcoming paper \cite{Krein_Coho}.

\paragraph{Higher-spin theories and dS gravity.} The framework developed in this paper suggests several natural directions for further investigation. An immediate extension is the study of higher-spin discrete-series representations, where richer gauge structures and more intricate indecomposable hierarchies are expected to arise. It would also be interesting to determine to what extent the boundary simplification observed here persists beyond the scalar case and whether analogous bulk-boundary correspondences can be established for the corresponding higher-spin theories.

A particularly compelling application concerns dS gravity. As discussed in Section \ref{Sec. Machinery}, the scalar representation $\Pi_{1,0}$ occupies a distinguished position in the representation-theoretic framework of dS-covariant quantum gravity, where it underlies one of the Gupta-Bleuler structures of the graviton field \cite{Pejhan2019}; for a relevant discussion, see Ref. \cite{Gibbons}. From this perspective, the holographic correspondence established in the present work for the family of representations $\Pi_{p,0}$ extends well beyond scalar field theory and becomes directly relevant to the representation-theoretic foundations of dS gravity. In particular, the existence of a reflection-positive holographic realization of $\Pi_{1,0}$ suggests a natural representation-theoretic pathway toward a holographic description of the dS graviton, opening a promising direction for future research.

\paragraph{Quantum field theory and antipodal duality.} The present representation-theoretic analysis is carried out at the level of ``first quantization'' and mode realizations. Extending this construction to a quantum field-theoretic (``second quantization'') framework for the general scalar discrete series remains an important problem; it is worth recalling that the scalar discrete-series representations have often been associated with dS ``tachyonic'' fields, primarily because they do not admit a dS-invariant Hilbert-space quantization \cite{GazeauFoundPhys}. The difficulty is already well known for the minimally coupled scalar field, corresponding to the $p=1$ case, for which the zero mode—corresponding to the gauge sector in the present formulation—obstructs the standard construction of a dS-invariant Fock vacuum \cite{Allen}. Several approaches have been proposed to circumvent this obstruction. One possibility is to relax the full dS-covariance requirement, for instance by retaining invariance only under a subgroup of the dS group, restricting the spacetime region under consideration, or requiring covariance only at the level of the dS Lie algebra rather than under the full group action. In particular, Allen and Follaci constructed a family of $\mathrm{O}(4)$-invariant Hadamard Fock vacua that break full dS invariance \cite{Allen-1987}; for a related construction encompassing the entire scalar discrete series, see Ref. \cite{Bros}, and for the two-dimensional case, see Ref. \cite{Gizem}. This symmetry reduction has a direct counterpart in the present analysis; the mixing responsible for the indecomposable dS structure disappears upon restriction to the maximal compact subgroup $\mathrm{SO}(4)$ (see Appendix \ref{App. Action}), under which the relevant mode sectors remain separately invariant. Kirsten and Garriga instead treated the zero mode separately from the ordinary oscillator sector and obtained a physically admissible state which they identified as dS invariant beyond the standard Fock construction \cite{Kirstenand-1993}. It was subsequently pointed out, however, that this vacuum is not invariant under the full dS group in the usual strong sense \cite{Gazeau2000, Bievre}.

More recently, a minimally coupled compact scalar with periodically identified field space, i.e., an axion, was canonically quantized in Ref. \cite{Letsios2026}. In contrast to the ordinary non-compact scalar considered in the present work, whose zero mode takes values in $\mathbb{R}$, the periodic identification makes the axion zero mode an angular variable on $\mathbb{S}^{1}$. Its quantization therefore yields an independent quantum-mechanical sector, with the full Hilbert space taking the form $\mathcal{H} = L^{2}(\mathbb{S}^{1}) \otimes \mathcal{F}_{\mathrm{osc}}$. Within this framework, the neutral zero-particle state is normalizable and invariant under the full dS group, whereas the non-zero-charge zero-particle states are normalizable but invariant only under the maximal compact subgroup.

A conceptually different route is provided by the Krein-Gupta-Bleuler quantization of the minimally coupled scalar \cite{Gazeau2000, Bievre}. Rather than reducing the symmetry group or modifying the zero-mode sector within a positive-definite quantization, this approach maintains covariance under the full dS group by enlarging the carrier space to a non-degenerate indefinite-metric Krein space carrying an indecomposable dS representation, with the physical Hilbert space canonically recovered through the Gupta-Bleuler quotient. The present work extends the representation-theoretic structure underlying this mechanism to the full scalar discrete series $\Pi_{p,0}$, $p=1,2,\cdots$. As in covariant gauge theories, preservation of dS covariance requires retaining the full indecomposable mode space, including the gauge sector, while the physical Hilbert space is recovered only after the Gupta-Bleuler quotient. Our construction therefore provides the natural one-particle foundation for developing a fully dS-covariant Krein-Gupta-Bleuler quantum field theory for the scalar discrete series with $p>1$.

Moreover, the explicit bulk-boundary correspondence established here suggests that the resulting quantum theory may admit a natural holographic formulation. Understanding the role of this correspondence in a fully dS-covariant quantum field theory, therefore, constitutes an important direction for future investigation.

Finally, the antipodal exchange of the physical and non-physical sectors, $V^\prime \longleftrightarrow \overline{V^\prime}$, revealed by the present construction, appears particularly suggestive. Since antipodally related regions of dS spacetime have been argued to admit complementary particle and antiparticle interpretations in an observer-dependent setting \cite{GazeauEPL}, the emergence of this exchange at the representation-theoretic level raises the possibility that such particle-antiparticle dualities are rooted in intrinsic properties of dS representations. Clarifying this connection within a dS-covariant quantum field theory may provide new insight into both the holographic and quantum aspects of dS spacetime.

\section*{Acknowledgement}

The authors would like to express their sincere gratitude to Petr Siegl (Institut f\"{u}r Angewandte Mathematik, Technische Universit\"{a}t Graz, A-8010 Graz, Austria) and Ahmed Youssef (Universit\'e Paris Cit\'e, CNRS, Astroparticule et Cosmologie, F-75013 Paris, France) for their valuable contributions to the early stages of this work. Maryam Bajalan and Hamed Pejhan are supported by the National Science Fund, Ministry of Education and Science of Bulgaria, under contract KP-06-N92/2. The authors are also grateful to the referee for the careful reading of the manuscript and for the insightful comments, which helped clarify its relation to the existing literature and sharpen the presentation of the main results.

\setcounter{equation}{0} 
\begin{appendix}

\section{Some Useful Relations}\label{Gauss-HF} 

For the convenience of the reader, we collect in this appendix several special-function identities and relations that underlie the calculations presented in the main text. Further details may be found in Refs. \cite{Olver2010, Magnus, Hua, Gradshteyn, Avery}.

\subsection{Gamma Function}\label{A-Gamma}

We collect here three standard identities involving the Gamma function $\Gamma(z)$ that are used throughout the paper:
\begin{align}\label{kha44}
    \Gamma(z+1) = z\, \Gamma(z)\,,
\end{align}
for $z\neq0,-1,-2,\cdots$, and Euler's Beta-function formula:
\begin{align}\label{beta function}
    \frac{\Gamma(z_1)\,\Gamma(z_2)} {\Gamma(z_1+z_2)} = \int_0^1 u^{z_1-1}(1-u)^{z_2-1}\,\mathrm{d}u\,, 
\end{align}
for $\Re(z_1)>0$ and $\Re(z_2)>0$, and Legendre's duplication formula: 
\begin{align}\label{Legendre}
    \Gamma(z) \,\Gamma\left(z+\tfrac{1}{2}\right) = 2^{1-2z}\sqrt{\pi}\,\Gamma(2z)\,,
\end{align}
for $2z\neq0,-1,-2,\cdots$.

\subsection{Hypergeometric Function}\label{A-Hyper-Fun}

Let $a,b,c\in\mathbb{C}$ with $c\notin\{0,-1,-2,\cdots\}$. The hypergeometric function is defined on the disk $|z|<1$ by 
\begin{align}\label{HG-def}
    {}_2F_1(a,\, b;\, c;\, z) = \sum_{j=0}^{\infty} \frac{(a)_j(b)_j}{(c)_j\,j!}\,z^j\,,
\end{align}
and elsewhere by analytic continuation in the distributional sense. Here, $(q)_j$ denotes the Pochhammer symbol,
\begin{align}\label{miriam8}
    (q)_j = 
    \begin{cases}
        1\,, & \text{if } j=0\,, \\[3pt]
        q(q+1)\cdots(q+j-1)\,, & \text{if } j\geqslant 1\,.
    \end{cases}
\end{align}
By definition, the hypergeometric function is symmetric under the interchange of its first two parameters,
\begin{align}
    {}_2F_1(a,\, b;\, c;\, z) = {}_2F_1(b,\, a;\, c;\, z) \,.
\end{align}

If $a$ or $b$ is a non-positive integer, the series terminates and reduces to a polynomial. For instance, if $b\in\{0,-1,-2,\cdots\}$, then $(b)_j=0$ for $j>-b$, and hence
\begin{align}\label{kha20}
    {}_2F_1(a,\, b;\, c;\, z) = \sum_{j=0}^{-b} \frac{(a)_j(b)_j}{(c)_j\,j!}\,z^j\,.
\end{align}

If $\Re(c-a-b)>0$, then
\begin{align}\label{Chu-Vandermonde Identity1}
    {}_2F_1(a,\, b;\, c;\, 1)= \frac{\Gamma(c) \,\Gamma(c-a-b)}{\Gamma(c-a) \,\Gamma(c-b)}\,.
\end{align}

The hypergeometric function satisfies Euler's transformation
\begin{align} \label{Euler's transformation}
     {}_2F_1(a,\,b;\,c;\,z) = (1-z)^{c-a-b} \, {}_2F_1(c-a,\, c-b;\, c;\, z)\,. 
\end{align}

The derivative of the hypergeometric function is
\begin{align}\label{deriv}
    \frac{\mathrm{d}}{\mathrm{d}z}\,{}_2F_1^{}(a,\, b;\, c;\, z)=\frac{ab}{c}\,{}_2F_1^{}(a+1,\,b+1;\,c+1;\,z)\,,
\end{align}
and, for $\Re(c) > \Re(b)>0$ and $|\arg(1-z)|<\pi$, its integral representation is
\begin{align}
    {}_2F_1^{}(a,\, b;\, c;\, z)=\frac{\Gamma(c)}{\Gamma(b) \,\Gamma(c-b)}\int_0^1 \frac{u_{}^{b-1}(1-u)_{}^{c-b-1}}{(1-zu)_{}^{a}}\,\mathrm{d}u\,.\label{Eu-int-Re}
\end{align}

The following differential-contiguous relations also hold
\begin{align} \label{contiguous-plus-app}
    & \frac{b(c-a)}{c}\,\,{}_2F_1(a,b+1;c+1;z)\nonumber\\
    &= b\,\,{}_2F_1(a,\, b;\, c;\, z) - (1-z)\frac{\mathrm{d}}{\mathrm{d}z}\,\,{}_2F_1(a,\, b;\, c;\, z)\,, 
\end{align}
\begin{align} \label{contiguous-minus-app}
    & (c-1)\,\,{}_2F_1(a,b-1;c-1;z)\nonumber\\[3pt]
    & \hspace{2cm} =(c-1-az)\,\,{}_2F_1(a,\, b;\, c;\, z) \nonumber\\[3pt]
    &\hspace{2.3cm}+ z(1-z)\frac{\mathrm{d}}{\mathrm{d}z}\; {}_2F_1(a,\, b;\, c;\, z)\,,
\end{align}
\begin{align} \label{contiguous-plus-app'}
    b\,{}_2F_1(a,b;c;z) + z\frac{\mathrm{d}}{\mathrm{d}z} \, {}_2F_1(a,b;c;z) = b\,{}_2F_1(a,b+1;c;z) \,,
\end{align}
and
\begin{align} \label{contiguous-minus-app'}
    &(b-c+az)\, {}_2F_1(a,b;c;z) + z(z-1) \frac{\mathrm{d}} {\mathrm{d}z}\, {}_2F_1(a,b;c;z) \nonumber\\[3pt]
    &\hspace{2cm} = (b-c)\, {}_2F_1(a,b-1;c;z) \,.
\end{align}

The hypergeometric differential equation is
\begin{align}\label{HG-eq}
    z(1-z)\frac{\mathrm{d}_{}^2\, w}{\mathrm{d}z_{}^2} + \left(c-(a+b+1)z\right)\frac{\mathrm{d}\, w}{\mathrm{d}z} - ab \,w= 0 \,.
\end{align}
This equation admits the following local solutions near $z=0$:
\begin{align}\label{o1}
    {}_2F_1(a,\, b;\, c;\, z)\,,
\end{align}
and
\begin{align}\label{o2}
    z^{1-c}\, {}_2F_1(a-c+1\,,b-c+1;\, 2-c;\, z)\,,
\end{align}
and the following local solutions near $z=1$:
\begin{align}\label{o3} 
    {}_2F_1(a,\, b;\, a+b+1-c;\, 1-z)\,,
\end{align}
and
\begin{align}\label{o4}
    (1-z)^{c-a-b}\, {}_2F_1(c-a,\, c-b;\, c-a-b+1;\, 1-z)\,.
\end{align}
If none of $c$, $c-a-b$, and $a-b$ is an integer, then the two functions in each pair are linearly independent.

\subsection{Gegenbauer Polynomials}\label{A-Ge-pol}

The Gegenbauer polynomials $C_n^\lambda(y)$ are defined by
\begin{align}
    C_n^\lambda(y) = \sum_{k=0}^{\lfloor n/2 \rfloor} (-1)^k \frac{\Gamma(n + \lambda - k)}{k! (n - 2k)!\, \Gamma(\lambda)} (2y)^{n-2k}\,,
\end{align}
where $n\in\mathbb N_0$. When $\lambda>-1/2$, the Gegenbauer polynomials form an orthogonal family on the interval $[-1,1]$. 

The Gegenbauer polynomials satisfy the following three-term recurrence relations:
\begin{align}
    &y\, C_n^\lambda(y) = \frac{n+1}{2(n+\lambda)} C_{n+1}^\lambda(y) + \frac{n+2\lambda-1}{2(n+\lambda)} C_{n-1}^\lambda(y)\,, \label{sec2-321}\\[3pt]
    &C_n^\lambda(y) = \frac{\lambda}{n+\lambda} \left( C_n^{\lambda+1}(y) - C_{n-2}^{\lambda+1}(y) \right)\,, \label{sec2-ne} \\[3pt]
    &(1-y^2) C_n^{\lambda+1}(y) = \frac{(n+2\lambda)(n+2\lambda+1)}{4\lambda(n+\lambda+1)} C_n^\lambda(y) \nonumber\\[3pt]
    &\hspace{2.9cm} - \frac{(n+1)(n+2)}{4\lambda(n+\lambda+1)} C_{n+2}^\lambda(y) \,, \label{miriamTR}
\end{align}
with the convention $C_m^\lambda(y) = 0$ for $m<0$.

\subsection{Associated Legendre Polynomials}\label{Sec. Ass. Legen. Pol.}

For non-negative order $m=0,1,\cdots,l$, with $l\in\mathbb N_0$, the associated Legendre polynomials are
\begin{align}
    P_l^m(x) = \frac{(-1)^{l+m}}{2^l \, l!} \,(1-x^2)^{m/2}\,  \frac{\mathrm{d}^{\,l+m}}{\mathrm{d}x^{\,l+m}} (1-x^2)^l \,,
\end{align}
and, for negative orders $m=-1,\cdots,-l$, with $-l\in\mathbb N$, they are defined by
\begin{align}
    P_l^{-m}(x) = (-1)^m \,\frac{(l-m)!}{(l+m)!} \,P_l^m(x) \,.
\end{align}
The associated Legendre polynomials satisfy
\begin{align} \label{legendre-recurrence}
    &(2l+1)\,x\, P_l^m(x) \nonumber\\[3pt]
    &\qquad = (l-m+1)P_{l+1}^m(x) + (l+m)P_{l-1}^m(x) \,.
\end{align}

\subsection{Hyperspherical Harmonics on $\mathbb{S}_{}^3$}\label{A-Hyp-Har}

The hyperspherical harmonics on $\mathbb{S}^3$ are defined by
\begin{align}
    Y_{Llm}^{}(u) = l!\,2_{}^{l+1}&  \left(\frac{(L+1)(L-l)!}{2\pi (L+l+1)!}\right)_{}^{\frac{1}{2}}(\sin \alpha)_{}^l \nonumber \\
    & \vphantom{\left(\frac{(L+1)(L-l)!}{2\pi (L+l+1)!}\right)^{\frac{1}{2}}} \times  C_{L-l}^{l+1}\left(\cos \alpha\right)\, Y_{lm}^{}\left(\theta, \psi\right)\,,
\end{align}
where $(L, l, m) \in \mathbb{N}_0^{} \times \mathbb{N}_0^{} \times \mathbb{Z}$ with $0 \leqslant l \leqslant L$ and $|m| \leqslant l$, the vector $u = (u_{}^1, u_{}^2, u_{}^3, u_{}^4)\in \mathbb{S}_{}^3 \subset \mathbb{R}_{}^4$ is given in Eq. \eqref{kha48}, $C_{L-l}^{l+1}$ denotes the Gegenbauer polynomial, and $Y_{lm}^{}$ denotes the ordinary spherical harmonic
\begin{align}
    Y_{lm}^{}(\theta, \psi) = \left({\frac{(2l + 1)(l - m)!}{4\pi(l + m)!}}\right)^{\frac{1}{2}} P_l^m(\cos \theta) \, e_{}^{\mathrm{i}m\psi}\,.
\end{align}

The hyperspherical harmonics satisfy the complex-conjugation property
\begin{align} \label{cop}
    \overline{Y_{Llm}(u)}=(-1)^m \, Y_{Ll,-m}(u)\,,
\end{align}
the antipodal symmetry
\begin{align}\label{antipodal-Yllm1}
    Y_{Llm}(-u) = (-1)^L\, Y_{Llm}(u) \,,
\end{align}
and the orthonormality relation
\begin{align}\label{or-re}
    \int_{{\mathbb{S}}^3} \overline{Y_{Llm}^{}(u)}\, Y_{L_{}^\prime l_{}^\prime m_{}^\prime}(u) \, \mathrm{d}u = \delta_{LL_{}^\prime}^{} \,\delta_{ll_{}^\prime} \,\delta_{mm_{}^\prime}\,, 
\end{align}
where $\mathrm{d}u$ denotes the invariant measure on ${\mathbb{S}}^3$ defined in Eq. \eqref{measure}.

\section{Analysis of the Radial Equation \eqref{eqrad}} \label{App. IndeSolu}

\subsection{Solutions of the Radial Equation \eqref{eqrad}}\label{details-on-se.3}

Since `$\cos\rho$' does not vanish on the interval
$\rho\in\left(-\frac{\pi}{2},\frac{\pi}{2}\right)$,
Eq. \eqref{eqrad} may be divided by $\cos^2\rho$, leading to
\begin{align} \label{rad_expanded}
    \frac{\mathrm{d}^2_{}\lambda}{\mathrm{d}\rho^2_{}} &+2\tan\rho\, \frac{\mathrm{d}\lambda}{\mathrm{d}\rho} \nonumber\\[3pt]
    &+ \left(L(L+2) - (p-1)(p+2)\sec^2_{}\rho\right)\lambda = 0\,,
\end{align}
where we recall that $L\in\mathbb{N}_0$, while $p=1,2,\cdots$ is fixed. Introducing the new variable $z:=-e_{}^{-2\mathrm{i}\rho}$, Eq. \eqref{rad_expanded} transforms into
\begin{align}
    \frac{\mathrm{d}_{}^2\lambda}{\mathrm{d}z_{}^2}&+\frac{2}{z(1-z)}\frac{\mathrm{d}\lambda}{\mathrm{d}z}\nonumber\\[3pt]
    &+\left(-\frac{L(L+2)}{4z_{}^2} - \frac{(p-1)(p+2)}{z(1-z)_{}^2}\right)\lambda=0\,.\label{eq:A27}
\end{align}

For later use, we analytically continue the variable $\rho$ into the lower complex half-plane according to
\begin{align}
    \rho \longmapsto \rho-\mathrm{i}\epsilon \,, \quad \epsilon>0 \,,
\end{align}
so that
\begin{align}
    z=-e^{-2\mathrm{i}\rho} \,e^{-2\epsilon} \,.
\end{align}
Consequently, the image of the continued coordinate $z$ lies strictly inside the unit disk,
\begin{align}
    0<|z|=e^{-2\epsilon}<1 \,.
\end{align}
Henceforth, we write $\rho$ for the analytically continued variable $\rho-\mathrm{i}\epsilon$, with $\epsilon>0$, suppressing the dependence on $\epsilon$. This prescription guarantees the convergence of the hypergeometric series appearing below and amounts to extending the dS coordinates into the backward tube; further details may be found in Ref. \cite{Gazeau2022}.

Equation \eqref{eq:A27} possesses regular singular points at $z=0$ and $z=1$ (see, e.g., Ref. \cite[Secs. 5.4-5.7]{Boyce-2021}). To isolate the characteristic behavior near these singularities, we introduce the Frobenius-type ansatz
\begin{align}\label{kha50}
    \lambda(z)=z_{}^\alpha\, (1-z)_{}^\beta\, F(z)\,,
\end{align}
where the exponents $\alpha$ and $\beta$ are chosen from the roots of the indicial equations associated with the singular points $z=0$ and $z=1$, respectively. The factors $z^\alpha$ and $(1-z)^\beta$ capture the leading local behavior of the solutions near the regular singular points $z=0$ and $z=1$, respectively, while $F(z)$ encodes the remaining regular part of the solution.

At the singular point $z=0$, we find
\begin{align}
    \lim_{z\longrightarrow 0}&\,z\frac{2}{z(1-z)}=2\,,\\[3pt]
    \lim_{z\longrightarrow 0}&\,z_{}^2 \left(-\frac{L(L+2)}{4z_{}^2} - \frac{(p-1)(p+2)}{z(1-z)_{}^2}\right) \nonumber\\[3pt]
    &\hspace{1.6cm} = -\frac{L(L+2)}{4}\,.
\end{align}
It follows that the corresponding indicial equation is
\begin{align}\label{indicial}
    r(r-1) + 2 r - \frac{L(L+2)}{4} = 0\,. 
\end{align}
The roots of Eq. \eqref{indicial} are $L/2$ and $-(L+2)/2$. The latter corresponds to the singular behavior $z^{-(L+2)/2}$ at $z=0$ and is therefore discarded. We thus choose $\alpha=L/2$. 

An analogous analysis at $z=1$ yields the indicial roots $1-p$ and $p+2$. We select $\beta=1-p$, since, as will be shown below (see Section \ref{Sec. AdamAshghal}), the alternative choice $\beta=p+2$ does not generate additional linearly independent solutions. Consequently, Eq. \eqref{kha50} reduces to
\begin{align}\label{ka60}
    \lambda(z)=z_{}^{L/2}\, (1-z)_{}^{1-p}\, F(z)\,.  
\end{align}

Substituting Eq. \eqref{ka60} into Eq. \eqref{eq:A27} and simplifying, we obtain
\begin{align}\label{Gauss}
    z(1-z)\frac{\mathrm{d}_{}^2F}{\mathrm{d}z_{}^2} + 
    \left(L+2-(L-2p+2)z\right)\frac{\mathrm{d}F}{\mathrm{d}z} & \nonumber\\[3pt] 
    +\, p(L-p+1)\, F &= 0\,,
\end{align}
which coincides with the hypergeometric equation \eqref{HG-eq}. It follows from the standard theory of the hypergeometric equation that $F(z)$ can be expressed in terms of the fundamental local solutions \eqref{o1}-\eqref{o4}. Taking into account the requirement that the third hypergeometric parameter must satisfy $c\notin \{0,-1,-2,\cdots\}$ (see Appendix \ref{A-Hyper-Fun}), the admissible local solutions are obtained via distributional analytic continuation. In a neighborhood of $z=0$, one finds
\begin{align}\label{o1-1}
    F(z) = {}_2F_1(-p,\, L-p+1;\, L+2;\, z)\,,
\end{align}
whereas in a neighborhood of $z=1$,
\begin{align}\label{o4-4}
    F(z) = (1-z)^{2p+1}\, {}_2F_1(L+p+2,\, p+1;\, 2p+2;\,1-z)\,.
\end{align}

Inserting the admissible local solutions \eqref{o1-1} and \eqref{o4-4} into Eq. \eqref{ka60} yields
\begin{align}
    \lambda^{(1)}_L(z) &:= z_{}^{L/2} \,(1-z)_{}^{1-p}\,{}_2F_1^{}\!\left(-p,\,L-p+1;\,L+2;\,z\right)\,, \label{kha65}\\[3pt]
    \lambda^{(2)}_L(z) &:= z^{L/2}\, (1-z)^{p+2} \nonumber\\[3pt]
    & \;\times {}_2F_1^{}\left(p+1,\,L+p+2;\,2p+2;\,1-z\right)\,.\label{kha65-1}
\end{align}
Finally, upon reverting to the variable $\rho$ through $z=-e^{-2\mathrm{i}\rho}$ and using the identity $1-z=2e^{-\mathrm{i}\rho}\cos\rho$, the solutions obtained above are seen to coincide with Eqs. \eqref{chi_p-1} and \eqref{chi_p-2}, respectively, up to an overall phase factor $\mathrm{i}^{L}$.

\subsubsection{Redundancy of the choice $\beta=p+2$.}\label{Sec. AdamAshghal}

We now examine the alternative choice $\beta=p+2$. In this case, Eq. \eqref{kha50} becomes
\begin{align}\label{Miriam}
    \lambda(z) = z^{L/2}\, (1-z)^{p+2} \,F(z)\,.
\end{align}
Substituting this expression into Eq. \eqref{eq:A27} and proceeding as before, we obtain the hypergeometric equation
\begin{align}\label{123} 
    z(1-z)\frac{\mathrm{d}^2F}{\mathrm{d}z^2} + \left(L+2-(L+2p+4)z\right)\frac{\mathrm{d}F}{\mathrm{d}z} & \nonumber\\[3pt] -\, (p+1)(L+p+2)F &= 0 \,. 
\end{align}

The corresponding admissible local solution in the neighborhood of $z=0$ is
\begin{align}\label{t1}
    F(z) = {}_2F_1(p+1,\, L+p+2;\, L+2;\, z)\,,
\end{align}
while, in the neighborhood of $z=1$, the admissible local solution is
\begin{align}\label{t3}
    F(z) = {}_2F_1(p+1,\, L+p+2;\, 2p+2;\, 1-z)\,,
\end{align}
both understood via distributional analytic continuation.

Substituting Eq. \eqref{t3} into Eq. \eqref{Miriam} yields
\begin{align}
    \lambda(z) &= z^{L/2} \,(1-z)^{p+2} \nonumber\\[3pt]
    & \quad \times {}_2F_1(p+1,\, L+p+2;\, 2p+2;\, 1-z)\,,
\end{align}
which is precisely the solution $\lambda_L^{(2)}(z)$ \eqref{kha65-1}. Likewise, substituting Eq. \eqref{t1} into Eq. \eqref{Miriam} gives
\begin{align}
    \lambda(z) = z^{L/2} \,(1-z)^{p+2} \,
    {}_2F_1(p+1,\, L+p+2;\, L+2;\, z)\,.
\end{align}
Applying Euler's transformation \eqref{Euler's transformation}, one readily verifies that this expression coincides with $\lambda_L^{(1)}(z)$ of Eq. \eqref{kha65}.

We therefore conclude that the choice $\beta=p+2$ yields no additional linearly independent solutions. Rather, it reproduces precisely the same pair of solutions obtained previously from the choice $\beta=1-p$.

\subsection{Linear Independence of the Solutions $\lambda_L^{(1)}(\rho)$ and $\lambda_L^{(2)}(\rho)$}\label{App. Linear Independence of the Two Solutions}

We now show that the two solutions \eqref{chi_p-1} and \eqref{chi_p-2}, equivalently \eqref{kha65} and \eqref{kha65-1}, are linearly independent. To this end, suppose that
\begin{align}\label{khass}
   \mathcal{A}\,\lambda_L^{(1)}(\rho) + \mathcal{B}\,\lambda_L^{(2)}(\rho) = 0 \,,
\end{align}
for all $\rho\in\left(-\tfrac{\pi}{2},\tfrac{\pi}{2}\right)$, where $\mathcal{A}$ and $\mathcal{B}$ are $\rho$-independent coefficients. Then, it suffices to show that Eq. \eqref{khass} necessarily implies $\mathcal{A}=0=\mathcal{B}$.

Using Eqs. \eqref{HG-def} and \eqref{Chu-Vandermonde Identity1}, one finds the following leading asymptotic behavior of the solutions as $\rho\longrightarrow\pm\tfrac{\pi}{2}$:
\begin{align}
    \lim_{\rho\longrightarrow\pm\tfrac{\pi}{2}} \, \lambda_L^{(1)}(\rho) &\;\cong\; C_L^{\,(1;\pm)}\,\left[(\cos\rho)^{1-p}\right]_{\rho\longrightarrow\pm\tfrac{\pi}{2}}\,, \label{asymptotically(1)}\\[3pt]
    \lim_{\rho\longrightarrow\pm\tfrac{\pi}{2}} \,\lambda_L^{(2)}(\rho) &\;\cong\; C_L^{\,(2;\pm)}\,\left[(\cos\rho)^{p+2}\right]_{\rho\longrightarrow\pm\tfrac{\pi}{2}} \,, \label{asymptotically(2)}
\end{align}
where the non-vanishing coefficients are given by
\begin{align}
     C_L^{\,(1;\pm)} &:= 2^{1-p}\,e^{\mp\mathrm{i}(L-p+1)\frac{\pi}{2}}\,
    \frac{\Gamma(L+2)\,\Gamma(2p+1)}{\Gamma(L+p+2)\,\Gamma(p+1)}\,,\\[3pt]
    C_L^{\,(2;\pm)} &:= 2^{p+2}\,e^{\mp\mathrm{i}(L+p+2)\frac{\pi}{2}}\,.
\end{align}

Substituting the asymptotic relations \eqref{asymptotically(1)} and \eqref{asymptotically(2)} into Eq. \eqref{khass}, and then dividing by the dominant factor $(\cos\rho)^{1-p}$ as $\rho\longrightarrow\tfrac{\pi}{2}$, we obtain
\begin{align}\label{lin-in}
    \mathcal{A}\,C_L^{\,(1;+)} + \mathcal{B}\,C_L^{\,(2;+)} \underbrace{\left[(\cos\rho)^{2p+1}\right]_{\rho\longrightarrow\tfrac{\pi}{2}}}_{\cong\; 0} = 0 \,.
\end{align}
Using the fact that $C_L^{(1;+)}\neq 0$, it follows immediately that $\mathcal{A}=0$. Substituting this result back into Eq. \eqref{khass} yields $\mathcal{B}\,\lambda_L^{(2)}(\rho)=0$. Since $\lambda_L^{(2)}(\rho)$ is not identically zero on the interval $\left(-\tfrac{\pi}{2},\tfrac{\pi}{2}\right)$, we must have $\mathcal{B}=0$. Hence, Eq. \eqref{khass} implies $\mathcal{A}=0=\mathcal{B}$, establishing the linear independence of $\lambda_L^{(1)}(\rho)$ and $\lambda_L^{(2)}(\rho)$ on $\left(-\tfrac{\pi}{2},\tfrac{\pi}{2}\right)$.

\subsection{Derivation of Eqs. \eqref{complex(1)} and \eqref{complex(2)}}\label{complex-con-sol}

Since Eq. \eqref{eqrad} has real-valued coefficients, the complex conjugates $\overline{\lambda^{(1)}_L(\rho)}$ and $\overline{\lambda^{(2)}_L(\rho)}$ are themselves solutions whenever $\lambda^{(1)}_L(\rho)$ and $\lambda^{(2)}_L(\rho)$ are. Moreover, since $\left\{ \lambda^{(1)}_L(\rho) \,,\, \lambda^{(2)}_L(\rho) \right\}$ forms a fundamental set of solutions of the second-order differential equation \eqref{eqrad}, every solution of Eq. \eqref{eqrad}, and in particular its complex-conjugate solutions, can be expressed as a linear combination of $\lambda^{(1)}_L(\rho)$ and $\lambda^{(2)}_L(\rho)$. It follows that there exist $\rho$-independent coefficients $\mathcal{A}$, $\mathcal{B}$, $\mathcal{C}$, and $\mathcal{D}$ such that
\begin{align}
    \overline{\lambda^{(1)}_L(\rho)} &= \mathcal{A}\,\lambda^{(1)}_L(\rho) + \mathcal{B}\,\lambda^{(2)}_L(\rho)\,,\label{Li-Cof 1}\\[3pt]
    \overline{\lambda^{(2)}_L(\rho)} &= \mathcal{C}\,\lambda^{(1)}_L(\rho) + \mathcal{D}\,\lambda^{(2)}_L(\rho) \label{Li-Cof 2}\,.
\end{align}

Using the asymptotic expansions \eqref{asymptotically(1)} and \eqref{asymptotically(2)}, we now analyze the behavior of Eq. \eqref{Li-Cof 2} in the limit $\rho\longrightarrow \frac{\pi}{2}$. Substituting these expansions into Eq. \eqref{Li-Cof 2} and dividing both sides by the leading factor $(\cos\rho)^{1-p}$, we obtain
\begin{align}
    &\overline{C_L^{\,(2;+)}}\, \left[(\cos\rho)^{2p+1}\right]_{\rho\longrightarrow \frac{\pi}{2}} \nonumber\\[3pt]
    &\quad\cong \mathcal{C}\, C_L^{\,(1;+)} + \mathcal{D}\, C_L^{\,(2;+)}\, \left[(\cos\rho)^{2p+1}\right]_{\rho\longrightarrow \frac{\pi}{2}}\,.
\end{align}
Since $2p+1>0$, taking the limit $\rho\longrightarrow \tfrac{\pi}{2}$ yields $\mathcal{C}\, C_L^{\,(1;+)}=0$. Since $C_L^{\,(1;+)}\neq 0$, we conclude that $\mathcal{C}=0$. Substituting this result back into Eq. \eqref{Li-Cof 2} and repeating the above asymptotic analysis, we obtain
\begin{align}\label{miriam1}
    \mathcal{D} = \frac{\overline{C_L^{\,(2;+)}}}{C_L^{\,(2;+)}} = e^{\mathrm{i}(L+p+2)\pi} = (-1)^{L+p}\,.
\end{align}
Consequently, Eq. \eqref{Li-Cof 2} simplifies to
\begin{align}\label{te-ra-be}
     \overline{\lambda^{(2)}_L(\rho)} = (-1)^{L+p}\, \lambda^{(2)}_L(\rho)\,,
\end{align}
which, together with Eq. \eqref{cop}, establishes Eq. \eqref{complex(2)}.

Applying the same asymptotic argument to Eq. \eqref{Li-Cof 1}, we similarly find that
\begin{align}
    \overline{C_L^{\,(1;+)}} \cong \mathcal{A}\, C_L^{\,(1;+)}\,+ \mathcal{B}\,C_L^{\,(2;+)}\,\underbrace{\left[(\cos\rho)^{2p+1}\right]_{\rho\longrightarrow\tfrac{\pi}{2}}}_{\cong\; 0} \,.
\end{align}
Then,
\begin{align}\label{B49}
    \mathcal{A} = \frac{\overline{C_L^{\,(1;+)}}}{C_L^{\,(1;+)}} = e^{\mathrm{i}(L-p+1)\pi} = (-1)^{L-p+1}=(-1)^{L+p+1}\,.
\end{align}
It should be noted that the final identity is trivially satisfied. We nevertheless record it in order to make explicit the relation between the coefficients $\mathcal{A}$ \eqref{B49} and $\mathcal{D}$ \eqref{miriam1}; that is, $\mathcal{D} = - \mathcal{A}$. This relation will prove useful when comparing Eqs. \eqref{complex(1)} and \eqref{complex(2)}.

To determine $\mathcal{B}$, we apply the bidirectional derivative \eqref{miriam2} with $\lambda^{(1)}_L(\rho)$ to both sides of Eq. \eqref{Li-Cof 1}. Making use of Eq. \eqref{help-re 41}, together with Eqs. \eqref{kha12-1} and \eqref{mar-mar} for the numerator, and Eqs. \eqref{asy} and \eqref{te-ra-be} for the denominator, we obtain
\begin{align}\label{Ko-1-3}
    \mathcal{B} &= \frac{\overline{\lambda^{(1)}_L}\stackrel{\leftrightarrow}{\partial}_{\rho=0}\lambda^{(1)}_L}{\lambda^{(2)}_L\stackrel{\leftrightarrow}{\partial}_{\rho=0}\lambda^{(1)}_L} \nonumber\\[3pt]
    &=
    \begin{cases}
        (-1)^{L+1} \frac{\Gamma(p+1) \,\Gamma(L+2)}{\Gamma(L-p+1) \,\Gamma(2p+2)} \,, & L\geqslant p\,, \\[7pt] 
        0 \,, & L<p \,,       
    \end{cases}
\end{align}
where $f\stackrel{\leftrightarrow}{\partial}_{\rho=0}g$ denotes the bidirectional derivative $f\stackrel{\leftrightarrow}{\partial}_{\rho}g$ evaluated at $\rho=0$; see Eq. \eqref{miriam3}.

Finally, inserting Eqs. \eqref{B49} and \eqref{Ko-1-3} into Eq. \eqref{Li-Cof 1}, we obtain
\begin{align}\label{miriamNB}
    \overline{\lambda_L^{(1)}(\rho)} \!= \!\!
    \begin{cases}
        (-1)^{L+p+1}\,\lambda_L^{(1)}(\rho) \\[8pt] 
        \;\;\;\; +\, (-1)^{L+1} \frac{\Gamma(p+1) \,\Gamma(L+2)}{\Gamma(L-p+1) \,\Gamma(2p+2)} \,\lambda_L^{(2)}(\rho)\,, \quad & L \geqslant p \,,\\[14pt]
        (-1)^{L+p+1}\, \lambda_L^{(1)}(\rho)\,, \quad &L < p \,.
    \end{cases}
\end{align}
which, together with Eq. \eqref{cop}, immediately yields Eq. \eqref{complex(1)}.

\section{KG Products and Mode Normalization}\label{App. KGNorm}

\subsection{Preliminaries}

Using the definition of the KG sesquilinear form in Eq. \eqref{KG.def}, together with the orthonormality relation \eqref{or-re}, the KG product of the separated modes $f(\rho)Y_{Llm}(u)$ and $g(\rho)Y_{L^\prime l^\prime m^\prime}(u)$ is
\begin{align} \label{KGN-general}
    &\left\langle f(\rho)Y_{Llm}(u) \;,\; g(\rho)Y_{L^\prime l^\prime m^\prime}(u) \right\rangle_{\vphantom{}_{\text{KG}}}\nonumber\\[3pt]
    &\qquad = \mathrm{i}R^2 \int_{\rho=0} \overline{f(\rho)\,Y_{Llm}(u)} \stackrel{\leftrightarrow}{\partial}_{\rho}
    \left( g(\rho)\,Y_{L^\prime l^\prime m^\prime}(u) \right) {\mathrm{d}}u\nonumber\\[3pt]
    &\qquad = \mathrm{i}R^2 \left( \overline{f(\rho)} \stackrel{\leftrightarrow}{\partial}_{\rho=0} g(\rho) \right) \int_{{\mathbb{S}}^3} \overline{Y_{Llm}(u)}\, Y_{L^\prime l^\prime m^\prime }(u)\,{\mathrm{d}}u \nonumber\\[3pt]
    &\qquad = \mathrm{i} R^2 \left(\overline{f(\rho)} \stackrel{\leftrightarrow}{\partial}_{\rho=0} g(\rho) \right) \,\delta_{LL^\prime} \,\delta_{ll^\prime} \,\delta_{mm^\prime}\,, 
\end{align}
where 
\begin{align}\label{miriam3}
    f(\rho) \stackrel{\leftrightarrow}{\partial}_{\rho=0} g(\rho) := \left( f(\rho) \stackrel{\leftrightarrow}{\partial}_{\rho} g(\rho) \right)\Big|_{\rho=0} \,.
\end{align}

Similarly, one finds
\begin{align}\label{KGN-general2}
    &\left\langle \overline{f(\rho)Y_{Llm}(u)} \;,\; \overline{g(\rho)Y_{L^\prime l^\prime m^\prime}(u)}\right\rangle_{\vphantom{}_{\text{KG}}}\nonumber\\[3pt]
    & \qquad = \mathrm{i} R^2 \left( f(\rho) \stackrel{\leftrightarrow}{\partial}_{\rho=0} \overline{g(\rho)} \right) \,\delta_{LL^\prime} \,\delta_{ll^\prime} \,\delta_{mm^\prime} \,.
\end{align}
Furthermore, using Eqs. \eqref{cop} and \eqref{or-re}, we obtain
\begin{align}\label{KGN-general3}
    &\left\langle \overline{f(\rho)Y_{Llm}(u)} \,,\, g(\rho)Y_{L^\prime l^\prime m^\prime}(u)\right\rangle_{\vphantom{}_{\text{KG}}} \nonumber \\[3pt]
    & = \mathrm{i} R^2 (-1)^{m} \left(f(\rho) \stackrel{\leftrightarrow}{\partial}_{\rho=0}g(\rho) \right) \,\delta_{LL^\prime} \,\delta_{ll^\prime} \,\delta_{m^\prime,-m} \,.
\end{align}

\paragraph{Selection rules for KG products.} The relations above impose strong restrictions on the possible non-vanishing KG products. In particular, since the sectors $L\geqslant p$ and $L<p$ correspond to disjoint ranges of the index $L$, all KG pairings between modes belonging to these sectors vanish identically. More generally, a necessary condition for a non-zero KG product is that the angular quantum numbers satisfy the Kronecker-delta constraints appearing in Eqs. \eqref{KGN-general}-\eqref{KGN-general3}. These conditions are not sufficient, however, as certain modes may still have vanishing KG products despite carrying the required quantum numbers. Consequently, in the remainder of this section, we restrict our attention to those pairings that are not excluded by the above selection rules.

\paragraph{Useful identities for the bidirectional derivative.}\label{bid_iden} The following identities hold:
\begin{align}
    \overline{f}\,
    \stackrel{\leftrightarrow}{\partial}_{\rho} f
    &= 2\mathrm{i}\, \Im\!\left(\overline{f}\,\frac{\mathrm{d}f}{\mathrm{d}\rho}\right)\,, \label{help-re 1}\\[2pt]
    f\stackrel{\leftrightarrow}{\partial}_{\rho}g
    &= -g\stackrel{\leftrightarrow}{\partial}_{\rho}f \,, \label{help-re 4}\\[2pt]
    f\stackrel{\leftrightarrow}{\partial}_{\rho}f
    &=0 \,, \label{help-re 41}\\[2pt]
    f\stackrel{\leftrightarrow}{\partial}_{\rho}\overline{g}
    & = \overline{\overline{f}\stackrel{\leftrightarrow}{\partial}_{\rho}g}\,. \label{help-re 45}
\end{align} 

\paragraph{Wronskian.} The Wronskian of two sufficiently regular functions $f(\rho)$ and $g(\rho)$ is defined by
\begin{align}
    W(f\,,\,g)\Big|_{\rho}:= f(\rho)\,\frac{\mathrm{d}g(\rho)}{\mathrm{d}\rho} - \frac{\mathrm{d}f(\rho)}{\mathrm{d}\rho}\,g(\rho) = f(\rho) \,\stackrel{\leftrightarrow}{\partial}_{\rho} g(\rho) \,.
\end{align}

\paragraph{Abel's identity.} Consider the homogeneous linear second-order ordinary differential equation
\begin{align} \label{second-order-lemma}
    \frac{\mathrm{d}^2 y}{\mathrm{d}\rho^2} + A(\rho)\frac{\mathrm{d}y}{\mathrm{d}\rho} + B(\rho)y = 0\,,
\end{align}
defined on an interval $I\subseteq\mathbb{R}$, where $A(\rho)$ and $B(\rho)$ are complex-valued functions. If $f(\rho)$ and $g(\rho)$ are two solutions of Eq. \eqref{second-order-lemma}, then Abel's identity asserts that
\begin{align}\label{Abel's identity}
    W(f\,,\,g)\Big|_{\rho=\rho_0} =  W(f\,,\,g)\Big|_{\rho}\, \exp\left(\int_{\rho_0}^{\rho} A(s)\,{\mathrm{d}}s \right)\,,
\end{align}
for any $\rho,\rho_0\in I$ \cite{Boyce-2021}. Note that
\begin{align}\label{miriam4}
    W(f\,,\,g)\Big|_{\rho=0} = f(\rho) \,\stackrel{\leftrightarrow}{\partial}_{\rho=0} g(\rho) \,.
\end{align}

\paragraph{Application of Abel's identity to KG products.} Equation \eqref{Abel's identity} plays a central role in the evaluation of KG products (see, for instance, Eq. \eqref{KGN-general}). Upon identifying Eq. \eqref{rad_expanded} with the general second-order differential equation \eqref{second-order-lemma}, Eq. \eqref{miriam4} implies that, for instance, the quantity $\overline{\lambda^{(1)}_L}\stackrel{\leftrightarrow}{\partial}_{\rho=0}\lambda^{(2)}_L$ is the Wronskian of $\overline{\lambda^{(1)}_L}$ and $\lambda^{(2)}_L$ evaluated at $\rho=0$. Consequently, Abel's identity allows this quantity to be expressed in terms of the Wronskian evaluated at any other point in the interval $\left(-\tfrac{\pi}{2},\tfrac{\pi}{2}\right)$. This flexibility is particularly useful when the solutions possess a simpler characterization away from $\rho=0$, for instance through their asymptotic behavior near the conformal boundaries $\rho\longrightarrow\pm\tfrac{\pi}{2}$.

\begin{widetext}

\subsection{Derivation of Eqs. \eqref{kha30}-\eqref{miriamDBD}, and \eqref{normalization constants1}}\label{norm-com}

We begin with Eq. \eqref{kha30}. By virtue of the relations \eqref{ModeNaive} and \eqref{KGN-general}, we have
\begin{align}\label{miriam6}
    \left\langle \phi^{(1)}_{L^{}_{\geqslant p}lm} \;,\; \phi^{(1)}_{L^\prime_{\geqslant p}l^\prime_{}m^\prime_{}}\right\rangle_{\vphantom{}_{\text{KG}}} &= \left\langle N^{(1)}_{L^{}_{\geqslant p}} \lambda_{L^{}_{\geqslant p}}^{(1)} Y^{}_{L^{}_{\geqslant p}lm} \;,\; N^{(1)}_{L^{\prime}_{\geqslant p}} \lambda_{L^{\prime}_{\geqslant p}}^{(1)} Y^{}_{L^\prime_{\geqslant p}l^\prime_{}m^\prime_{}}\right\rangle_{\vphantom{}_{\text{KG}}} \nonumber\\[3pt]
    &= \mathrm{i} R^2 \, \left(N^{(1)}_{L^{}_{\geqslant p}}\right)^2 \left( \overline{\lambda_{L^{}_{\geqslant p}}^{(1)}} \stackrel{\leftrightarrow}{\partial}_{\rho=0} \lambda_{L^{}_{\geqslant p}}^{(1)} \right) \,\delta_{LL^\prime} \,\delta_{ll^\prime} \,\delta_{mm^\prime}\,.
\end{align}
Hence, it remains to evaluate $\overline{\lambda_{L^{}_{\geqslant p}}^{(1)}} \stackrel{\leftrightarrow}{\partial}_{\rho=0} \lambda_{L^{}_{\geqslant p}}^{(1)}$, thereby determining the real normalization constant $N^{(1)}_{L^{}_{\geqslant p}}$. Using Eqs. \eqref{help-re 1}, \eqref{deriv}, and \eqref{chi_p-1}, we find
\begin{align}\label{miriam5}
    \overline{\lambda_{L^{}_{\geqslant p}}^{(1)}} \stackrel{\leftrightarrow}{\partial}_{\rho=0} \lambda_{L^{}_{\geqslant p}}^{(1)} &= -\mathrm{i} 2^{3-2p}\, {}_2F_1(-p, L+1-p; L+2; -1)\nonumber \\
    &\times \left( (L-p+1)\, {}_2F_1(-p, L+1-p; L+2; -1) + \frac{2p(L-p+1)}{L+2}\, {}_2F_1(1-p, L+2-p; L+3; -1) \right)\,.
\end{align}

To simplify this expression, we apply Eqs. \eqref{Eu-int-Re} and \eqref{beta function} (recall that $L\geqslant p$). After an appropriate change of integration variable, we obtain
\begin{align}\label{sum-a}
    -\mathrm{i} 2^{3-2p} {}_2F_1^{}(-p,\, L+1-p;\, L+2;\, -1) &= -\mathrm{i} 2^{3-2p}\, \frac{\Gamma(L+2)}{2\, \Gamma(L-p+1)\, \Gamma(p+1)}\int_0^1 v^{\frac{L-p+1}{2}-1}\, (1-v)^p\,\mathrm{d}v\nonumber\\[3pt]
    &= -\mathrm{i} 2^{3-2p}\, \frac{\Gamma(L+2)}{2\,\Gamma(L-p+1)} \,\frac{\Gamma\left(\frac{L-p+1}{2}\right)}{\Gamma\left(\frac{L+p+3}{2}\right)} \,.
\end{align}
Furthermore, applying Eqs. \eqref{Eu-int-Re} and \eqref{kha44}, followed by integration by parts and the use of Eq. \eqref{beta function}, yields
\begin{align}\label{kha11}
    & (L-p+1)\, {}_2F_1(-p, L+1-p; L+2; -1) + \frac{2p(L-p+1)}{L+2}\, {}_2F_1(1-p, L+2-p; L+3; -1) \nonumber\\[3pt]
    &= \frac{\Gamma(L+2)}{\Gamma(L-p+1)\, \Gamma(p+1)}
    \left((L-p+1)\int_0^1 u^{L-p} (1-u)^p (1+u)^p\,\mathrm{d}u + 2p\int_0^1 u^{L-p+1}(1-u)^p(1+u)^{p-1}\,\mathrm{d}u\right) \nonumber\\[3pt]
    &= \frac{\Gamma(L+2)}{\Gamma(L-p+1)\, \Gamma(p+1)}\int_0^1u^{L-p}(1-u)^p\left((L-p+1)(1+u)^p+2pu(1+u)^{p-1}\right)\,\mathrm{d}u \nonumber\\[3pt]
    &= \frac{\Gamma(L+2)}{\Gamma(L-p+1)\, \Gamma(p+1)}\left( \underbrace{\int_0^1(1-u)^p\frac{\mathrm{d}}{\mathrm{d}u}\left(u^{L-p+1}(1+u)^p\right)\,\mathrm{d}u}_{=\, p\int_0^1 u^{L-p+1}(1-u)^{p-1}(1+u)^p\,\mathrm{d}u} \;+\; p\int_0^1 u^{L-p+1}(1-u)^p(1+u)^{p-1}\, \mathrm{d}u\right) \nonumber\\[3pt]
    &= \frac{p\,\Gamma(L+2)}{\Gamma(L-p+1)\, \Gamma(p+1)}\int_0^1 u^{L-p+1}(1-u)^{p-1}(1+u)^{p-1}\left((1+u)+(1-u)\right)\,\mathrm{d}u \nonumber\\[3pt]
    &= \frac{2p\,\Gamma(L+2)}{\Gamma(L-p+1)\, \Gamma(p+1)}\int_0^1 u^{L-p+1}(1-u^2)^{p-1}\,\mathrm{d}u \nonumber\\[3pt]
    &= \frac{p\,\Gamma(L+2)}{\Gamma(L-p+1)\, \Gamma(p+1)}\int_0^1 v^{\frac{L-p+2}{2}-1}(1-v)^{p-1}\,\mathrm{d}v \qquad (v:=u^2) \nonumber\\[3pt]
    &= \frac{\Gamma(L+2)}{\Gamma(L-p+1)}\, \frac{\Gamma\!\left(\frac{L-p+2}{2}\right)}{\Gamma\!\left(\frac{L-p+2}{2}+p\right)}\,.
\end{align}
Substituting Eqs. \eqref{sum-a} and \eqref{kha11} into Eq. \eqref{miriam5}, followed by an application of Eq. \eqref{Legendre}, yields
\begin{align}\label{kha12-1}
    \overline{\lambda_{L^{}_{\geqslant p}}^{(1)}} \stackrel{\leftrightarrow}{\partial}_{\rho=0} \lambda_{L^{}_{\geqslant p}}^{(1)} &= \left(-\mathrm{i} 2^{3-2p} \frac{\Gamma(L+2)}{2\,\Gamma(L-p+1)}\,\,\frac{\Gamma\left(\frac{L-p+1}{2}\right)}{\Gamma\left(\frac{L-p+1}{2}+p+1\right)}\right)\left(\frac{\Gamma(L+2)}{\Gamma(L-p+1)}
    \frac{\Gamma\!\left(\frac{L-p+2}{2}\right)}{\Gamma\!\left(\frac{L-p+2}{2}+p\right)}\right) \nonumber\\[3pt]
    &= -\mathrm{i} 2^{2-2p} \,\frac{\left(\Gamma(L+2)\right)^2}{\left(\Gamma(L-p+1)\right)^2} \,\frac{2^{p-L}\sqrt{\pi}\,\Gamma(L-p+1)}{2^{-L-p-1}\sqrt{\pi}\,\Gamma(L+p+2)} \nonumber\\[3pt]
    &= -8\mathrm{i}\,\frac{\left(\Gamma(L+2)\right)^2}{\Gamma(L-p+1)\, \Gamma(L+p+2)}\,.
\end{align}
\end{widetext}
Combining Eqs. \eqref{kha12-1} and \eqref{miriam6}, we immediately recover Eq. \eqref{kha30}, provided that the normalization constant $N^{(1)}_{L^{}_{\geqslant p}}$ is chosen as in Eq. \eqref{normalization constants1}.

Equation \eqref{kha30'} is obtained by combining the auxiliary relation \eqref{KGN-general2} with Eqs. \eqref{kha12-1} and \eqref{help-re 45}. Equivalently, it follows directly from the identity \eqref{miriamCNK} satisfied by the KG product, together with Eq. \eqref{kha30}. Indeed,
\begin{align}
     \left\langle \overline{\phi^{(1)}_{L^{}_{\geqslant p}lm}} \;,\; \overline{\phi^{(1)}_{L^\prime_{\geqslant p}l^\prime_{}m^\prime_{}}}\right\rangle_{\vphantom{}_{\text{KG}}} &= - \left\langle {\phi^{(1)}_{L^{\prime}_{\geqslant p}l^\prime m^\prime}} \;,\; {\phi^{(1)}_{L^{}_{\geqslant p}lm}}\right\rangle_{\vphantom{}_{\text{KG}}}\nonumber\\[3pt] 
     &= - \delta_{LL_{}^\prime} \,\delta_{ll_{}^\prime} \,\delta_{mm_{}^\prime}\,,
\end{align}
which is precisely Eq. \eqref{kha30'}. Thus, Eq. \eqref{kha30'} is not an independent KG-product relation, but rather a direct consequence of Eq. \eqref{kha30} and the general identity \eqref{miriamCNK}.

By contrast, Eq. \eqref{miriamDBD} does not follow from any previously established KG-product relation. Instead, it is obtained independently from Eqs. \eqref{KGN-general3}, \eqref{help-re 41}, and \eqref{help-re 45}.

\subsection{Derivation of Eqs. \eqref{kha27}-\eqref{thelastone'} and  \eqref{normalization constants2}-\eqref{normalization constants3}}\label{norm-com1}

\subsubsection{Derivation of Eqs. \eqref{kha27}, \eqref{kha27'}, \eqref{normalization constants2}, and \eqref{normalization constants3}}

Proceeding as in the derivation of Eq. \eqref{miriam6}, Eq. \eqref{kha27} takes the explicit form
\begin{align}\label{miriam9}
    &\left\langle \phi^{(1)}_{L^{}_{< p}lm} \;,\; \phi^{(2)}_{L^\prime_{< p}l^\prime_{}m^\prime_{}}\right\rangle_{\vphantom{}_{\text{KG}}} \nonumber\\[3pt]
    &= \left\langle N^{(1)}_{L^{}_{< p}} \lambda_{L^{}_{< p}}^{(1)} Y^{}_{L^{}_{< p}lm} \;,\; N^{(2)}_{L^{\prime}_{< p}} \lambda_{L^{\prime}_{< p}}^{(2)} Y^{}_{L^\prime_{< p}l^\prime_{}m^\prime_{}}\right\rangle_{\vphantom{}_{\text{KG}}} \nonumber\\[3pt]
    &= \mathrm{i} R^2 \, \left(N^{(1)}_{L^{}_{< p}} N^{(2)}_{L^{}_{< p}}\right) \left( \overline{\lambda_{L^{}_{< p}}^{(1)}} \stackrel{\leftrightarrow}{\partial}_{\rho=0} \lambda_{L^{}_{< p}}^{(2)} \right) \,\delta_{LL^\prime} \,\delta_{ll^\prime} \,\delta_{mm^\prime}\,.
\end{align}
Hence, the computation of Eq. \eqref{miriam9} and the determination of the normalization constants $N^{(1)}_{L^{}_{< p}}$ and $N^{(2)}_{L^{}_{< p}}$ reduce to the evaluation of $\overline{\lambda_{L^{}_{<p}}^{(1)}} \stackrel{\leftrightarrow}{\partial}_{\rho=0} \lambda_{L^{}_{<p}}^{(2)}$. Rather than treating this case separately, we consider the more general quantity $\overline{\lambda_L^{(2)}}\stackrel{\leftrightarrow}{\partial}_{\rho=0}\lambda_L^{(1)}$ (for any admissible value of $L$), which encompasses both the present calculation and the derivation of Eq. \eqref{Ko-1-3}. The desired result then follows as a particular case of the general expression obtained below. 

By identifying Eq. \eqref{rad_expanded} with the general second-order differential equation \eqref{second-order-lemma}, Abel's identity \eqref{Abel's identity} takes the form
\begin{align}\label{Va-Az}
    \overline{\lambda_{L}^{(2)}} \stackrel{\leftrightarrow}{\partial}_{\rho=0}\lambda_{L}^{(1)} &= \left( \overline{\lambda_{L}^{(2)}}\stackrel{\leftrightarrow}{\partial}_{\rho}\lambda_{L}^{(1)} \right)\exp\left(\int_{0}^{\rho} 2\tan s\,{\mathrm{d}}s \right)\,,\nonumber\\[3pt]
    & = \left( \overline{\lambda_{L}^{(2)}}\stackrel{\leftrightarrow}{\partial}_{\rho}\lambda_{L}^{(1)} \right) \frac{1}{\cos^2\rho} \,.
\end{align}
This relation holds for every $\rho\in\left(-\tfrac{\pi}{2},\tfrac{\pi}{2}\right)$. It is therefore sufficient to evaluate the right-hand side in the limit $\rho\longrightarrow\tfrac{\pi}{2}$. To this end, making use of Eqs. \eqref{kha44}, \eqref{Chu-Vandermonde Identity1}, and \eqref{deriv}, we obtain
\begin{align}\label{asy}
    \overline{\lambda_{L}^{(2)}} \stackrel{\leftrightarrow}{\partial}_{\rho=0}\lambda_{L}^{(1)} &= \left[ \left( \overline{\lambda_{L}^{(2)}}\stackrel{\leftrightarrow}{\partial}_{\rho}\lambda_{L}^{(1)} \right) \frac{1}{\cos^2\rho} \right]_{\rho\longrightarrow \tfrac{\pi}{2}} \nonumber\\[3pt]
    &\cong 8\mathrm{i}\,(-1)^p \, \frac{\Gamma(2p+2)\, \Gamma(L+2)}{\Gamma(p+1)\, \Gamma(L+p+2)} \,.
\end{align}

On the other hand, Eqs. \eqref{help-re 4} and \eqref{help-re 45} yield
\begin{align}\label{miriam7}
    \overline{\lambda_{L}^{(1)}} \stackrel{\leftrightarrow}{\partial}_{\rho=0} \lambda_{L}^{(2)} = -\,\overline{\left(\overline{\lambda_{L}^{(2)}} \stackrel{\leftrightarrow}{\partial}_{\rho=0}\lambda_{L}^{(1)}\right)} = \overline{\lambda_{L}^{(2)}} \stackrel{\leftrightarrow}{\partial}_{\rho=0}\lambda_{L}^{(1)} \,.
\end{align}
Combining Eqs. \eqref{miriam9}, \eqref{asy}, and \eqref{miriam7} for $L<p$, one recovers Eq. \eqref{kha27}, provided that
\begin{align}\label{miriam10}
    N^{(1)}_{L^{}_{< p}} N^{(2)}_{L^{}_{< p}} = \frac{(-1)^{p+1}}{8 R^2}\, \frac{\Gamma(p+1) \,\Gamma(L+p+2)}{\Gamma(2p+2) \, \Gamma(L+2)} \,.
\end{align}
This requirement agrees with the product of the coefficients specified in Eqs. \eqref{normalization constants2} and \eqref{normalization constants3}. Nevertheless, it constrains only their product, leaving each coefficient individually arbitrary.

In Section \ref{ap-59-norm}, we shall prove that the modes
\begin{align}
    \phi^{(g)}_{L_{<p}^{}lm} &:= \phi^{(1)}_{L_{<p}^{}lm} = N^{(1)}_{L^{}_{< p}}\, \lambda^{(1)}_{L^{}_{< p}}(\rho)\, Y^{}_{L^{}_{< p}lm}(u) \,, \nonumber\\[3pt] \phi^{(s)}_{L_{<p}^{}lm} &:= \phi^{(2)}_{L_{<p}^{}lm} = N^{(2)}_{L^{}_{< p}}\, \lambda^{(2)}_{L^{}_{< p}}(\rho)\, Y^{}_{L^{}_{< p}lm}(u)
\end{align}
are null with respect to the KG sesquilinear form, thereby establishing Eqs. \eqref{thelastone} and  \eqref{thelastone'}. Since vanishing KG norm provides no intrinsic normalization criterion, the freedom left by Eq. \eqref{miriam10} persists.

We therefore exploit this freedom and adopt the conventions \eqref{normalization constants2} and \eqref{normalization constants3}. As will become apparent in the subsequent analysis, this choice is particularly natural from the perspective of the dS group action on the mode space.

Finally, the derivation of Eq. \eqref{kha27'} relies on the identity \eqref{miriamCNK} satisfied by the KG product, together with Eqs. \eqref{complex(1)}, \eqref{complex(2)}, and \eqref{kha27}. Indeed,
\begin{align}
    &\left\langle \phi^{(2)}_{L_{<p}lm} \,,\, \phi^{(1)}_{L^\prime_{<p}l^\prime m^\prime}\right\rangle_{\vphantom{}_{\text{KG}}} \nonumber\\[3pt]
    &\quad = -\left\langle \overline{\phi^{(1)}_{L^\prime_{<p}l^\prime m^\prime}} \,,\, \overline{\phi^{(2)}_{L_{<p}l,m}} \right\rangle_{\vphantom{}_{\text{KG}}} \nonumber\\[3pt]
    &\quad = (-1)^{L+L^\prime+m+m^\prime}\left\langle \phi^{(1)}_{L_{<p}^\prime l^\prime,-m^\prime} \,,\, \phi^{(2)}_{L_{<p}l, -m}\right\rangle_{\vphantom{}_{\text{KG}}} \nonumber\\[3pt]
    &\quad = \delta_{LL^\prime}\, \delta_{ll^\prime}\, \delta_{mm^\prime} \,,
\end{align}
where, in the last step, we used Eq. \eqref{kha27} together with the fact that the prefactor $(-1)^{L+L^\prime+m+m^\prime}$ reduces to unity on the support of the Kronecker deltas. This is precisely Eq. \eqref{kha27'}, showing that Eq. \eqref{kha27'} is not independent of Eq. \eqref{kha27}.

\subsubsection{Derivation of Eqs. \eqref{thelastone} and \eqref{thelastone'}}\label{ap-59-norm}

Proceeding as in the derivation of Eq. \eqref{miriam6}, the proof of Eqs. \eqref{thelastone} and \eqref{thelastone'} reduces to the evaluation of the quantities $\overline{\lambda^{(1)}_{L^{}_{<p}}}\stackrel{\leftrightarrow}{\partial}_{\rho=0}\lambda^{(1)}_{L^{}_{<p}}$ and $\overline{\lambda^{(2)}_{L^{}_{<p}}}\stackrel{\leftrightarrow}{\partial}_{\rho=0}\lambda^{(2)}_{L^{}_{<p}}$, respectively. We begin by collecting the auxiliary results required for their evaluation.

For convenience, let
\begin{align}
    c_j := \frac{(-p)_j\,\left(-(p-L-1)\right)_j}{(L+2)_j \,j!}\,,
\end{align}
where $(q)_j$ is the Pochhammer symbol defined in Eq. \eqref{miriam8}. For $L<p$ and $0\leqslant j\leqslant p-L-1$, one readily finds
\begin{align}
    (-p)_j &= (-1)^j\frac{p!}{(p-j)!} \,, \nonumber\\[3pt]
    \left(-(p-L-1)\right)_j &= (-1)^j\frac{(p-L-1)!}{(p-L-1-j)!} \,, \nonumber\\[3pt]
    (L+2)_j &= \frac{(L+j+1)!}{(L+1)!} \,,\nonumber\\[3pt]
    &\Longrightarrow c_{p-L-1-j} = c_j \,.
\end{align}

Accordingly, with respect to Eq. \eqref{kha20}, we have
\begin{align}\label{hyper-kha}
    & \left(-e^{-2\mathrm{i}\rho}\right)^{p-L-1}{}_2F_1^{}\left(-p,\,L+1-p;\,L+2;\,-e^{2\mathrm{i}\rho}\right) \nonumber\\[3pt]
    &\quad = \sum_{j=0}^{p-L-1} c_j \times \left(-e^{-2\mathrm{i}\rho}\right)^{p-L-1-j} \nonumber\\[3pt]
    &\quad = \sum_{j=0}^{p-L-1} c_{p-L-1-j} \times \left(-e^{-2\mathrm{i}\rho}\right)^{p-L-1-j} \nonumber\\[3pt]
    &\quad = \sum_{s=p-L-1}^0 c_{s} \times \left(-e^{-2\mathrm{i}\rho}\right)^{s} \quad\quad (s:=p-L-1-j \geqslant 0)\nonumber\\[3pt]
    &\quad = {}_2F_1^{} \left(-p,\,L+1-p;\,L+2;\,-e^{-2\mathrm{i}\rho}\right) \,.
\end{align}
From the above result and Eq. \eqref{chi_p-1}, it follows that
\begin{align}\label{lambda1-kha}
    &\lambda_{L<p}^{(1)}(\rho) \nonumber\\[3pt]
    &= 2^{1-p}\, e^{-\mathrm{i}(L-p+1)\rho}\, (\cos\rho)^{1-p}\left(-e^{-2\mathrm{i}\rho}\right)^{p-L-1} \nonumber\\[3pt]
    &\qquad \times {}_2F_1^{}\left(-p,\,L+1-p;\,L+2;\,-e^{2\mathrm{i}\rho}\right) \nonumber\\[3pt]
    & = (-1)^{p-L-1}\, 2^{1-p}\, e^{\mathrm{i}(L-p+1)\rho}\, (\cos\rho)^{1-p} \nonumber\\[3pt]
    &\qquad\times {}_2F_1^{} \left(-p,\,L+1-p;\,L+2;\,-e^{2\mathrm{i}\rho}\right) \nonumber\\[3pt]
    & = (-1)^{p-L-1}\, \overline{\lambda_{L<p}^{(1)}(\rho)} \,.
\end{align}
The above equation, together with Eq. \eqref{help-re 41}, immediately yields
\begin{align}\label{mar-mar}
    \overline{\lambda^{(1)}_{L<p}} \stackrel{\leftrightarrow}{\partial}_{\rho=0}\lambda^{(1)}_{L<p} = 0 \,, 
\end{align}
which is precisely the equality given in Eq. \eqref{thelastone}.

The proof of Eq. \eqref{thelastone'} follows directly from
\begin{align}\label{c24}
    \overline{\lambda_{L<p}^{(2)}} \stackrel{\leftrightarrow}{\partial}_{\rho=0} \lambda_{L<p}^{(2)} = 0 \,,
\end{align}
while Eq. \eqref{c24} itself is an immediate consequence of Eq. \eqref{te-ra-be} together with Eq. \eqref{help-re 41}.

As a concluding remark in this appendix, we emphasize that Eq. \eqref{miriamNB} implies that, for $L<p$,
\begin{align}
    \overline{\lambda_{L<p}^{(1)}(\rho)} = (-1)^{L+p+1}\lambda_{L<p}^{(1)}(\rho) \,.
\end{align}
This is fully consistent with Eq. \eqref{lambda1-kha}, since $(-1)^{L+p+1}=(-1)^{L-p+1}$, the two exponents differing by the even integer $2p$. 

It should be stressed, however, that Eq. \eqref{lambda1-kha} is not merely a reformulation of Eq. \eqref{miriamNB}. Although both equations ultimately express the same conjugation property, Eq. \eqref{lambda1-kha} is established independently and, through Eq. \eqref{mar-mar}, enters directly into the determination of the coefficient $\mathcal{B}$, which forms an essential part of the proof of Eq. \eqref{miriamNB}. Consequently, Eq. \eqref{lambda1-kha} constitutes an intermediate step in the derivation of Eq. \eqref{miriamNB}, and its inclusion is necessary to avoid any appearance of circular reasoning.

The situation is different for Eq. \eqref{c24}. Once Eq. \eqref{te-ra-be} has been established, Eq. \eqref{c24} follows immediately and plays no role in the derivation of Eq. \eqref{te-ra-be}. Thus, unlike Eq. \eqref{lambda1-kha}, which provides an essential intermediate input to the proof of Eq. \eqref{miriamNB}, Eq. \eqref{c24} is simply a corollary of Eq. \eqref{te-ra-be} and does not contribute independently to the argument.

\section{dS Action on Bulk and Boundary Modes and Covariance of the Fourier-Type Transform}\label{App. Action}

This appendix presents the explicit calculations underlying the action of the dS group on the bulk and boundary mode bases introduced in Sections \ref{Sec. Modes} and \ref{Sec. Boundary}, respectively. These computations make transparent the representation-theoretic structure of the corresponding mode spaces, including the emergence of invariant subspaces and indecomposable structures.

To this end, it is sufficient to analyze the action of the non-compact generators $M_{0j}$, $j=1,\cdots,4$. Indeed, the generators $M_{ij}$, $i,j=1,\cdots,4$, span the Lie algebra of the maximal compact subgroup $\mathrm{SO}(4)$ and act irreducibly on each fixed-$L$ hyperspherical-harmonic multiplet. Consequently, they preserve the quantum number $L$ and only mix the degeneracy labels $(l,m)$. Since the representation-theoretic structures of interest are encoded in the way the dS action connects sectors with different values of $L$, no essential information is lost by restricting the analysis to the generators $M_{0j}$.

Furthermore, it is not necessary to treat all four generators $M_{0j}$ independently. The commutation relations
\begin{align}
    \left[M_{ab},M_{cd}\right] = \mathrm{i}\left( \eta_{ac}M_{bd} + \eta_{bd}M_{ac} - \eta_{ad}M_{bc} - \eta_{bc}M_{ad} \right)
\end{align}
show that the generators $M_{0j}$ transform among themselves under the adjoint action of the compact subalgebra $\mathfrak{so}(4)$. Therefore, once the action of a representative pair, say $M_{03}$ and $M_{04}$, has been determined, the action of the remaining generators $M_{01}$ and $M_{02}$ follows from the above commutation relations together with the known $\mathrm{SO}(4)$ action. Accordingly, it suffices to carry out the explicit calculations for $M_{03}$ and $M_{04}$.

Besides revealing the structure of the bulk and boundary realizations, the calculations presented below provide the ingredients needed to establish the covariance of the Fourier-type transform $\mathcal{F}$ introduced in Section \ref{Sec. Fourier}. Specifically, we show that $\mathcal{F}$ intertwines the dS actions on the bulk and boundary mode spaces, thereby furnishing a dS-covariant correspondence between the corresponding realizations.

\subsection{Preliminaries}

\subsubsection{Bulk and Boundary Realizations of the Non-Compact Generators}

In the global conformal coordinates $X=X(\rho,u)$ introduced in Eq. \eqref{confo-coor}, the non-compact generators of the dS group admit, respectively, the following realizations on the bulk and on the conformal boundary obtained in the limit $\rho\longrightarrow \pm\frac{\pi}{2}$:
\begin{align}
    \mbox{bulk}&:\;\; M^{(\rho,u)}_{0j} = -\mathrm{i} \left( u^j \cos\rho\, \partial_\rho + \sin\rho\, D^j \right) \,, \label{miriambulk}\\[3pt]
    \mbox{boundary}&:\;\; M^{(u)}_{0j} = -\mathrm{i} \left( (p-1) u^j + D^j \right) \label{miriamboundary}\,,
\end{align}
where $j = 1, \cdots, 4$,
\begin{align}\label{def.D}
    D^j = (\partial_\alpha u^j) \partial_\alpha + \frac{(\partial_\theta u^j)}{\sin^2\alpha} \partial_\theta + \frac{(\partial_\psi u^j)}{\sin^2\alpha \sin^2\theta} \partial_\psi \,,   
\end{align}
and $u^j = u^j(\alpha,\theta,\psi)$ are defined in Eq. \eqref{kha48}.

\subsubsection{Mode Reduction at $L = p$}

Using Eq. \eqref{miriamNB} with $L=p$, we obtain
\begin{align}
    \lambda^{(2)}_p(\rho) = (-1)^{1+p}\, \frac{\Gamma(2p+2)}{\Gamma(p+1) \,\Gamma(p+2)}\left(\lambda^{(1)}_p(\rho) + \overline{\lambda^{(1)}_p(\rho)}\right) \,.
\end{align}
Therefore, using Eqs. \eqref{ModeNaive} and \eqref{cop}, one finds
\begin{align} \label{second-sol-First}
    \lambda_p^{(2)}(\rho)\,Y_{plm}(u) &= (-1)^{1+p} \frac{\Gamma(2p+2)} {N_p^{(1)}\,\Gamma(p+1)\,\Gamma(p+2)}\nonumber\\
    &\times \left(\phi^{}_{plm}(\rho,u)+(-1)^m\, \overline{\phi^{}_{pl,-m}(\rho,u)}\right) \,.
\end{align}

\subsubsection{Laplace-Beltrami Operator on ${\mathbb{S}}^3$}\label{A-Lap-Belt}

The Laplace-Beltrami operator on ${\mathbb{S}}^3$, with respect to the global conformal coordinates \eqref{confo-coor}, is given by
\begin{align}
    \Delta_3 =& \left(\partial_\alpha\right)^2 + 2 \cot\alpha\; \partial_\alpha \nonumber\\
    & + \frac{1}{\sin^2 \alpha} \left(\left(\partial_\theta\right)^2 + \cot\theta\,\partial_\theta + \frac{1}{\sin^2 \theta}\,\left(\partial_\psi\right)^2 \right)\,.
\end{align}

Given two sufficiently smooth functions $f$ and $g$ on ${\mathbb{S}}^3$, the following identity holds
\begin{align}
\Delta_{3}(fg) &= f\,\Delta_{3}g + g\,\Delta_{3}f \nonumber \\
&+2\left( (\partial_{\alpha}f)(\partial_{\alpha}g) + \frac{(\partial_{\theta}f)(\partial_{\theta}g)}{\sin^{2}\alpha} + \frac{(\partial_{\psi}f)(\partial_{\psi}g)}{\sin^{2}\alpha\,\sin^{2}\theta} \right)\,. \label{LB-product-rule}
\end{align}
Moreover, for all $j = 1,\cdots,4$, we have
\begin{align}\label{delta on u}
    \Delta_3 u^j = -3\,u^j \,.
\end{align}

\subsubsection{Action of $u^4 \mp \mathrm{i}u^3$ on $Y_{Llm}$}\label{sec2-60}

Applying Eq. \eqref{sec2-321}, the product of $u^4$ with a hyperspherical harmonic decomposes as
\begin{align}\label{sec2-13-1}
    u^4 \, Y_{Llm} = \sum_{\sigma=\pm 1} A^{\sigma,0}_{Llm} \, Y_{L+\sigma,l,m} \,,
\end{align}
where the coefficients are given by
\begin{align} \label{eq:coeff-A}
    A^{\sigma,0}_{Llm} = \frac{1}{2}\, \left( \frac{(L-l+r_\sigma)\, (L+l+1+r_\sigma)}{(L+r_\sigma)\, (L+1+r_\sigma)} \right)^{1/2},
\end{align}
with $r_\sigma = (1+\sigma)/2$. 

Equations \eqref{sec2-ne} and \eqref{miriamTR}, together with the standard spherical-harmonic identity derived from \eqref{legendre-recurrence},
\begin{align}\label{cos-theta-Ylm}
    \cos\theta\, Y_{lm}(\theta,\psi) = a_{l+1,m}\, Y_{l+1,m}(\theta,\psi) + a_{lm}\, Y_{l-1,m}(\theta,\psi) \,,
\end{align}
where
\begin{align}
    a_{lm} = \left( \frac{l^2-m^2}{(2l-1)(2l+1)} \right)^{1/2} \,,
\end{align}
implies a decomposition of $u^3\, Y_{Llm}$ into hyperspherical harmonics. Explicitly,
\begin{align}\label{sec2-14}
    u^3\, Y_{Llm} = \sum_{\sigma=\pm 1} \sum_{\tau=\pm 1} B^{\sigma,\tau}_{Llm}\, Y_{L+\sigma,l+\tau,m} \,,
\end{align}
where 
\begin{align} \label{eq:A-sigma-tau}
    & B^{\sigma,\tau}_{Llm} = \frac{\sigma\tau}{2} \nonumber\\
    & \times \sqrt{\frac{(L+\sigma\tau l+c_{\sigma\tau})(L+\sigma\tau l+c_{\sigma\tau}+1)\left((l+s_\tau)^2-m^2\right)}{(L+r_\sigma)(L+1+r_\sigma)(2l+\tau)(2l+\tau+2)}}\,,
\end{align}
with 
\begin{align}
    r_\sigma = \frac{1+\sigma}{2} \,, \quad s_\tau = \frac{1+\tau}{2} \,, \quad c_{\sigma\tau} = \frac{1+2\sigma+\sigma\tau}{2} \,.
\end{align}

Therefore, 
\begin{align}\label{eq:qY}
    &\left(u^4 \mp \mathrm{i}u^3\right)Y_{Llm} \nonumber\\[3pt]
    &= \sum_{\sigma=\pm 1} A^{\sigma,0}_{Llm} Y_{L+\sigma,l,m} \mp \mathrm{i} \sum_{\sigma=\pm 1} \sum_{\tau=\pm 1} B^{\sigma,\tau}_{Llm}Y_{L+\sigma,l+\tau,m}.
\end{align}

\subsubsection{Action of $D^4 \mp \mathrm{i}D^3$ on $Y_{Llm}$.}\label{sec2-61}

By Eqs. \eqref{def.D} and \eqref{LB-product-rule}, we have
\begin{align}\label{eq:Dj-as-grad}
    \left(D^4 \mp \mathrm{i}D^3\right)Y_{Llm} =&\, \frac{1}{2} \Delta_3\left((u^4 \mp \mathrm{i}u^3)\,Y_{Llm}\right) \nonumber\\[3pt]
    - & \,\frac{1}{2} (u^4 \mp \mathrm{i}u^3)\, \Delta_3 Y_{Llm} \nonumber\\[3pt]
    -& \,\frac{1}{2} Y_{Llm}\, \Delta_3(u^4 \mp \mathrm{i}u^3) \vphantom{\Delta_3 \left((u^4 \mp \mathrm{i}u^3)\, Y_{Llm}\right)} \,.
\end{align}
Subsequently, applying Eqs. \eqref{S3-eigenvalue} and \eqref{delta on u}, we get
\begin{align} \label{Dq-Y-final}
    (D^4 \mp \mathrm{i}D^3)Y_{Llm} &= \frac12  \Delta_3\left( (u^4 \mp \mathrm{i}u^3)Y_{Llm} \right) \nonumber\\[3pt]
    & + \frac12 (L(L+2)+3) (u^4 \mp \mathrm{i}u^3)Y_{Llm} \,.
\end{align}
Substituting Eq. \eqref{eq:qY} into the preceding expression and making use once again of Eq. \eqref{S3-eigenvalue} results in
\begin{align}\label{eq:DqY-final}
    &\left(D^4 \mp \mathrm{i}D^3\right)Y_{Llm} \nonumber\\[3pt]
    &=\sum_{\sigma=\pm 1} \left(1-\sigma(L+1)\right) A^{\sigma,0}_{Llm} Y_{L+\sigma,l,m} \nonumber\\[3pt]
    &\quad \mp \mathrm{i} \sum_{\sigma=\pm 1} \sum_{\tau=\pm 1} \left(1-\sigma(L+1)\right) B^{\sigma,\tau}_{Llm} Y_{L+\sigma,l+\tau,m}\,.
\end{align}
The above expansion follows upon using the identity $\sigma^2=1$.

\subsubsection{Useful Recurrence Relations for the Radial Mode Functions}

Using Eqs. \eqref{chi_p-1}, \eqref{contiguous-plus-app}, and \eqref{contiguous-minus-app}, one readily obtains
\begin{align}
    &\cos\rho\,\frac{\mathrm{d}\lambda_L^{(1)}}{\mathrm{d}\rho} -L \sin\rho\,\lambda_L^{(1)} \nonumber\\[3pt]
    &\qquad\quad = -\mathrm{i}\frac{(L-p+1)(L+p+2)}{L+2}\lambda_{L+1}^{(1)}(\rho)\,, \\[3pt]
    &\cos\rho\,\frac{\mathrm{d}\lambda_L^{(1)}}{\mathrm{d}\rho} + (L+2) \sin\rho\,\lambda_L^{(1)} = -\mathrm{i} (L+1) \lambda_{L-1}^{(1)}(\rho) \,,
\end{align}
while Eqs. \eqref{chi_p-2}, \eqref{contiguous-plus-app'}, and \eqref{contiguous-minus-app'} similarly yield
\begin{align}
    &\cos\rho\,\frac{\mathrm{d}\lambda_L^{(2)}}{\mathrm{d}\rho} -L \sin\rho\,\lambda_L^{(2)} = -\mathrm{i}(L+p+2)\lambda_{L+1}^{(2)}(\rho) \,, \\[3pt]
    &\cos\rho\,\frac{\mathrm{d}\lambda_L^{(2)}}{\mathrm{d}\rho} + (L+2) \sin\rho\,\lambda_L^{(2)} = -\mathrm{i}(L-p)\lambda_{L-1}^{(2)}(\rho) \,.
\end{align}

\subsection{dS Action on Bulk and Boundary Modes and Covariance of the Fourier-Type Transform}\label{subsec:ladder}

As noted at the beginning of this appendix, it suffices to determine the dS action through the generators $M_{03}$ and $M_{04}$, or equivalently through the combinations $M_{03}\pm \mathrm{i}M_{04}$, whose expressions in the conformal coordinates \eqref{confo-coor} are given in Eqs. \eqref{miriambulk} and \eqref{miriamboundary}, respectively. Using the preliminary materials established above, we carry this out below, step by step, for the bulk and boundary modes.

\subsubsection{$\mathrm{dS}$ Action on $V^\prime$ and $\overline{V^\prime}$} 

Restricting to the interior physical levels $L>p$, the action of the dS generators on the bulk physical-mode basis of $V^\prime$ \eqref{V'} (see also \eqref{true>=p}) reads
\begin{align}
    &\!\!\!\!\!\!\!\left(M_{03}^{(\rho, u)} \pm \mathrm{i}M_{04}^{(\rho, u)}\right) \phi^{}_{L^{}_{> p}lm}(\rho,u) \nonumber\\[3pt]
    =&\, \left(M_{03}^{(\rho, u)} \pm \mathrm{i} M_{04}^{(\rho, u)}\right) N^{(1)}_{L_{> p}}\, \lambda_{L_{> p}}^{(1)}(\rho)\, Y^{}_{L^{}_{> p}lm}(u) \nonumber\\[3pt]
    =&\, \mp\mathrm{i} \sqrt{(L-p+1)(L+p+2)} \; A^{1,0}_{L^{}_{> p}lm}\;\phi^{}_{L^{}_{> p}+1,l,m} \nonumber\\[3pt]    
    &\, - \sqrt{(L-p+1)(L+p+2)} \; \sum_{\tau=\pm1} B^{1,\tau}_{L^{}_{> p}lm} \, \phi^{}_{L^{}_{> p}+1,l+\tau,m} \nonumber\\[3pt]
    &\, \mp \mathrm{i} \sqrt{(L-p)(L+p+1)} \; A^{-1,0}_{L^{}_{> p}lm}\, \phi^{}_{L^{}_{> p}-1,l,m} \nonumber\\[3pt]
    &\, - \sqrt{(L-p)(L+p+1)} \; \sum_{\tau=\pm1} B^{-1,\tau}_{L^{}_{> p}lm}\, \phi^{}_{L^{}_{> p}-1,l+\tau,m} \,.
\end{align}
It follows immediately that, for $L> p$, the dS action preserves the physical-mode space \eqref{V'}:
\begin{align}
    \left\{\phi^{}_{L^{}_{> p}lm}\right\}_{\mathrm{span}} \;\;\xrightarrow[]{\text{ dS action }}\;\; V^\prime \,.
\end{align}

Restricting to the lowest physical level $L=p$, the action of the dS generators on the corresponding bulk physical modes in $V^\prime$ \eqref{V'} (see also \eqref{true>=p}) reads
\begin{align}
    &\!\!\!\!\!\!\!\left(M_{03}^{(\rho, u)} \pm \mathrm{i}M_{04}^{(\rho, u)}\right) \phi^{}_{plm}(\rho,u) \nonumber\\[3pt]
    =&\, \left(M_{03}^{(\rho, u)} \pm \mathrm{i} M_{04}^{(\rho, u)}\right) N^{(1)}_{p}\, \lambda_{p}^{(1)}(\rho)\, Y^{}_{plm}(u) \nonumber\\[3pt]
    =&\, \sqrt{2p+2} \left(\mp\mathrm{i} \, A^{1,0}_{plm} \;\phi^{}_{p+1,l,m} - \sum_{\tau=\pm1} B^{1,\tau}_{plm} \, \phi^{}_{p+1,l+\tau,m}\right) \nonumber\\[3pt]
    +& \frac{\sqrt{\Gamma(2p+2)}}{2\, \Gamma(p+1)} \! \left(\!\mp \mathrm{i} A^{-1,0}_{plm}\, \phi^{(g)}_{p-1,l,m} - \!\!\sum_{\tau=\pm1}\!\! B^{-1,\tau}_{plm}\, \phi^{(g)}_{p-1,l+\tau,m} \right) .
\end{align}
For the lowest physical level $L=p$, the lowering terms generate modes with $L=p-1$, which belong to the gauge sector $V_g$ \eqref{Vg} (see also \eqref{gauge}). Consequently,
\begin{align}
    \left\{\phi^{}_{plm}\right\}_{\mathrm{span}} \;\;\xrightarrow[]{\text{ dS action }}\;\; V^\prime\oplus V_{\mathrm{g}} \,,
\end{align}
showing that $V^\prime$ is not dS-invariant and thereby verifying Eq. \eqref{mix.action}.

A completely analogous argument applies to the complex-conjugate physical sector $\overline{V^\prime}$. By complex conjugating the above relations, one finds that the dS action preserves $\overline{V^\prime}$ for $L>p$, while the lowest physical level $L=p$ is mapped partly into the conjugate gauge sector $\overline{V_g}$. Thus,
\begin{align}
    \overline{V^\prime} \;\;\xrightarrow[]{\text{ dS action }}\;\; \overline{V^\prime}\oplus \overline{V_g} \,,
\end{align}
and hence $\overline{V^\prime}$ is not dS-invariant. Using Eq. \eqref{barVgs=Vgs}, namely $\overline{V_g}=V_g$, one sees that the same gauge sector appears in the dS extensions of both $V^\prime$ and $\overline{V^\prime}$.

\subsubsection{$\mathrm{dS}$ Action on $H_{p-1}$} 

Similarly, restricting to the interior physical levels $L>p$, the action of the dS generators on the boundary physical-mode basis of $H_{p-1}$ \eqref{eq:Hp-1-def} (see also \eqref{eq:5.78}) reads
\begin{align}
    &\!\!\!\!\!\!\! \left(M_{03}^{(u)} \pm \mathrm{i}M_{04}^{(u)}\right) \psi^{}_{L^{}_{>p}lm}(u) \nonumber\\[3pt]
    =&\, \left(M_{03}^{(u)} \pm \mathrm{i}M_{04}^{(u)}\right)
    \sqrt{\frac{\Gamma(L-p+1)}{\Gamma(L+p+2)}}\,
    Y_{L^{}_{>p}lm}(u) \nonumber\\[3pt]
    =&\, \mp \sqrt{(L-p+1)(L+p+2)}
    \;A^{1,0}_{L^{}_{>p}lm}\,
    \psi^{}_{L^{}_{>p}+1,l,m} \nonumber\\[3pt]
    &\, +\mathrm{i}\sqrt{(L-p+1)(L+p+2)}
    \sum_{\tau=\pm1}
    B^{1,\tau}_{L^{}_{>p}lm}\,
    \psi^{}_{L^{}_{>p}+1,l+\tau,m} \nonumber\\[3pt]
    &\, \pm\sqrt{(L-p)(L+p+1)}
    \;A^{-1,0}_{L^{}_{>p}lm}\,
    \psi^{}_{L^{}_{>p}-1,l,m} \nonumber\\[3pt]
    &\, -\mathrm{i} \sqrt{(L-p)(L+p+1)}
    \sum_{\tau=\pm1}
    B^{-1,\tau}_{L^{}_{>p}lm}\,
    \psi^{}_{L^{}_{>p}-1,l+\tau,m}\,.
\end{align}
It follows immediately that, for $L>p$, the dS action preserves the boundary physical space $H_{p-1}$:
\begin{align}
    \left\{\psi^{}_{L^{}_{>p}lm}\right\}_{\mathrm{span}}
    \;\;\xrightarrow[]{\text{dS action}}\;\;
    H_{p-1}\,.
\end{align}

Restricting to the lowest physical level $L=p$, the action of the dS generators on the corresponding boundary physical modes in $H_{p-1}$ \eqref{eq:Hp-1-def} (see also \eqref{eq:5.78}) reads
\begin{align}
    &\!\!\!\!\!\!\! \left(M_{03}^{(u)} \pm \mathrm{i}M_{04}^{(u)}\right) \psi^{}_{plm}(u) \nonumber\\[3pt]
    =&\, \left(M_{03}^{(u)} \pm \mathrm{i}M_{04}^{(u)}\right)
    \sqrt{\frac{1}{\Gamma(2p+2)}}\;
    Y_{plm}(u) \nonumber\\[3pt]
    =&\, \sqrt{2p+2} \left( \mp A^{1,0}_{plm}\, \psi^{}_{p+1,l,m} + \mathrm{i} \sum_{\tau=\pm1} B^{1,\tau}_{plm}\, \psi^{}_{p+1,l+\tau,m} \right) \nonumber\\[3pt]
    +& \sqrt{\!\frac{2p+1}{\Gamma(2p+1)}} \!\!\left(\!\pm \, A^{-1,0}_{plm}\, Y_{p-1,l,m} -\mathrm{i}\!\! \sum_{\tau=\pm1}\!\! B^{-1,\tau}_{plm}\, Y_{p-1,l+\tau,m} \right) .
\end{align}
For the lowest physical level $L=p$, the lowering terms generate modes with $L=p-1$, which belong to the sector $V_{p-1}$ \eqref{actionVp-1}. Consequently,
\begin{align}
    \left\{\psi^{}_{plm}\right\}_{\mathrm{span}}
    \;\;\xrightarrow[]{\text{ dS action }}\;\;
    H_{p-1}\oplus V_{p-1}\,,
\end{align}
showing that $H_{p-1}$ is not dS-invariant by itself and thereby verifying Eq. \eqref{actionHp-1}.

\subsubsection{Covariance of the Fourier-Type Transform} 

The preceding calculations establish not only the representation-theoretic structure of the bulk and boundary realizations, but also the dS covariance of the Fourier-type transform $\mathcal{F}$ \eqref{Fourier} introduced in Section \ref{Sec. Fourier}. Indeed, the explicit actions of the dS generators on the bulk and boundary mode families are mapped into one another under $\mathcal{F}$, up to an overall phase factor. For $L>p$, this correspondence holds directly at the level of individual physical modes. At the lowest physical level $L=p$, however, the lowering part of the dS action generates gauge modes with $L=p-1$. Consequently, the intertwining property is realized only after passing to the physical quotients $V/V_g$ and $H_{p-1}/V_{p-1}$. In this sense, $\mathcal{F}$ furnishes an intertwining operator between the bulk and boundary realizations of the dS algebra. More precisely, if $\mathcal{F}$, $M_{ab}^{(\rho,u)}$, and $M_{ab}^{(u)}$ denote the operators induced on the quotient spaces, then
\begin{align}
    \mathcal{F}\, M_{ab}^{(\rho, u)} = M_{ab}^{(u)}\, \mathcal{F} \,.
\end{align}
This intertwining structure is summarized in Fig. \ref{fig:intertwining}. 

\begin{figure}[h]
    \begin{center}
    \begin{tikzpicture}[scale=1.0,
        lbl/.style ={font=\small, inner sep=4pt}]
    \node (A) at (0, 3) {$\psi^{}_{L^{}_{\geqslant p}lm}(u)$};
    \node (B) at (4, 3) {$\psi^{\prime}_{L^{}_{\geqslant p}lm}(u)$};
    \node (C) at (0, 0) {$\phi^{}_{L^{}_{\geqslant p}lm}(\rho,u)$};
    \node (D) at (4, 0) {$\phi^{\prime}_{L^{}_{\geqslant p}lm}(\rho,u)$};

    \draw[<->, thick, line width=1pt] (A) -- (B)
        node[lbl, midway, above] {$M^{(u)}_{ab}$};

    \draw[<->, thick, line width=1pt] (C) -- (D)
        node[lbl, midway, below] {$M^{(\rho,u)}_{ab}$};

    \draw[<->, thick, line width=1pt] (A) -- (C)
        node[lbl, midway, left] {$\mathcal{F}$};

    \draw[<->, thick, line width=1pt] (B) -- (D)
        node[lbl, midway, right] {$\mathcal{F}$};

    \end{tikzpicture}
    \end{center}
    \caption{Intertwining of the boundary and bulk realizations of the scalar discrete-series representation $\Pi_{p,0}$ by the Fourier-type transform $\mathcal{F}$. The horizontal arrows denote the action of the dS generators in the respective realizations, and the vertical arrows denote $\mathcal{F}$. The commutativity of the diagram on the physical quotient spaces $V/V_g$ and $H_{p-1}/V_{p-1}$ expresses the dS covariance of $\mathcal{F}$.}
    \label{fig:intertwining}
\end{figure}

\subsubsection{$\mathrm{dS}$ Action on $V_g$}

Restricting attention to the interior gauge levels $L<p-1$, the action of the dS generators on the bulk gauge-mode basis spanning $V_g$ \eqref{Vg} (see also \eqref{gauge}) takes the form
\begin{align}
    &\!\!\!\!\!\!\!\left(M_{03}^{(\rho, u)} \pm \mathrm{i}M_{04}^{(\rho, u)}\right) \phi^{(g)}_{L^{}_{<p-1}lm}(\rho,u) \nonumber\\[3pt]
    =&\, \left(M_{03}^{(\rho, u)} \pm \mathrm{i} M_{04}^{(\rho, u)}\right) N^{(1)}_{L_{<p-1}}\, \lambda_{L_{<p-1}}^{(1)}(\rho)\, Y^{}_{L^{}_{<p-1}lm}(u) \nonumber\\[3pt]
    =&\, \mp\mathrm{i} \frac{(L-p+1)(L+p+2)}{L+2} \; A^{1,0}_{L^{}_{<p-1}lm}\;\phi^{(g)}_{L^{}_{<p-1}+1,l,m} \nonumber\\[3pt]    
    &\, -\! \frac{(L-p+1)(L+p+2)}{L+2} \!\! \sum_{\tau=\pm1} \!\!\! B^{1,\tau}_{L^{}_{<p-1}lm} \, \phi^{(g)}_{L^{}_{<p-1}+1,l+\tau,m} \nonumber\\[3pt]
    &\, \mp \mathrm{i} (L+1) \; A^{-1,0}_{L^{}_{<p-1}lm}\, \phi^{(g)}_{L^{}_{<p-1}-1,l,m} \nonumber\\[3pt]
    &\, - (L+1) \; \sum_{\tau=\pm1} B^{-1,\tau}_{L^{}_{<p-1}lm}\, \phi^{(g)}_{L^{}_{<p-1}-1,l+\tau,m} \,.
\end{align}
It follows immediately that, for $L<p-1$, the dS action preserves the gauge sector $V_g$:
\begin{align}
    \left\{\phi^{(g)}_{L^{}_{<p-1}lm}\right\}_{\mathrm{span}} \;\;\xrightarrow[]{\text{ dS action }}\;\; V_g \,.
\end{align}

Restricting now to the highest gauge level $L=p-1$, the action of the dS generators on the corresponding bulk gauge modes in $V_g$ \eqref{Vg} (see also \eqref{gauge}) reads
\begin{align}
    &\!\!\!\!\!\!\!\left(M_{03}^{(\rho, u)} \pm \mathrm{i}M_{04}^{(\rho, u)}\right) \phi^{(g)}_{p-1,lm}(\rho,u) \nonumber\\[3pt]
    =&\, \left(M_{03}^{(\rho, u)} \pm \mathrm{i} M_{04}^{(\rho, u)}\right) N^{(1)}_{p-1}\, \lambda_{p-1}^{(1)}(\rho)\, Y^{}_{p-1,lm}(u) \nonumber\\[3pt]
    =&\, p \left(\!\mp \mathrm{i} A^{-1,0}_{p-1,lm}\, \phi^{(g)}_{p-2,l,m} - \!\!\sum_{\tau=\pm1}\!\! B^{-1,\tau}_{p-1,lm}\, \phi^{(g)}_{p-2,l+\tau,m} \right) .
\end{align}
For the highest gauge level $L=p-1$, the raising terms vanish identically, while the lowering terms generate only gauge modes with $L=p-2$. Consequently,
\begin{align}
    \left\{\phi^{(g)}_{p-1,lm}\right\}_{\mathrm{span}} \;\;\xrightarrow[]{\text{ dS action }}\;\; V_g \,.
\end{align}
Combining the above result with the analysis of the interior gauge levels $L<p-1$, one obtains Eq. \eqref{gau-ds}, confirming that $V_g$ furnishes a dS-invariant subspace.

\subsubsection{$\mathrm{dS}$ Action on $V_{p-1}$}

Restricting attention to the interior ``gauge'' levels $L<p-1$, the action of the dS generators on the boundary mode basis spanning $V_{p-1}$ \eqref{eq:Vp-1-def} (see also \eqref{gaugeBoundaryModes}) reads
\begin{align}
    &\!\!\!\!\!\!\!\left(M_{03}^{(u)} \pm \mathrm{i}M_{04}^{(u)}\right)\psi^{}_{L^{}_{<p-1}lm}(u)
    \nonumber\\[3pt]
    =&\, \mp(L-p+1)\,A^{1,0}_{L^{}_{<p-1}lm}\,
    \psi^{}_{L^{}_{<p-1}+1,l,m}
    \nonumber\\[3pt]
    &\, +\mathrm{i}(L-p+1)\sum_{\tau=\pm1}
    B^{1,\tau}_{L^{}_{<p-1}lm}\,
    \psi^{}_{L^{}_{<p-1}+1,l+\tau,m}
    \nonumber\\[3pt]
    &\, \pm(L+p+1)\,A^{-1,0}_{L^{}_{<p-1}lm}\,
    \psi^{}_{L^{}_{<p-1}-1,l,m}
    \nonumber\\[3pt]
    &\, -\mathrm{i}(L+p+1)\sum_{\tau=\pm1}
    B^{-1,\tau}_{L^{}_{<p-1}lm}\,
    \psi^{}_{L^{}_{<p-1}-1,l+\tau,m} \,.
\end{align}
It follows immediately that, for $L<p-1$, the dS action preserves the boundary sector:
\begin{align}
    \left\{\psi^{}_{L^{}_{<p-1}lm}\right\}_{\mathrm{span}} \;\;\xrightarrow[]{\text{ dS action }}\;\; V_{p-1} \,.
\end{align}

Restricting now to the highest level $L=p-1$, one finds
\begin{align}
    &\!\!\!\!\!\!\!\left(M_{03}^{(u)} \pm \mathrm{i}M_{04}^{(u)}\right)\psi^{}_{p-1,lm}(u) \nonumber\\[3pt]
    =&\, 2p \left( \pm\, A^{-1,0}_{p-1,lm}\, \psi^{}_{p-2,l,m} - \mathrm{i}\!\!\sum_{\tau=\pm1}\!\! B^{-1,\tau}_{p-1,lm}\, \psi^{}_{p-2,l+\tau,m} \right) \,.
\end{align}
Since the raising terms vanish identically, while the lowering terms generate only boundary modes with $L=p-2$, it follows that
\begin{align}
    \left\{\psi^{}_{p-1,lm}\right\}_{\mathrm{span}} \;\;\xrightarrow[]{\text{ dS action }}\;\; V_{p-1}\,.
\end{align}
Combining this result with the preceding analysis for $L<p-1$, one obtains Eq. \eqref{actionVp-1}, showing that $V_{p-1}$ is invariant under the dS action.

\subsubsection{dS Action on $V_s$}

Restricting to the interior supplementary levels $L<p-1$, the action of the dS generators on the bulk supplementary-mode basis of $V_s$ \eqref{Vs} (see also \eqref{supp}) takes the form
\begin{align}
    &\!\!\!\!\!\!\!\left(M_{03}^{(\rho, u)} \pm \mathrm{i}M_{04}^{(\rho, u)}\right) \phi^{(s)}_{L^{}_{<p-1}lm}(\rho,u) \nonumber\\[3pt]
    =&\, \left(M_{03}^{(\rho, u)} \pm \mathrm{i} M_{04}^{(\rho, u)}\right) N^{(2)}_{L_{<p-1}}\, \lambda_{L_{<p-1}}^{(2)}(\rho)\, Y^{}_{L^{}_{<p-1}lm}(u) \nonumber\\[3pt]
    =&\, \mp\mathrm{i} (L+2) \; A^{1,0}_{L^{}_{<p-1}lm}\;\phi^{(s)}_{L^{}_{<p-1}+1,l,m} \nonumber\\[3pt]    
    &\, - (L+2) \sum_{\tau=\pm1} B^{1,\tau}_{L^{}_{<p-1}lm} \, \phi^{(s)}_{L^{}_{<p-1}+1,l+\tau,m} \nonumber\\[3pt]
    &\, \mp \mathrm{i} \frac{(L+p+1)\, (L-p)}{L+1} \; A^{-1,0}_{L^{}_{<p-1}lm}\, \phi^{(s)}_{L^{}_{<p-1}-1,l,m} \nonumber\\[3pt]
    &\, - \frac{(L+p+1)\, (L-p)}{L+1} \! \sum_{\tau=\pm1} \! B^{-1,\tau}_{L^{}_{<p-1}lm}\, \phi^{(s)}_{L^{}_{<p-1}-1,l+\tau,m} \,.
\end{align}
It follows immediately that, for $L<p-1$, the dS action preserves the supplementary-mode space \eqref{Vs}:
\begin{align}
    \left\{\phi^{(s)}_{L^{}_{<p-1}lm}\right\}_{\mathrm{span}} \;\;\xrightarrow[]{\text{ dS action }}\;\; V_s \,.
\end{align}

Restricting now to the highest supplementary level $L=p-1$, the action of the dS generators on the corresponding bulk supplementary modes in $V_s$ \eqref{Vs} (see also \eqref{supp}) reads
\begin{align}
    &\!\!\!\!\!\!\!\left(M_{03}^{(\rho, u)} \pm \mathrm{i}M_{04}^{(\rho, u)}\right) \phi^{(s)}_{p-1,lm}(\rho,u) \nonumber\\[3pt]
    =&\, \left(M_{03}^{(\rho, u)} \pm \mathrm{i} M_{04}^{(\rho, u)}\right) N^{(2)}_{p-1}\,\lambda_{p-1}^{(2)}(\rho)\,Y^{}_{p-1,lm}(u) \nonumber\\[3pt]
    =& \mp \mathrm{i} \frac{\sqrt{\Gamma(2p+2)}}{2\, \Gamma(p+1)} \, A^{1,0}_{p-1, lm} \left( \phi^{}_{plm} + (-1)^m \overline{\phi^{}_{pl,-m}} \right) \nonumber\\[3pt]
    & - \frac{\sqrt{\Gamma(2p+2)}}{2\, \Gamma(p+1)} \, \sum_{\tau=\pm 1} B^{1,\tau}_{p-1, lm} \left( \phi^{}_{plm} + (-1)^m \overline{\phi^{}_{pl,-m}} \right) \nonumber\\[3pt]
    & \pm 2\mathrm{i} A^{-1,0}_{p-1, lm}\, \phi^{(s)}_{p-2,lm} + 2 \sum_{\tau=\pm 1} \! B^{-1,\tau}_{p-1, lm} \, \phi^{(s)}_{p-2, l+\tau, m}  \,.
\end{align}
For the highest supplementary level $L=p-1$, the raising terms connect the supplementary sector to the physical sectors $V^\prime$ and $\overline{V^\prime}$ through modes with $L=p$, while the lowering terms remain within $V_s$ by generating modes with $L=p-2$. Consequently,
\begin{align}
    \left\{\phi^{(s)}_{p-1,lm}\right\}_{\mathrm{span}} \;\;\xrightarrow[]{\text{ dS action }}\;\; V^\prime \oplus \overline{V^\prime} \oplus V_s \,.
\end{align}
Combining this result with the preceding analysis for $L<p-1$ shows that the dS action maps the supplementary sector into
\begin{align}
    V_s \;\;\xrightarrow[]{\text{ dS action }}\;\; V^\prime \oplus \overline{V^\prime} \oplus V_s \,.
\end{align}  

\end{appendix}


\end{document}